\documentclass[twocolumn]{aastex62}

\usepackage{amsmath}
\usepackage{amstext}
\usepackage{apjfonts}
\usepackage{graphicx}

\usepackage{amssymb}
\usepackage{latexsym}
\usepackage{longtable}

\defcitealias{Narayan16}{N16}
\defcitealias{Narayan19}{N19}
\defcitealias{Calamida19}{C19}

\renewcommand{\deg}{\mbox{$^{\circ}$}}

\shorttitle{Perfecting our set of spectrophotometric standards} 
\shortauthors{Calamida et al.}

\begin{document}

\title{Perfecting our set of spectrophotometric standard DA white dwarfs}

\correspondingauthor{Annalisa Calamida}
\email{calamida@stsci.edu}

\author[0000-0002-0882-7702]{Annalisa Calamida}
\affiliation{Space Telescope Science Institute - AURA, 3700 San Martin Drive, Baltimore, MD 21218, USA}

\author[0000-0001-6685-0479]{Thomas Matheson}
\affiliation{NSF's National Optical-Infrared Astronomy Research Laboratory, 
950 North Cherry Avenue, 
Tucson, AZ 85749, USA}

\author[0000-0002-7157-500X]{Edward W. Olszewski}
\affiliation{Steward Observatory, The University of Arizona, 933 N Cherry Ave, Tucson, AZ, 85719, USA}

\author[0000-0002-6839-4881]{Abhijit Saha}
\affiliation{NSF's National Optical-Infrared Astronomy Research Laboratory, 
950 North Cherry Avenue, 
Tucson, AZ 85749, USA}

\author[0000-0002-5722-7199]{Tim Axelrod}
\affiliation{Steward Observatory, The University of Arizona, 933 N Cherry Ave, Tucson, AZ, 85719, USA}

\author{Clare Shanahan}
\affiliation{Space Telescope Science Institute - AURA, 3700 San Martin Drive, Baltimore, MD 21218, USA}

\author{Jay Holberg}
\affiliation{Lunar and Planetary Laboratory, The University of Arizona, 1629 E University Blvd, Tucson, AZ 85721, USA}

\author[0000-0002-4596-1337]{Sean Points}
\affiliation{Cerro Tololo Inter-American Observatory, Casilla 603, La Serena, Chile}

\author[0000-0001-6022-0484]{Gautham Narayan}
\affiliation{University of Illinois Urbana-Champaign, 1002 W Green St. M/C 221 Urbana, IL 61801 USA}
\affiliation{Center for AstroPhysical Surveys, National Center for Supercomputing Applications, 1205 W. Clark St., Urbana, IL 61801 USA}

\author[0000-0001-7179-7406]{Konstantin Malanchev}
\affiliation{University of Illinois Urbana-Champaign, 1002 W Green St. M/C 221 Urbana, IL 61801 USA}
\affiliation{Sternberg Astronomical Institute, Lomonosov Moscow State University, Universitetsky pr. 13, Moscow 119234 Russia}

\author[0000-0003-1724-2885]{Ryan Ridden-Harper}
\affiliation{University of Canterbury, 20 Kirkwood Avenue, Upper Riccarton, 8041 Christchurch, NZ}

\author[0000-0002-6428-4378]{Nicola Gentile-Fusillo}
\affiliation{European Southern Observatory, Karl Schwarzschild Straße 2, D-85748 Garching, Germany}

\author[0000-0002-9090-9191]{Roberto Raddi}
\affiliation{Universitat Politècnica de Catalunya, Departament de Física, c/ Esteve Terrades 5, 08860 Castelldefels, Spain}

\author[0000-0001-9806-0551]{Ralph Bohlin}
\affiliation{Space Telescope Science Institute - AURA, 3700 San Martin Drive, Baltimore, MD 21218, USA}

\author[0000-0002-4410-5387]{Armin Rest}
\affiliation{Space Telescope Science Institute - AURA, 3700 San Martin Drive, Baltimore, MD 21218, USA}
\affiliation{Department of Physics and Astronomy, Johns Hopkins University, Baltimore, MD 21218, USA}

\author[0000-0001-8816-236X]{Ivan Hubeny}
\affiliation{Steward Observatory, University of Arizona, 933 N Cherry Ave, Tucson, AZ, 85719, USA}

\author[0000-0003-2823-360X]{Susana Deustua}
\altaffiliation{Formerly at Space Telescope Science Institute, 3700 San Martin Drive, Baltimore, MD 21218, USA}
\affiliation{Sensor Science Division, National Institute of Standards and Technology, 100 Bureau Drive, MS 8444, Gaithersburg, MD 20899, USA}

\author{John Mackenty}
\affiliation{Space Telescope Science Institute - AURA, 3700 San Martin Drive, Baltimore, MD 21218, USA}

\author[0000-0003-2954-7643]{Elena Sabbi}
\affiliation{Space Telescope Science Institute - AURA, 3700 San Martin Drive, Baltimore, MD 21218, USA}

\author[0000-0003-0347-1724]{Christopher W. Stubbs}
\affiliation{Department of Physics, Department of Astronomy, Harvard University, 17 Oxford St. Cambridge MA 02138, USA}


\begin{abstract}
We verified for photometric stability a set of DA white dwarfs with \emph{Hubble Space Telescope} magnitudes from the near-ultraviolet 
to the near-infrared and ground-based spectroscopy by using time-spaced observations from the Las Cumbres Observatory network of telescopes. 
The initial list of 38 stars was whittled to 32 final ones which comprise a high quality set of spectrophotometric standards.
These stars are homogeneously distributed around the sky and are all fainter than $r \sim$ 16.5 mag. Their distribution is such that at least two of them would be available to be observed from any observatory on the ground at any time at airmass less than two.
Light curves and different variability indices from the Las Cumbres Observatory data were used to determine the 
stability of the candidate standards. When available, Pan-STARRS1, Zwicky Transient Facility and \emph{TESS} 
data were also used to confirm the star classification. 
Our analysis showed that four DA white dwarfs may exhibit evidence of photometric variability, while a fifth is cooler than our established lower temperature limit, and a sixth star might be a binary.
In some instances, due to the presence of faint nearby red sources, care should be used when observing a few of the spectrophotometric standards with ground-based telescopes.
Light curves and finding charts for all the stars are provided.
\end{abstract}



\section{Introduction}\label{sec:intro}
An era of deep imaging surveys of large areas of the sky has started with 
projects such as the Sloan Digital Sky Survey (SDSS), Pan-STARRS (PS), the Dark Energy Survey (DES), Skymapper, the Asteroid Terrestrial-impact Last Alert System (ATLAS), the All-Sky Automated Survey for Supernovae (ASAS-SN), the Zwicky Transient Facility (ZTF). Missions such as the Vera Rubin Observatory (VRO)
and the Nancy Grace Roman Space Telescope are only a few of years away.
Other facilities such as \emph{Gaia}, \emph{Kepler} and \emph{TESS} report photometry of millions of 
stars to very high internal accuracy. All these projects have their own 
native photometric system, with some of them differing significantly. 
To make the astrophysical information across these missions and surveys commensurate with one another, they must be put on a common photometric system by
relying on a set of calibration references.

Sub-percent global photometric standardization has been challenging in the past,
but is now in high demand for several ongoing scientific studies. 
For instance, photometric calibration is the major source of uncertainty in the use of Type Ia supernovae as probes of the history of cosmic expansion to infer 
the properties of the dark energy \citep{Betoule2014,scolnic2015,stubbs2015,scolnic2019,scolnic2021,brout2021}. 
Experiments that require accurate and reliable photo-redshift determination, 
such as weak lensing tomography and baryonic acoustic oscillation analysis with the
Vera Rubin Observatory \citep{Gorecki2014}, are also limited by systematic uncertainties arising from their relative photometric calibration. 


\begin{deluxetable*}{llllccccccc}
\tabletypesize{\scriptsize}
\tablecaption{\emph{Gaia} DR3 astrometry and photometry for the candidate spectrophotometric standard DA white dwarfs. 
Stars are divided in the Northern and Equatorial and Southern samples and listed in order of increasing $RA$. \label{table:1}}
\tablehead{
\colhead{Star}&
\colhead{Name}&
\colhead{Alternative name}&
\colhead{\emph{Gaia} DR3 ID}&
\colhead{RA\tablenotemark{a}}&
\colhead{DEC\tablenotemark{a}}&
\colhead{$PM_{RA}$}&
\colhead{$PM_{DEC}$}&
\colhead{$G$}&
\colhead{$G_{RP}$}&
\colhead{$G_{BP}$}\\
\colhead{}&
\colhead{}&
\colhead{}&
\colhead{}&
\colhead{(hh:mm:ss.s)}&
\colhead{(dd:mm:ss.s)}&
\colhead{(mas/yr)}&
\colhead{(mas/yr)}&
\colhead{mag}&
\colhead{mag}&
\colhead{mag}
}
\startdata
\multicolumn{11}{c}{Northern and Equatorial DAWDs} \\
\hline                    
 J0103-0020 & WDFS0103-00  & SDSSJ010322.19-002047.7	& 2536159496590552704  & 01:03:22.201 & -00:20:47.800   &   6.196$\pm$0.382  &  -6.550$\pm$0.355 & 19.30 & 19.67 & 19.16 \\
 J0228-0827 & WDFS0228-08  & SDSSJ022817.16-082716.4	& 5176546064064586624  & 02:28:17.183 & -08:27:16.301 &  10.916$\pm$0.783  &   3.151$\pm$0.539 & 19.97 & 20.07 & 19.82 \\
 J0248+3345 & WDFS0248+33  & SDSSJ024854.96+334548.3	& 139724391470489472   & 02:48:54.965 &  33:45:48.244 &   4.093$\pm$0.253  &  -4.759$\pm$0.205 & 18.52 & 18.74 & 18.42 \\
 J0410-0630\tablenotemark{*} &   \ldots     & SDSSJ041053.632-063027.580 & 3196384966004896640 & 04:10:53.641 & -06:30:27.677 &   8.577$\pm$0.279  &  9.719$\pm$0.185  & 18.99 & 19.22 & 19.02  \\
 J0557-1635\tablenotemark{*} &   \ldots     & WD0554-165		 & 2991789869534666240  & 05:57:01.292 & -16:35:12.159 &  -6.747$\pm$0.099  &  4.272$\pm$0.101  & 17.94 & 18.40 & 17.83  \\
 J0727+3214 & WDFS0727+32  & SDSSJ072752.76+321416.1	& 892231562565363072   & 07:27:52.752 &  32:14:16.046 & -13.151$\pm$0.168  &  -6.923$\pm$0.128 & 18.19 & 18.45 & 18.04  \\
 J0815+0731 & WDFS0815+07  & SDSSJ081508.78+073145.7	& 3097940536010212736  & 08:15:08.782 &  07:31:45.775 &   5.519$\pm$0.811  &  -0.190$\pm$0.733 & 19.93 & 20.25 & 19.79  \\
 J1024-0032 & WDFS1024-00  & SDSSJ102430.93-003207.0	& 3830980604624181376  & 10:24:30.912 & -00:32:07.160  & -21.301$\pm$0.388  &  -5.670$\pm$0.590 & 19.08 & 19.23 & 19.00  \\
 J1110-1709 & WDFS1110-17  & SDSSJ111059.42-170954.2	& 3559181712491390208  & 11:10:59.436 & -17:09:54.308 &   5.454$\pm$0.162  &  -8.015$\pm$0.136 & 18.05 & 18.37 & 17.91  \\
 J1111+3956 & WDFS1111+39  & SDSSJ111127.30+395628.0	& 765355922242992000   & 11:11:27.313 &  39:56:28.105 &   2.734$\pm$0.231  &   2.933$\pm$0.255 & 18.64 & 19.07 & 18.48  \\
 J1206+0201 & WDFS1206+02  & SDSSJ120650.504+020143.810 & 3891742709551744640  & 12:06:50.410 &  02:01:42.138 &  -5.061$\pm$0.300  & -23.367$\pm$0.149 & 18.85 & 19.07 & 18.75  \\
 J1214+4538 & WDFS1214+45  & SDSSJ121405.11+453818.5	& 1539041748873288704  & 12:14:05.111 &  45:38:18.626 &   0.278$\pm$0.088  &  13.925$\pm$0.104 & 17.98 & 18.23 & 17.84  \\
 J1302+1012 & WDFS1302+10  & SDSSJ130234.43+101238.9	& 3734528631432609920  & 13:02:34.422 &  10:12:38.717 & -12.856$\pm$0.132  & -16.837$\pm$0.122 & 17.24 & 17.54 & 17.10  \\
 J1314-0314 & WDFS1314-03  & SDSSJ131445.050-031415.588 & 3684543213630134784  & 13:14:45.046 & -03:14:15.685 &  -3.930$\pm$0.404  &  -5.659$\pm$0.265 & 19.31 & 19.74 & 19.25  \\
 J1514+0047 & WDFS1514+00  & SDSSJ151421.27+004752.8	& 4419865155422033280  & 15:14:21.277 &  00:47:52.380 &   4.350$\pm$0.059  & -26.855$\pm$0.053 & 15.88 & 16.11 & 15.77  \\
 J1557+5546 & WDFS1557+55  & SDSSJ155745.40+554609.7	& 1621657158502507520  & 15:57:45.380 &  55:46:09.361 & -11.677$\pm$0.112  & -21.478$\pm$0.126 & 17.69 & 18.04 & 17.53  \\
 J1638+0047 & WDFS1638+00  & SDSSJ163800.360+004717.822 & 4383979187540364288  & 16:38:00.352 &  00:47:17.739 &  -9.171$\pm$0.320  &  -2.737$\pm$0.239 & 19.02 & 19.36 & 18.91  \\
 J1721+2940\tablenotemark{*} &   \ldots	   & SDSSJ172135.97+294016.0	& 4599419007715436928 &  17:21:35.951 &  29:40:16.178 & -20.919$\pm$0.230  &  10.536$\pm$0.26  & 19.60 & 19.50 & 19.69  \\
 J1814+7854 & WDFS1814+78  & SDSSJ181424.075+785403.048 & 2293913930823813888  & 18:14:24.078 &  78:54:03.084 & -10.738$\pm$0.060  &  11.535$\pm$0.057 & 16.74 & 17.03 & 16.61  \\
 J2037-0513\tablenotemark{*} &   \ldots	   & SDSSJ203722.169-051302.964 & 6908492038494775680 & 20:37:22.173  & -05:13:03.023 &   3.118$\pm$0.267  &  -2.000$\pm$0.206 & 19.11 & 19.40 & 19.04 \\
 J2101-0545 & WDFS2101-05  & SDSSJ210150.65-054550.9	& 6910475935427725824  & 21:01:50.667 & -05:45:51.159 &   9.984$\pm$0.218  & -11.694$\pm$0.210 & 18.83 & 19.10 & 18.74  \\
 J2329+0011 & WDFS2329+00  & SDSSJ232941.330+001107.755 & 2644572064644349952  & 23:29:41.321 &  00:11:07.565 &  -7.982$\pm$0.189  & -14.919$\pm$0.162 & 18.29 & 18.42 & 18.24  \\
 J2351+3755 & WDFS2351+37  & SDSSJ235144.29+375542.6	& 2881271732415859072  & 23:51:44.274 &  37:55:42.569 & -16.412$\pm$0.145  &  -9.941$\pm$0.107 & 18.23 & 18.50 & 18.12  \\
\hline                    
\multicolumn{11}{c}{Southern DAWDs} \\
\hline                    
 J0122-3052 & WDFS0122-30 & ATLAS020.503022 & 5028544686500198144 & 01:22:00.725 & -30:52:03.950  &  20.621$\pm$0.14  & -12.303$\pm$0.135 & 18.66 & 19.01 & 18.53  \\
 J0238-3602 & WDFS0238-36 & SSSJ023824     & 4953936951336477440 & 02:38:24.969 & -36:02:23.222   &  57.993$\pm$0.078 &  13.747$\pm$0.119 & 18.24 & 18.39 & 18.19  \\
 J0419-5319\tablenotemark{*} & \ldots& WD0418-534 & 4779427928974390272 & 04:19:24.608  & -53:19:16.659  & -17.587$\pm$0.048 &  27.166$\pm$0.063 & 16.42 & 16.69 & 16.30  \\
 J0458-5637 & WDFS0458-56 & SSSJ045822 & 4764189621230467584 & 04:58:23.133 & -56:37:33.434  & 143.596$\pm$0.118 &  66.486$\pm$0.13  & 17.96 & 18.25 & 17.85  \\
 J0541-1930 & WDFS0541-19 & SSSJ054114 & 2967083052984612736 & 05:41:14.759 & -19:30:38.896  &  19.248$\pm$0.126 & -26.954$\pm$0.142 & 18.43 & 18.61 & 18.35  \\
 J0639-5712 & WDFS0639-57 & SSSJ063941 & 3486471764460448512 & 06:39:41.468 & -57:12:31.164  &  17.513$\pm$0.126 &  43.576$\pm$0.151 & 18.37 & 18.70 & 18.27  \\
 J0757-6049\tablenotemark{*} & \ldots& WD0757-606 & 5484605140287436416 & 07:57:50.637 & -60:49:54.634  &  -4.590$\pm$0.287 &  11.067$\pm$0.223 & 18.95 & 19.15 & 18.89  \\
 J0956-3841 & WDFS0956-38 & SSSJ095657 & 5290720695823013376 & 09:56:57.009 & -38:41:30.269  &  -8.269$\pm$0.084 & -46.075$\pm$0.092 & 18.00 & 18.16 & 17.94  \\
 J1055-3612 & WDFS1055-36 & SSSJ105525 & 5421579652019276160 & 10:55:25.356 & -36:12:14.731  & -21.353$\pm$0.124 &  46.134$\pm$0.119 & 18.20 & 18.45 & 18.12  \\
 J1206-2729 & WDFS1206-27 & WD1203-272 & 5401230062610609920 & 12:06:20.354 & -27:29:40.639  &   3.019$\pm$0.074 &   2.796$\pm$0.081 & 16.67 & 16.93 & 16.54  \\
 J1434-2819 & WDFS1434-28 & SSSJ143459 & 6222123588482712832 & 14:34:59.528 & -28:19:03.295  & -48.559$\pm$0.206 &  18.600$\pm$0.195 & 18.10 & 18.35 & 18.07  \\
 J1535-7724 & WDFS1535-77 & WD1529-772 & 5779908502946006784 & 15:35:45.179 & -77:24:44.832  & -26.881$\pm$0.055 & -43.749$\pm$0.058 & 16.76 & 17.09 & 16.60  \\
 J1837-7002 & WDFS1837-70 & SSSJ183717 & 6431766714636858240 & 18:37:17.906 & -70:02:52.513  &  10.378$\pm$0.072 & -75.989$\pm$0.106 & 17.91 & 18.08 & 17.85  \\
 J1930-5203 & WDFS1930-52 & SSSJ193018 & 6646236009641999488 & 19:30:18.995 & -52:03:46.550  &  21.546$\pm$0.123 & -33.286$\pm$0.102 & 17.67 & 17.94 & 17.55  \\
 J2317-2903 & WDFS2317-29 & WD2314-293 & 2378059688840742912 & 23:17:20.294 & -29:03:21.647  &   3.991$\pm$0.146 &  25.051$\pm$0.196 & 18.53 & 18.81 & 18.44  \\
\enddata
\tablenotetext{a}{Coordinates are from \emph{Gaia} DR3 at epoch $2016.0$ precessed to $J2000.0$, no proper motion applied. To get current coordinates apply the proper motions from 2016 and precess.}
\tablenotetext{*}{This star was excluded from the final network of spectrophotometric standard DAWDs. See text for more details. }
\end{deluxetable*}


A few years ago we started a project, led by A. Saha, to create a network of all-sky spectrophotometric standard DA white dwarfs (DAWDs).
The majority of these stars are fainter than V $\approx$ 18 mag, i.e., bright enough to provide a good signal-to-noise ratio while still avoiding saturation in existing and future deep surveys.
We used Wide Field Camera 3 (WFC3) on the \emph{Hubble Space Telescope} (\emph{HST}) photometry collected in six filters from the near-ultra-violet (NUV) to the near-infrared (NIR) regime, ground-based spectroscopy and hydrogen atmosphere 
white dwarf (WD) models to provide theoretical Spectral Energy Distributions (SEDs) for all the DAWDs. 
19 stars out of 23 candidates (blue stars in Fig.~\ref{fig:allsky}), distributed around the celestial equators and in the Northern hemisphere, were established as standards in \citet[hereafter CA19]{calamida2019} and \citet[hereafter NA19]{narayan2019}.
Their SEDs agree with the multi-band \emph{HST} photometry to better than 1\% (see Fig.~16 in NA19).
These standards are tied to the \emph{HST} photometric system, based on the spectrophotometry of the 
three CALSPEC primary DAWDs\footnote{http://www.stsci.edu/hst/observatory/crds/calspec.html}
\citep{bohlin2014b, bohlin2020}. 

More recently, we collected ground-based spectroscopy, presented in the current paper, for candidate WDs in the Southern celestial hemisphere;
on the basis of their spectra, 15 of them were selected to be subsequently observed with \emph{HST}\footnote{GO program 15113 (PI: Saha)}.
A companion paper describes how 13 of these DAWDs will be established as spectrophotometric standards (Axelrod et al.\ 2022, in prep.). 

Our final goal is providing an all-sky network of spectrophotometric standards so that at least two of these stars will be visible at any time from any observatory at an airmass less than two.
The distribution of the established standards on the sky (19 in Northern hemisphere and around the celestial equators, and 13 in the 
Southern hemisphere) is shown in the Hammer-Aitoff projection of Fig.~\ref{fig:allsky}.

In addition to verifying consistency of the spectroscopy and HST photometry with WD atmosphere models, we have monitored all the 
candidate spectrophotometric standard DAWDs for variability by collecting time-spaced data with the Las Cumbres Observatory (LCO) 
network of telescopes\footnote{Proposals LCO2016B-007, LCO2017AB-002, LCO2018A-002, LCO2018B-001, LCO2019-B004 (PI: Matheson).}
In CA19 we already showed how these data allowed us to identify two candidate standard DAWDs in the Northern hemisphere, namely 
SDSSJ203722.169-051302.964 and WD0554-165, as not stable. These were then excluded from our network of standards (CA19, NA19).
This manuscript will illustrate the detailed analysis of the photometric monitoring data collected with LCO for all the candidate 
spectrophotometric standard DAWDs.

The structure of the paper is as follows. In \S 2 we present our sample of spectrophotometric standard DAWDs and in \S 3 we illustrate the time-monitoring observations and the 
photometric reduction procedures.
In \S 4 we describe the variability analysis and \S 5 lists the details for each of the DAWDs. We summarize the results in \S 6. The Appendix shows the light curves for all the DAWDs observed with LCO and finding charts based on \emph{HST} NIR images.

\section{The candidate spectrophotometric standard stars}\label{sec:cand}
\begin{figure*}
\hspace{1.0cm}
\includegraphics[width=0.9\textwidth]{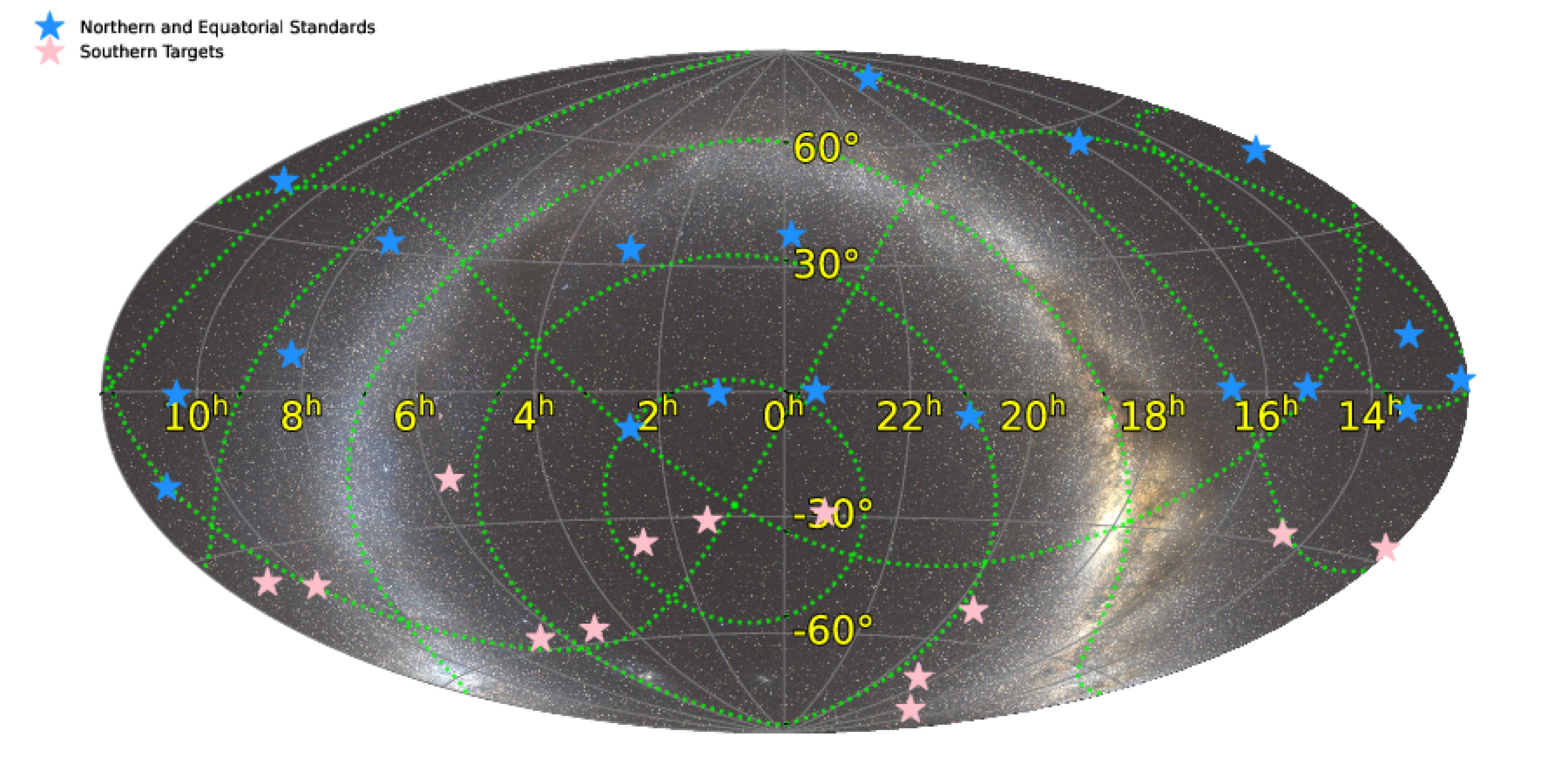} 
\caption{Hammer--Aitoff projection of our network of spectrophotometric standard DAWDs. This includes
19 DAWDs distributed in the Northern hemisphere and around the celestial equators (blue stars) 
and 13 in the Southern hemisphere (pink). \label{fig:allsky} }
\end{figure*}

DAWDs were selected from the SDSS \citep{Adelman-McCarthy2008, girven2012, kleinman2013} 
and the Villanova catalog \citep{mccook1999} to uniformly cover the sky around the celestial equators and in the Northern hemisphere (Fig.~\ref{fig:allsky}).
For more details on the selection of these stars we refer the reader to CA19.
In the Southern hemisphere, there are relatively few faint WDs with prior spectroscopic identifications. Therefore, we relied on published lists of probable WDs based on color and proper motions selection such as those of \citet{gentilefusillo2017} and \citet{raddi2016, raddi2017}, to establish candidate stars with suitable brightness and celestial placement. These works used photometry collected with the Very Large Telescope Survey Telescope (VST) ATLAS and VPHAS$+$ surveys and proper motions from the Absolute Proper motions Outside the Plane \citep[APOP]{qi2015} catalogue to identify WD candidates. For more details on the selection process please see the aforementioned manuscripts.

The original sample for the Northern hemisphere consisted of 23 DAWDs
for which we presented the analysis of the \emph{HST} photometry and ground-based spectroscopy in CA19. 
Of these, 19 DAWDs were established as standards in NA19.
For the Southern hemisphere, we collected spectroscopy for a sample of 48 candidate DAWDs and selected 
15 of them to be observed with \emph{HST} and LCO. 13 of these will be established as standards in Axelrod et al..

Table~\ref{table:1} lists the 38 (23 + 15) DAWDs of the original sample, with \emph{HST}, ground-based spectra and 
LCO observations, and provides their name (with the $J$ designation plus the 4 digits of \emph{Gaia} 
Data Release 3 (DR3, \emph{Gaia} Collaboration et al.\ 2022) $RA$ and $DEC$ coordinates at epoch $2016.0$ precessed to $J2000.0$), 
a new name for the 32 (19 + 13) stars established as spectrophotometric standards, the alternative name, and the IDs, 
coordinates, proper motions and photometry from \emph{Gaia} DR3.
\emph{Gaia} magnitudes are provided to illustrate the brightness range of our candidate spectrophotometric standards. 
A discussion on the comparison between \emph{Gaia} DR3 magnitudes and synthetic \emph{Gaia} magnitudes derived for our entire sample of DAWDs is deferred to Axelrod et al..

We assigned new names to the 32 established spectrophotometric standard DAWDs: these are composed of the {\it White Dwarf Flux Standard} 
(WDFS) designation and 4 digits of the $RA$ and 2 of the $DEC$ coordinates from \emph{Gaia} DR3 as 
their ID number. The new names are listed in the third column of Table~\ref{table:1}.

In the Appendix we provide finding charts for all the 38 DAWDs. These are based on \emph{HST} images collected 
in the $F160W$ filter during our observing programs.
We selected this filter since some possibly contaminant faint red sources become
visible in the infrared regime.

\subsection{Spectroscopy of the candidate spectrophotometric DAWDs}
Spectra of the Northern sample of DAWDs were collected with the Gemini Multi-Object
Spectrograph \citep[GMOS,][]{hook04} mounted on the Gemini North and South telescopes.
However, due to issues with the quality of the GMOS spectra, further observations were
collected with the Blue Channel spectrograph at the MMT Observatory. Details on these
data and their reduction and analysis were presented in CA19.

Spectra of 48 Southern candidate WDs were obtained with the 
Goodman spectrograph \citep{clemens04} on the 4m-SOAR telescope (NOIRLab). Exposures were 
collected for each star and some were observed multiple times between 2016 February
and 2017 February. The log of the observations 
is provided in Table~\ref{table:2} and includes the star name (with the $J$ designation plus the 
4 digits of \emph{Gaia} DR3 $RA$ and $DEC$), the alternative name, \emph{Gaia} DR3 IDs and coordinates.

All the spectra were visually inspected and 
non-DA WD stars and stars with obvious spectroscopic peculiarities 
or magnetic activity were rejected. We ended up with 15 candidate spectrophotometric standard DAWDs that we observed with WFC3/\emph{HST} and monitored for stability with LCO.

\begin{figure*}
\includegraphics[height=0.95\textheight,width=\textwidth]{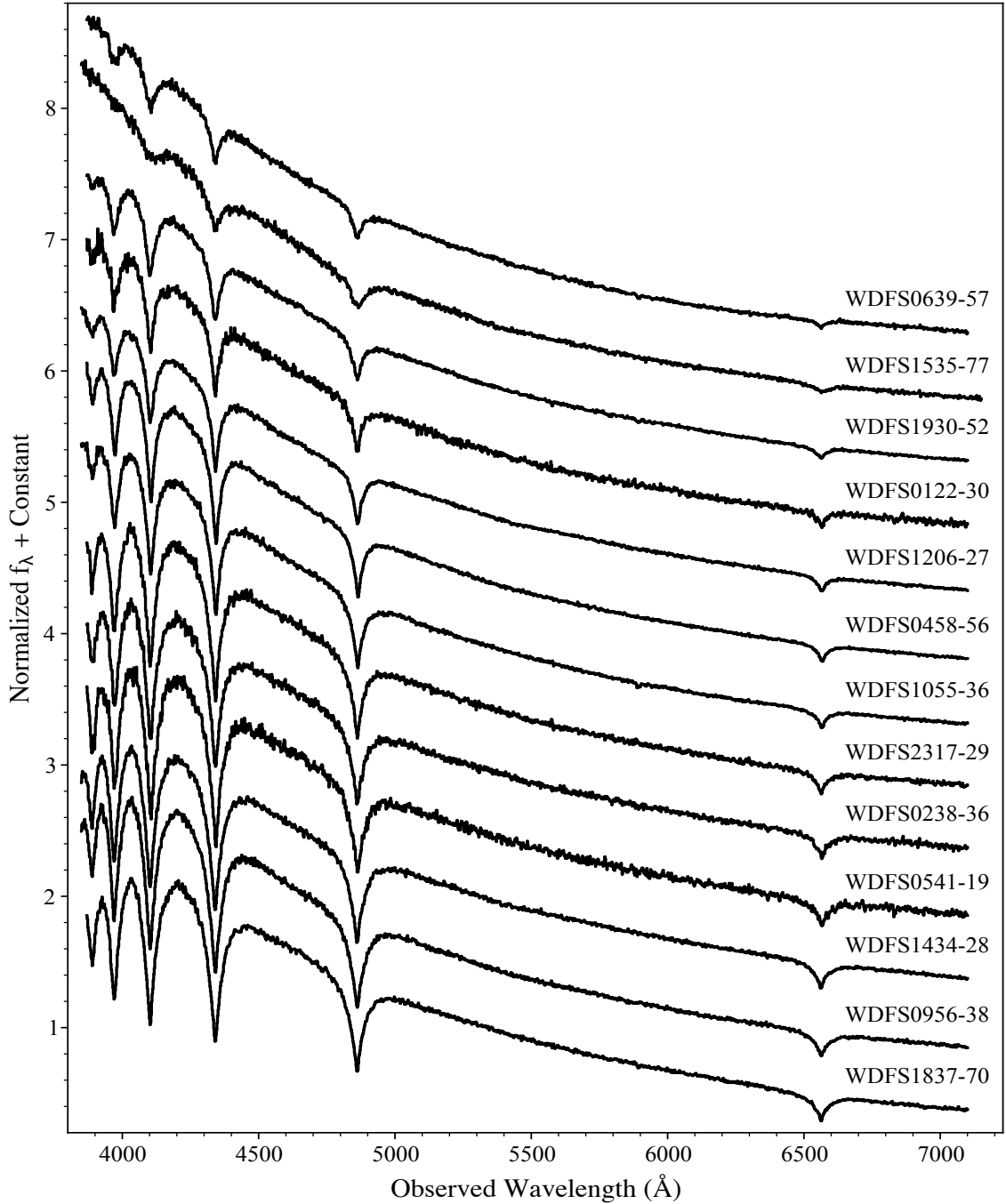} 
\caption{Spectra collected with the Goodman spectrograph on the 4m-SOAR telescope (NOIRLab) for the 13 confirmed spectrophotometric standard DAWDs with \emph{HST} and LCO follow-up observations. \label{fig:good}}
\end{figure*}

\begin{figure*}
\includegraphics[height=0.95\textheight,width=\textwidth]{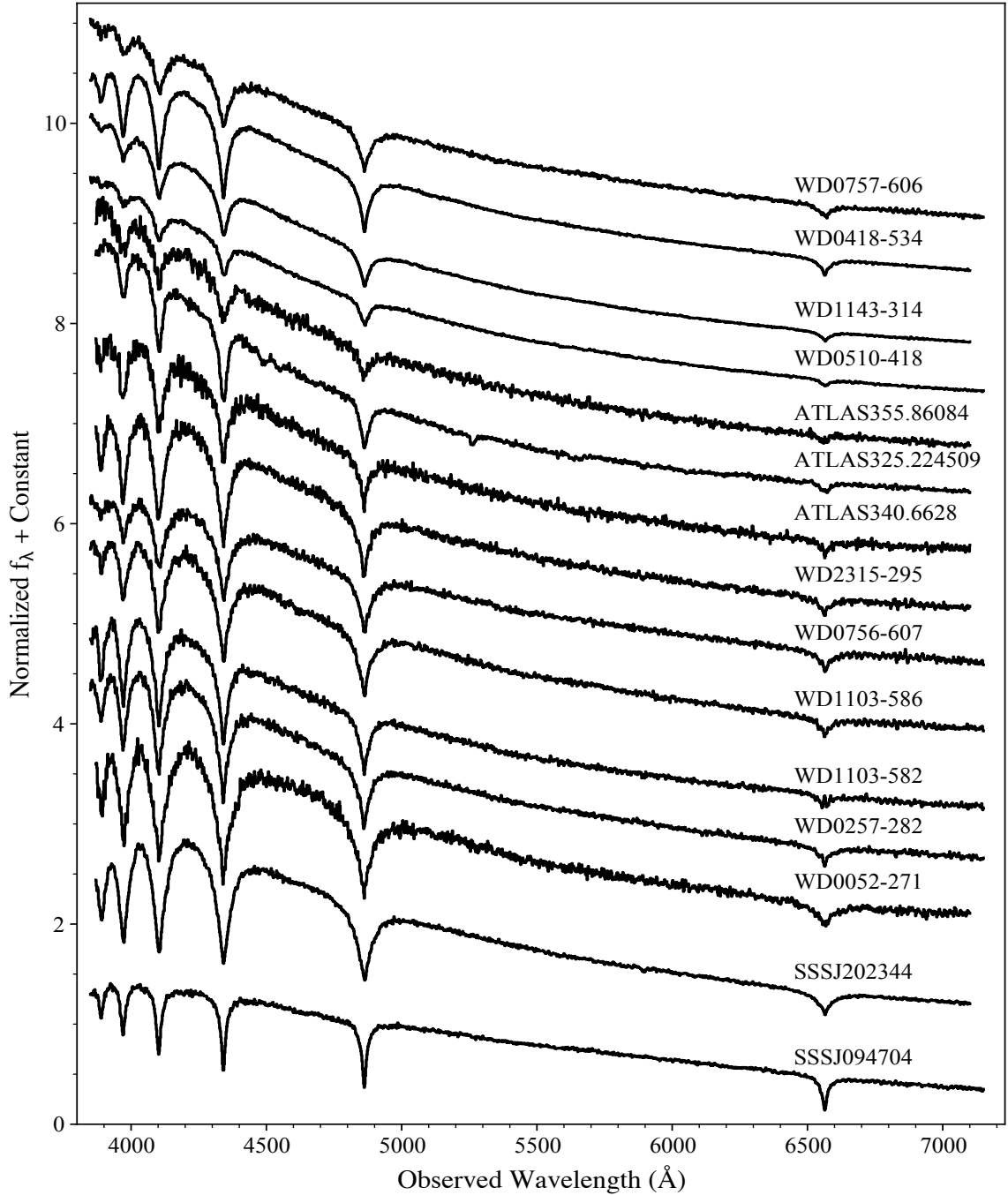} 
\caption{Spectra collected with the Goodman spectrograph on the 4m-SOAR telescope (NOIRLab) for candidate DAWDs that were discarded from the Southern hemisphere sample (continues in Fig.~\ref{fig:other2}.) 
The first and second star, WD0756-607 and WD0418-534, were excluded from out network of standards due to their variability and other reasons. 
See text for more details. \label{fig:other1}}
\end{figure*}

\begin{figure*}
\includegraphics[height=0.95\textheight,width=\textwidth]{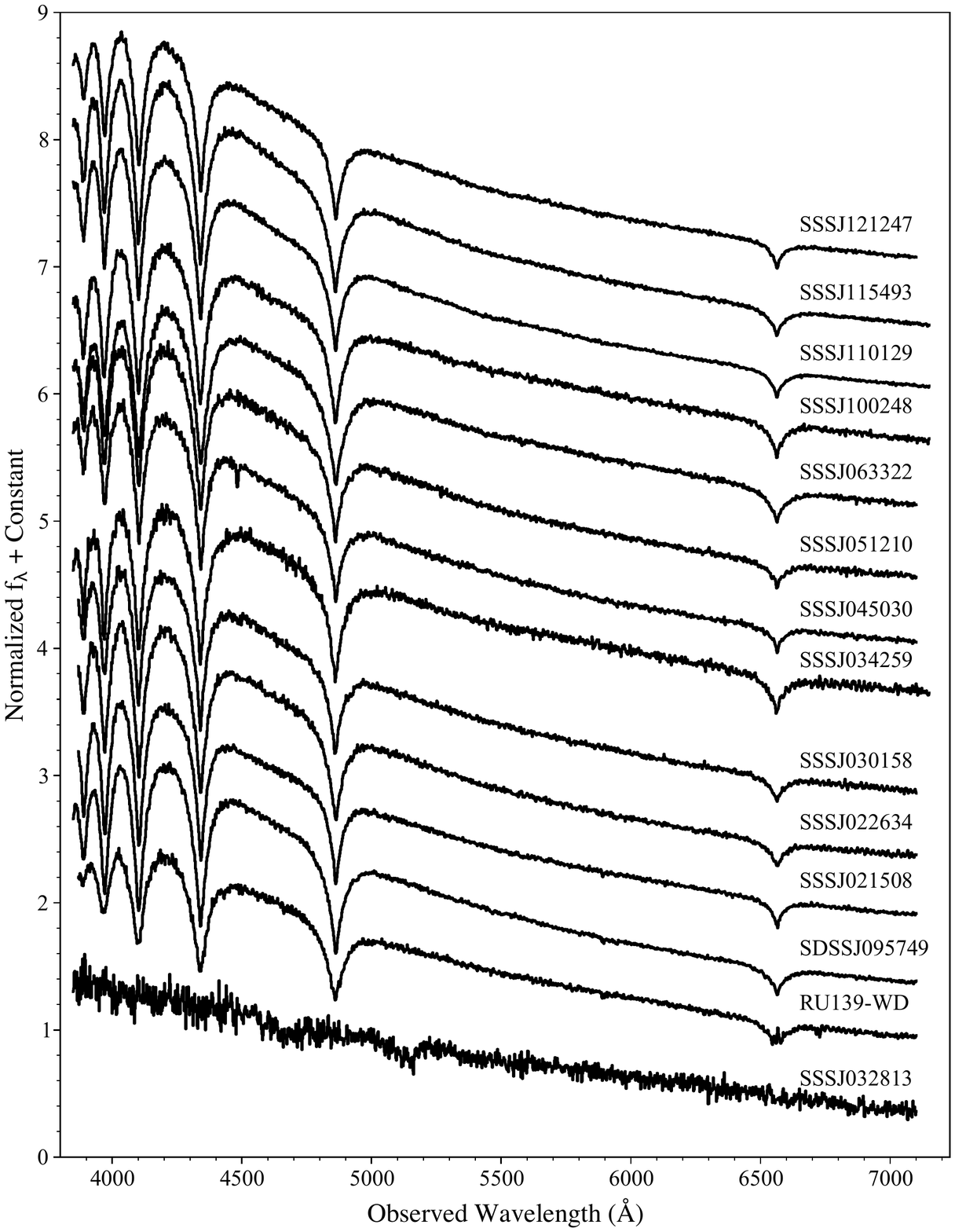} 
\caption{Same as Fig.~\ref{fig:other1} but for different candidate DAWDs discarded from our Southern hemisphere sample. 
The last star could be a DQ white dwarf, showing the $C2$ Swan band at $\lambda \approx 5150$ \AA. \label{fig:other2}}
\end{figure*}

For the SOAR spectroscopic observations, we used the 1\farcs07 slit, oriented at the parallactic angle, where possible. Spectra were reduced and processed by following the same technique used for the Northern DAWD sample as described in CA19. 
The final spectra have a typical range of 3850\AA\ to 7100\AA\ with an
intrinsic dispersion of 1.99\AA\ per pixel, re-binned to 2\AA\ per pixel for the final spectra.  Figs.~\ref{fig:good}, \ref{fig:other1}, \ref{fig:other2}, \ref{fig:odd} show the spectra of all the candidate WDs that were observed in the Southern hemisphere. 
The first figure shows the 13 DAWDs with HST and LCO observations that were established as spectrophotometric standards, 
while the other three figures show objects that were discarded for various reasons. In particular, Fig.~\ref{fig:odd} displays six objects with unusual spectra: the top plot shows a strong blue featureless spectrum, possibly a hot DC WD, while the second and third ones are blue He-rich spectra, possibly DB degenerates. 
The remaining spectra in the figure show Zeeman splitting of the Balmer lines indicating magnetic degenerates.

\begin{deluxetable*}{lllcccccccc}
\tabletypesize{\scriptsize}
\tablecaption{Log of the observations collected with the Goodman spectrograph on the 4m-SOAR telescope NOAO programs 2017A-0052 
(PI: Olszewski).  \label{table:2}}
\tablehead{
\colhead{Star}&
\colhead{Alternative name}&
\colhead{Gaia DR3 ID}&
\colhead{RA\tablenotemark{a}}& 
\colhead{DEC\tablenotemark{a}}& 
\colhead{UT Date}&
\colhead{Range}&
\colhead{P.A.}&
\colhead{Air.}& 
\colhead{Flux Std.}&
\colhead{Exp.}\\
\colhead{}&
\colhead{}&
\colhead{}&
\colhead{(hh:mm:ss.s)}&
\colhead{(dd:mm:ss.s)}&
\colhead{}&
\colhead{(\AA)}&
\colhead{($^\circ$)}&
\colhead{}&
\colhead{}&
\colhead{(s)}
}
\startdata
J0122-3052  & ATLAS020.503022     & 5028544686500198144 & 01:22:00.725 & -30:52:03.950 & 2016-10-07 & 3870-7100 & 263.0 & 1.1 & Feige110 & 4x900 \\
J2140-3231  & ATLAS325.224509-32.5& 6592388973858801920 & 21:40:53.887 & -32:31:17.381 & 2016-10-05 & 3870-7100 & 90.0 & 1.1 & Feige110 & 3x900 \\
J2214-2954  & ATLAS333.686598     & 6614900207422023168 & 22:14:44.788 & -29:54:37.641 & 2016-10-06 & 3870-7100 & 261.0 & 1.1 & Feige110 & 900 \\
J2242-2913  & ATLAS340.6628	   & 6608532725132112768 & 22:42:39.087 & -29:13:16.348 & 2016-10-07 & 3870-7100 & 260.0 & 1.1 & Feige110 & 4x900 \\
J2343-3732  & ATLAS355.86084	   & 6538080044408150784 & 23:43:26.600 & -37:32:36.724 & 2016-10-06 & 3870-7100 & 60.0 & 1.0 & Feige110 & 4x900 \\
J0025-2840  & ATLAS6.345142	   & 2321781186172842624 & 00:25:22.838 & -28:40:33.944 & 2016-10-06 & 3870-7100 & 100.0 & 1.1 & Feige110 & 3x900 \\
J1800-2332  & RU139-WD   & 4069124858877707776 & 18:00:42.030 & -23:32:38.652 & 2016-10-05 & 3870-7100 & 112.0 & 1.3 & Feige110 & 4x900 \\
J0141-6140  & SSSJ014151 & 4712803636068739456 & 01:41:51.827 & -61:40:48.350 & 2016-10-07 & 3870-7100 & 298.0 & 1.3 & Feige110 & 900 \\
J0215-6127  & SSSJ021508 & 4713619542415902336 & 02:15:08.335 & -61:27:30.571 & 2016-10-05 & 3870-7100 & 321.0 & 1.2 & Feige110 & 4x900 \\
J0226-2214  & SSSJ022634 & 5120586282330070784 & 02:26:34.644 & -22:14:22.824 & 2016-10-06 & 3870-7100 & 130.0 & 1.0 & Feige110 & 3x900 \\
J0238-3602  & SSSJ023824 & 4953936951336477440 & 02:38:24.969 & -36:02:23.222 & 2016-10-07 & 3870-7100 & 273.0 & 1.1 & Feige110 & 4x900 \\
J0301-2450  & SSSJ030158 & 5074738090560529792 & 03:01:58.400 & -24:50:44.068 & 2016-10-07 & 3870-7100 & 243.0 & 1.0 & Feige110 & 3x900 \\
J0328-2839  & SSSJ032813 & 5056979122346336384 & 03:28:13.113 & -28:39:25.621 & 2017-02-22 & 3850-7100 & 102.0 & 1.3 & Feige67 & 1000 \\
J0358-7559  & SSSJ035817 & 4628635093249821952 & 03:58:17.886 & -75:59:29.089 & 2016-10-07 & 3870-7100 & 243.0 & 1.0 & Feige110 & 3x900 \\
J0343-2556  & SSSJ034259 & 5082086886979911424 & 03:43:04.171 & -25:56:55.110 & 2016-02-12 & 3850-7150 & 60.0 & 1.1 & Feige67 & 2x900 \\
J0343-2556  & SSSJ034259 & 5082086886979911424 & 03:43:04.171 & -25:56:55.110 & 2016-02-12 & 3850-7150 & 60.0 & 1.2 & Feige67 & 3x1200 \\
J0358-7559  & SSSJ035817 & 4628635093249821952 & 03:58:17.886 & -75:59:29.089 & 2016-02-11 & 3850-7150 & 58.0 & 1.5 & Feige67 & 2x1200 \\
J0450-2846  & SSSJ045030 & 4879984623886489600 & 04:50:30.966 & -28:46:02.222 & 2017-02-23 & 3850-7100 & 100.0 & 1.1 & GD71 & 2x1200 \\
J0458-5637  & SSSJ045822 & 4764189621230467584 & 04:58:23.133 & -56:37:33.434 & 2016-10-06 & 3870-7100 & 339.0 & 1.1 & Feige110 & 3x900 \\
J0458-5637  & SSSJ045822 & 4764189621230467584 & 04:58:23.133 & -56:37:33.434 & 2017-02-22 & 3850-7100 & 58.0 & 1.2 & Feige67 & 3x1000 \\
J0512-3112  & SSSJ051210 & 4827685043344935168 & 05:12:10.898 & -31:12:59.861 & 2016-10-05 & 3870-7100 & 265.0 & 1.1 & Feige110 & 5x900 \\
J0541-1930  & SSSJ054114 & 2967083052984612736 & 05:41:14.759 & -19:30:38.896 & 2016-10-07 & 3870-7100 & 244.0 & 1.2 & Feige110 & 3x900 \\
J0633-7858  & SSSJ063322 & 5211173052478517120 & 06:33:22.458 & -78:58:19.125 & 2017-02-22 & 3850-7100 & 31.0 & 1.6 & Feige67 & 3x1200 \\
J0639-5712  & SSSJ063941 & 5484605140287436416 & 06:39:41.468 & -57:12:31.164 & 2016-10-07 & 3870-7100 & 303.0 & 1.2 & Feige110 & 3x900 \\
J0639-5712  & SSSJ063941 & 5484605140287436416 & 06:39:41.468 & -57:12:31.164 & 2017-02-23 & 3850-7100 & 24.0 & 1.2 & GD71 & 3x1200 \\
J0947-8458  & SSSJ094704 & 5191604494282635136 & 09:47:02.983 & -84:58:39.558 & 2016-02-10 & 3850-7150 & 5.0 & 1.7 & Feige67 & 4x1200 \\
J0956-3841  & SSSJ095657 & 5421579652019276160 & 09:56:57.009 & -38:41:30.269 & 2017-02-22 & 3850-7100 & 334.0 & 1.0 & Feige67 & 3x1200 \\
J0957-0019  & SSSJ095749 & 3833430797566676352 & 09:57:49.372 & -00:19:49.381 & 2017-02-23 & 3850-7100 & 136.0 & 1.5 & GD71 & 3x1200 \\
J1002-8031  & SSSJ100248 & 5201701240840634624 & 10:02:49.314 & -80:31:10.614 & 2016-02-12 & 3850-7150 & 348.0 & 1.6 & Feige67 & 2x1200 \\
J1002-8031  & SSSJ100248 & 5201701240840634624 & 10:02:49.314 & -80:31:10.614 & 2016-02-12 & 3850-7150 & 30.0 & 1.6 & Feige67 & 2x900 \\
J1055-3612  & SSSJ105525 & 5401230062610609920 & 10:55:25.356 & -36:12:14.731 & 2017-02-23 & 3850-7100 & 11.0 & 1.0 & GD71 & 3x1200 \\
J1101-3621  & SSSJ110129 & 5400333861851343616 & 11:01:29.850 & -36:21:04.362 & 2017-03-18 & 3850-7150 & 75.0 & 1.1 & \ldots & 2x1200 \\
J1159-5008  & SSSJ115943 & 5370694601787409152 & 11:59:43.451 & -50:08:18.370 & 2017-03-18 & 3850-7150 & 334.0 & 1.1 & \ldots & 2x1200 \\
J1212-4029  & SSSJ121247 & 6149314478246925056 & 12:12:47.130 & -40:29:46.827 & 2017-02-22 & 3850-7100 & 295.0 & 1.1 & Feige67 & 3x1200 \\
J1434-2819  & SSSJ143459 & 6222123588482712832 & 14:34:59.528 & -28:19:03.295 & 2017-02-22 & 3850-7100 & 257.0 & 1.1 & Feige67 & 3x1200 \\
J1837-7002  & SSSJ183717 & 6431766714636858240 & 18:37:17.906 & -70:02:52.513 & 2016-10-06 & 3870-7100 & 44.0 & 1.4 & Feige110 & 3x900 \\
J1930-5203  & SSSJ193018 & 6646236009641999488 & 19:30:18.995 & -52:03:46.550 & 2016-10-06 & 3870-7100 & 64.0 & 1.2 & Feige110 & 3x900 \\
J1947-3100  & SSSJ194736 & 6751474223204472576 & 19:47:36.361 & -31:00:39.385 & 2016-10-04 & 3870-7100 & 95.0 & 1.1 & Feige110 & 3x900 \\
J2023-4015  & SSSJ202344 & 6680866227868812288 & 20:23:44.526 & -40:15:21.092 & 2016-10-06 & 3870-7100 & 84.0 & 1.2 & Feige110 & 3x900 \\
J2220-4645  & SSSJ222035 & 6518394383932274432 & 22:20:36.053 & -46:45:52.384 & 2016-10-06 & 3870-7100 & 60.0 & 1.1 & Feige110 & 900 \\
J0054-2650  & WD0052-271 & 2344195005582967040 & 00:54:57.994 & -26:50:23.221 & 2016-10-05 & 3870-7100 & 106.0 & 1.0 & Feige110 & 3x900 \\
J0259-2805  & WD0257-282 & 5071554695160551040 & 02:59:23.254 & -28:05:33.327 & 2016-02-11 & 3850-7150 & 105.0 & 1.2 & Feige67 & 3x1200 \\
J0419-5319  & WD0418-534 & 4779427928974390272 & 04:19:24.680 & -53:19:16.659 & 2017-02-22 & 3850-7100 & 62.0 & 1.2 & Feige67 & 3x600 \\
J0512-4145  & WD0510-418 & 4812859061053900928 & 05:12:23.053 & -41:45:26.057 & 2016-02-10 & 3850-7150 & 40.0 & 1.0 & Feige67 & 4x600 \\
J0757-6054  & WD0756-607 & 5290719287073728128 & 07:57:03.112 & -60:54:52.622 & 2016-02-11 & 3850-7150 & 27.0 & 1.2 & Feige67 & 3x1200 \\
J0757-6049  & WD0757-606.2 & 5290720695823013376 & 07:57:50.637 & -60:49:54.634 & 2016-02-10 & 3850-7150 & 338.0 & 1.2 & Feige67 & 4x1200 \\
J1105-5852  & WD1103-586.1 & 5338652084186678400 & 11:05:35.811 & -58:52:26.385 & 2016-02-11 & 3850-7150 & 345.0 & 1.2 & Feige67 & 3x1200 \\
J1105-5829  & WD1103-582 & 5340167657872411520 & 11:05:53.071 & -58:29:31.090 & 2016-02-10 & 3850-7150 & 8.0 & 1.2 & Feige67 & 4x1200 \\
J1146-3141  & WD1143-314 & 3479327447141240320 & 11:46:18.107 & -31:41:01.612 & 2016-02-11 & 3850-7150 & 85.0 & 1.0 & Feige67 & 3x600 \\
J1146-3141  & WD1143-314 & 3479327447141240320 & 11:46:18.107 & -31:41:01.612 & 2017-02-23 & 3850-7100 & 96.0 & 1.2 & GD71 & 2x420 \\
J1206-2729  & WD1203-272 & 3486471764460448512 & 12:06:20.354 & -27:29:40.639 & 2016-02-11 & 3850-7150 & 105.0 & 1.0 & Feige67 & 3x600 \\
J1206-2729  & WD1203-272 & 3486471764460448512 & 12:06:20.354 & -27:29:40.639 & 2017-02-22 & 3850-7100 & 103.0 & 1.1 & Feige67 & 2x300 \\
J1535-7724  & WD1529-772 & 5779908502946006784 & 15:35:45.179 & -77:24:44.832 & 2016-02-12 & 3850-7150 & 320.0 & 1.6 & Feige67 & 2x900 \\
J2317-2903  & WD2314-293 & 2378059688840742912 & 23:17:20.294 & -29:03:21.647 & 2016-10-05 & 3870-7100 & 81.0 & 1.0 & Feige110 & 3x900 \\
J2317-2918  & WD2315-295 & 2330002990527676800 & 23:17:58.479 & -29:18:19.535 & 2016-10-07 & 3870-7100 & 261.0 & 1.1 & Feige110 & 3x900 \\
\enddata
\tablecomments{Every row corresponds to a different visit. Observations are sorted by survey and by increasing $RA$ values.}
\tablenotetext{a}{Coordinates are from \emph{Gaia} DR3 at epoch $2016.0$ precessed to $J2000.0$, no proper motion applied. To get current coordinates apply the proper motions from 2016 and precess.}
\end{deluxetable*}

\subsection{Stability of the spectrophotometric standard DAWDs}
In order to assess the DAWDs as stable standards we monitored them 
by collecting time-spaced data with the LCO network of telescopes.
WDs can vary due to several reasons, depending on their 
effective temperature, atmosphere abundance and presence of 
magnetic activity or of an unseen faint companion star.

\begin{figure*}
\includegraphics[height=0.95\textheight,width=\textwidth]{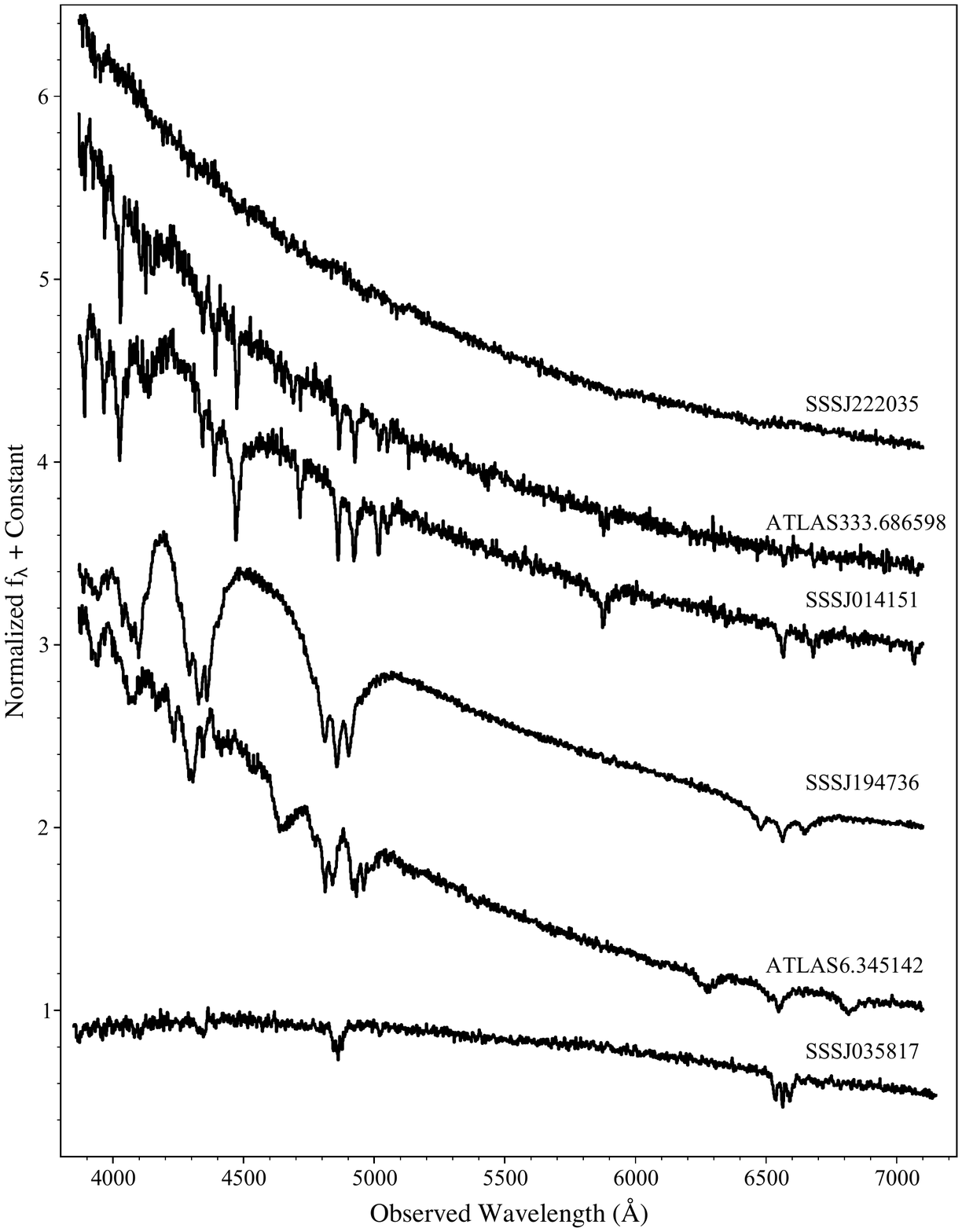} 
\caption{Spectra collected with the Goodman spectrograph on the 4m-SOAR telescope (NOIRLab) for candidate DAWDs that were discarded from our Southern hemisphere sample. 
The first star is featureless and could be a DC WD, the second and third could be DB degenerates. 
The last three are magnetic WDs. \label{fig:odd}}
\end{figure*}

Hydrogen-rich atmosphere WDs might present gravity-mode
pulsations around $T_{eff}$ $\sim$ 12,000 K \citep[ZZ Ceti pulsators]{fontaine2008}. 
Our DAWDs were selected to have temperatures ($T_{eff} \gtrsim$ 20,000 K) outside the ZZ Ceti instability strip, so we do not expect them to be pulsators 
(note that SDSSJ172135.97+294016.0 was removed from the network of standards as its
temperature is $T_{eff}$ = 9,261 K, see CA19, NA19).
Strong magnetic fields can also cause flux variations in WDs
with a time scale from hours to days. 
These variations can be due to magnetically confined "spots" of
higher opacity modulating the stellar flux via stellar rotation \citep{dupuis2000, 
holberg2011}. Alternately, magnetic variations
can be due to spots in the convective atmosphere \citep{brinkworth2004, brinkworth2013}.
However, our candidate standard DAWDs have effective temperatures above
$\sim$ 20,000 K, and their atmosphere are fully radiative, so
they should not vary due to the presence of spots. 
Furthermore, we excluded candidates with spectra showing Zeeman 
splitting of the Balmer lines indicative of the presence of a strong magnetic field (see Fig.~\ref{fig:odd}).

On the other hand, the selected DAWDs could still vary due to the presence of an unseen faint companion star, for example, and we need to characterize
the amount of flux variation, if present, before setting these stars as spectrophotometric standards.

A study by \citet{hermes2017}, based on precise \emph{Kepler} time-series
photometry, showed that $\sim$ 97\% of apparently isolated WDs are stable, or 
show less than 1\% flux variations, and they can be used as spectrophotometric standards. 
Hermes et al.\ sample included mostly DAWDs but also several helium- or carbon-dominated
atmosphere WDs, with temperatures hotter than $\sim$ 8,000 K.

By comparing observations and binary population synthesis models, \citet{toonen2017} 
studied the binarity of the almost complete local WD sample ($d \lesssim$ 20 pc). 
Assuming an initial binary fraction of 50\%, these models show that most systems
undergo a common envelope phase and subsequent merger with the final outcome of
$\sim$ 70-80\% of WDs being isolated. The remaining fraction of systems are probably 
on a wide orbit, as these binaries can more easily avoid the common envelope-merger phase, 
and can be observed as resolved binaries. These systems are usually separated by a few arcseconds. Our DAWDs are more distant than the local sample, 100 $\lesssim d \lesssim$ 1000 pc, but WFC3 detectors' spatial resolution would easily allow us to resolve the companions (pixel scales of $\sim$ 0\farcs04 and 0\farcs13 for the UVIS and the IR camera, respectively, resulting in separations of more than a dozen pixels in both cases). The same models from \citet{toonen2017} predict a fraction of $\sim$ 0.5-1\% of unresolved binaries. For most of these, the companion would be identified through spectroscopy, such as for SDSSJ203722.169-051302.964, where spectra showed an emission feature in the cores of the Balmer absorption lines, probably the result of a low-luminosity companion or some other activity associated with the DAWD. 
The variability of this star was also confirmed by the analysis of the LCO time-spaced observations (see discussion in \S 4).
However, if the companion to the DAWD is a faint red dwarf, then current spectroscopy is not able to detect it, and not even imaging at the spatial resolution of \emph{HST}. That is why we started a photometric monitoring campaign.

On the basis of the criteria used to select our set of DAWDs and the evidence from the WFC3 and spectroscopic data, we do not expect a large fraction of our candidate spectrophotometric standards to vary. However, these DAWDs have not yet been subject to a consistent and well-defined 
observational campaign to demonstrate a lack of variability at a wide range of time scales. 
WFC3 observations are obtained within a short time frame for each target, and so they are unsuitable as tests of variation. 
Ground-based surveys (SDSS, PanSTARRS, ATLAS) and space facilities (\emph{Kepler}, \emph{TESS}) also do not have the necessary temporal coverage or spatial resolution,
and \emph{Gaia} does not provide variability constraints on these stars yet.

\section{Time-spaced observations}\label{sec:series}
Time-spaced data for 23 candidate spectrophotometric standard DAWDs in the Northern hemisphere and around the celestial equators were collected with LCO starting in the fall of 2016 until the summer of 2017, for a total of $\sim$ 1 year of observations\footnote{LCO2016B-007 and LCO2017AB-002 (PI: Matheson)}. 
A few exposures were also collected in the first semester of 2018\footnote{LCO2018A-002 (PI: Matheson)}, to add more epochs to some targets. 
Observations for the 15 candidates in the Southern hemisphere were collected in a semester in 2018 and one in 2019\footnote{LCO2018B-001 and LCO2019-B004  (PI: Matheson)}.

Data consist of a sequence of exposures in the Sloan $g$ filter, separated by minutes up to 
a month in time. A minimum of 20 exposures for each target were collected, spread over 2-3 months at different time intervals, for a total of $\approx$ 1,400 images. 
The log of the observations is shown in Table~\ref{table:3}.

In order to schedule our observations at the LCO observatory, we developed a \texttt{python} routine that calculates the observing window for each star on the different network telescopes during a selected semester and the 
optimal exposure time to reach a $S/N \approx$ 100.
This code is available on Github at the following URL\footnote{https://github.com/gnarayan/LCO\_scheduler}.

\begin{deluxetable*}{lccccc}
\tablecaption{Log of the observations collected with LCO
during programs LCO2016B-007, LCO2017AB-002, LCO2018A-002, LCO2018B-001, and LCO2019-B004 (PI: T. Matheson). \label{table:3}}  
\tablehead{
\colhead{Star}&
\colhead{Code\tablenotemark{a}}&
\colhead{RA\tablenotemark{b}}&
\colhead{DEC\tablenotemark{b}}&
\colhead{MJD}&
\colhead{Exposure time}\\
\colhead{}&  
\colhead{}&  
\colhead{(hh:mm:ss.s)}&
\colhead{(dd:mm:ss.s)}&
\colhead{}& 
\colhead{(s)}
}
\startdata
\hline
WDFS1314-03 & fl15 & 13:14:45.050 & -03:14:15.64 & 57832.3508 & 304 \\ 
WDFS1314-03 & fl06 & 13:14:45.050 & -03:14:15.64 & 57832.1 & 304   \\ 
WDFS1314-03 & fl06 & 13:14:45.047 & -03:14:15.65 & 57960.7696 & 423 \\ 
WDFS1314-03 & fl04 & 13:14:45.050 & -03:14:15.64 & 57832.2853 & 304 \\ 
WDFS1314-03 & fl04 & 13:14:45.050 & -03:14:15.64 & 57832.2163 & 304 \\ 
WDFS1314-03 & fl14 & 13:14:45.050 & -03:14:15.64 & 57842.9781 & 304 \\ 
WDFS1314-03 & fl03 & 13:14:45.047 & -03:14:15.65 & 58143.3084 & 349 \\ 
WDFS1314-03 & fl03 & 13:14:45.047 & -03:14:15.65 & 58160.2109 & 349 \\ 
WDFS1314-03 & fl15 & 13:14:45.050 & -03:14:15.64 & 57832.3548 & 304 \\ 
WDFS1314-03 & fl15 & 13:14:45.050 & -03:14:15.64 & 57832.2018 & 304 \\
WDFS1314-03 & fl04 & 13:14:45.050 & -03:14:15.64 & 57832.2918 & 304 \\ 
WDFS1314-03 & fl06 & 13:14:45.050 & -03:14:15.64 & 57832.096 & 304 \\ 
\enddata
\tablenotetext{a}{Telescope code}
\tablenotetext{b}{Coordinates are at epoch J2000.}
\tablecomments{Table~\ref{table:3} is published in its entirety in the machine readable format.  A portion is shown here for guidance regarding its form and content.}
\end{deluxetable*}

\subsection{Data processing and reduction}\label{sec:redu}
We downloaded all images collected for our programs from the LCO archive. These are processed by the BANZAI
pre-reduction pipeline\footnote{https://github.com/LCOGT/banzai}. The pipeline performs a bad-pixel masking, bias and dark subtraction and a
flat-field correction. It also provides an astrometric solution for the images and extracts aperture photometry for
the sources by using Source Extractor \citep{bertin1996}. For more details please refer to the LCO BANZAI pipeline 
web page\footnote{https://lco.global/observatory/data/BANZAIpipeline/}.

Images were collected under different conditions, effectively being spread over different nights and months and utilizing different telescopes and observatories.
All data were collected with the Sinistro 4K$\times$4K cameras mounted on the 1m-class
network of telescopes. This includes Siding Spring (observatory code, COJ), Sutherland (CPT), Cerro Tololo Cerro Tololo Inter-American Observatory (LSC),
McDonald (ELP). The Sinistro cameras provide a total field of view (FoV) of $\sim$ 26\arcmin$\times$26\arcmin ~with a pixel scale of 0\farcs389.
Seeing for the different observations for all DAWDs ranged between 1\farcs7~ to 2\farcs5~ on the images, 
with an average seeing $\approx$ 2\arcsec.

As a first step, the average Full-Width Half Maximum (FWHM) for each frame was derived from the available Source Extractor photometry.  To exclude observations affected by poor observing conditions or bad focus, all the images with FWHM $\ge$ 7.5 pixels (2\farcs9) were discarded.

We then performed aperture photometry with DAOPHOTIV \citep{stetson1987}, using an aperture radius of 5 pixels, and the sky background calculated in an annulus with radii 7 and 20 pixels.
Starting from the aperture photometry, we performed Point-Spread Function (PSF) photometry with DAOPHOTIV/ALLSTAR.
An automatic pipeline was developed in \texttt{python} to run the different routines of DAOPHOT and derive a PSF for each image.
The result is a sample of moderately bright, isolated and 
well-measured PSF stars per image. The pipeline also runs ALLSTAR, i.e., the PSF fitting routine, on all the images and produces a catalog with identified sources, coordinates and magnitudes for every single image.
Images that failed the automatic procedure were 
individually visually inspected and checked for problems. Most frequently they were out of focus or affected by clouds. 



As a second step, we identified the best image (smallest average FWHM) for a set of exposures for each target
and established it as a reference frame. This process is needed to flux scale all the photometric catalogs and to register all the exposures to the same coordinate system. 
In order to derive transformations between the images we used DAOMATCH/DAOMASTER \citep{stetson1994}
and created a master catalog for each FoV.
We then used the code ALLFRAME \citep{stetson1994} to
perform simultaneous PSF-fitting photometry on all images available for a target. 
ALLFRAME output catalogs were matched to derive light curves for all the stars in the observed FoV, including the DAWDs. Note that exposures for each target were flux scaled
to the reference images to take into account effects due to the differences in the PSF, observing conditions and exposure times. 
It is important to note that we are interested in relative and not absolute photometry, and the derived light curves for our DAWDs and all stars in the same FoV are not calibrated.

We ended up with 38 final photometric catalogs, one for each DAWD star. On average, catalogs include $\approx$ 300-1,000 stars distributed over the FoV. The DAWDs were identified in the final photometric catalogs by searching around their position ($RA, DEC$) within a radius of 0.001\deg.

The data were also independently measured with the DoPHOT program \citep{schechter1993}, using a process described in \citet{saha2019}. This produces a list of aperture corrected instrumental magnitudes for each image. The lists for all images in the $g$ band centered on any given target DAWD star were then matched by position ($RA, DEC$). A single zero-point adjustment in instrumental magnitudes was applied to each list, so that the error-weighted ensemble average instrumental magnitudes of all matched objects were made the same across all images/epochs centered on the DAWD star. This assumes that the majority of stars on the frame are non-variable, thus putting all instrumental magnitudes for any given DAWD star field on the same footing. The results were written into an SQL data-base. Then, for any star in the field, the variability can be tested by extracting its measurements at all epochs, as described in the following section.

\section{Variability analysis}\label{sec:varia} 
Measurements obtained with DoPHOT were used to calculate a reduced $\chi^2$ for each star in a given FoV as: 

\begin{equation}
\chi ^{2} =  \frac{\sum_{i=1}^{n}\frac{(m_{i}- \bar{m} )^{2}}{err_{i}^{2}}}{n-1}
\end{equation}

where $m_i$ are the individual measurements, $\bar{m}$ is the mean weighted magnitude of each identified object, 
$err_{i}$ is the error on the individual measurements.
The error estimates were those furnished by DoPHOT and propagated through the ensuing process. For robustness, for each star, multiple values of $\chi^2$s were calculated using a bootstrap process.  If, for instance, there were $n$ measurements available, we constructed a sample of $n$ measurements by randomly picking from the available measurements with replacement, and take the average of all the ensuing $\chi^2$ values. The resulting reduced $\chi^2$s for each object in a given field was plotted against instrumental magnitude to visually de-trend the effects of miss-estimation of the errors and make variable objects stand out in this diagram.

Since the data were taken at various telescopes of the LCO network, we expect minor differences in the actual transmission curves and detector responses from telescope to telescope.  Our DAWDs are expected to be much hotter than other stars in the FoV, so the consequent differences in the color response from one telescope to another can induce an excess variation to be seen for the DAWD stars, since the instrumental magnitudes from epoch to epoch were adjusted by matching their ensemble averages across the different exposures. 
In practice, this does appear as a large effect for observations obtained by one particular telescope at the Cerro Tololo Inter-American Observatory (observatory code LSC), identified as {\it lsc1m004} in the image header.
The DoPHOT based analysis described here clearly showed this discrepancy. Subsequently, measurements from this telescope were discarded from all final DoPHOT and DAOPHOT photometric catalogs.
With this exclusion in place, the results from the different analysis methods are in general agreement, and lead to the star by star evaluations presented in Section~\ref{sec:results}.

\begin{figure}
\hspace{-0.4cm}
\includegraphics[height=0.35\textheight,width=0.5\textwidth]{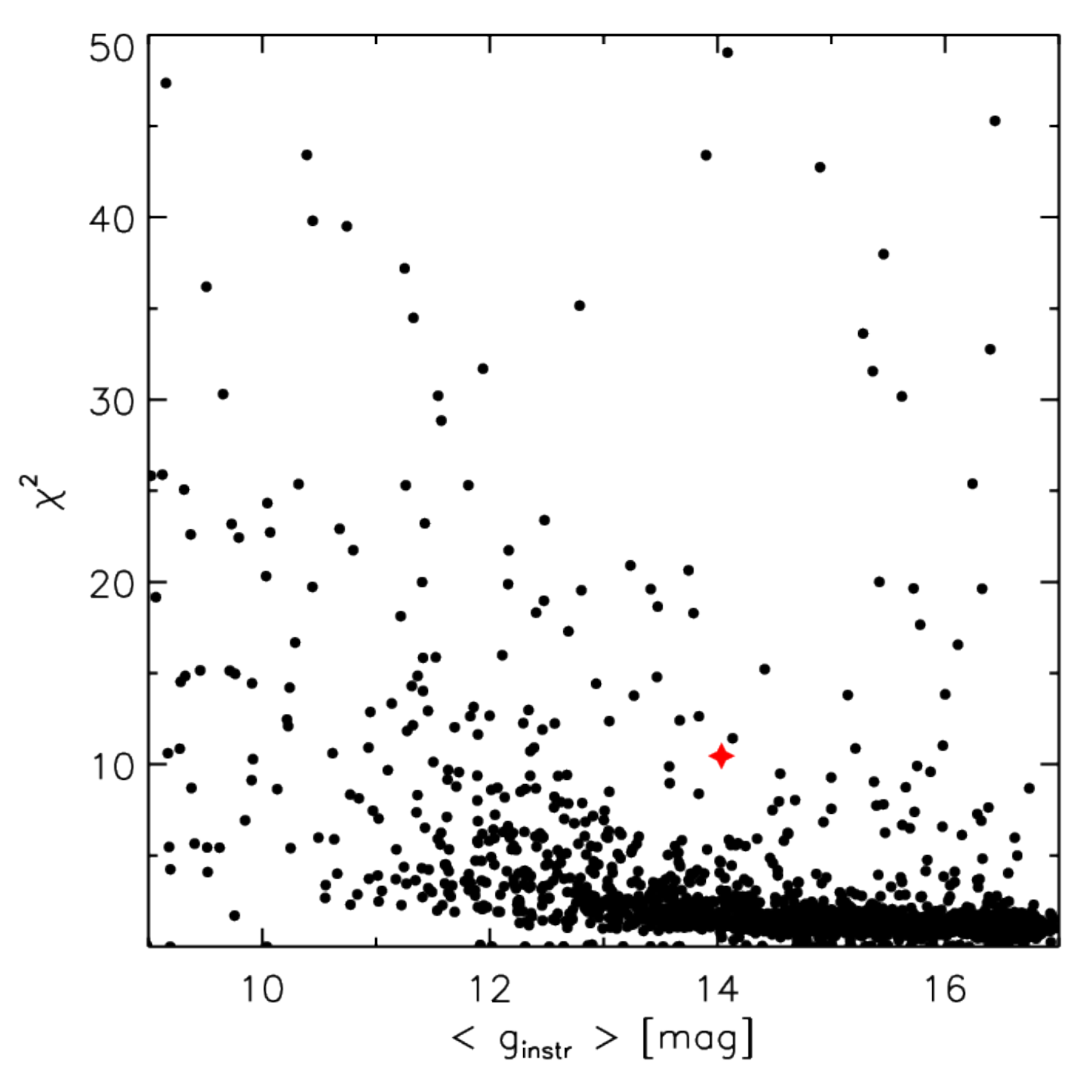}
\caption{$\chi^2$ index versus instrumental average $g$ magnitude for all stars observed
in the FoV towards WD0554-165 (red star). \label{fig:wd0554ind}}
\end{figure}

We also used the final DAOPHOT photometric catalogs to calculate a reduced $\chi^2$
for all the stars in the FoV. Fig.~\ref{fig:wd0554ind} shows the
$\chi^2$ plotted versus instrumental magnitude for the FoV observed towards WD0554-165. 
The position of the DAWD on the diagram is shown with a red star.
This plot shows that WD0554-165 has a larger $\chi^2$ compared to most of the stars observed in the same field, assumed not to be variables. However, to establish this DAWD as not stable, a more detailed analysis is needed. 

Therefore, a sample of $\approx$ 10-20 {\it stable} comparison stars was selected for each of the DAWD set of observations.
As a requirement, {\it stable} stars have a detection in every frame, a $\chi^2$ index
smaller than the median $\chi^2$ of all stars in the FoV,
sharpness of the PSF in the range -0.5 $< sharpness <$ 0.5 (to exclude extended objects and cosmic rays), and a proximity in instrumental $g$ magnitude to the target DAWD within $\approx$ 0.2 mag.

An absolute calibration of the photometry was not performed as described in the previous section. However, we need to take into account spurious flux variations due to instrumental and atmospheric effects (observations are performed with different telescopes and detectors and from different sites in different conditions). The light curves of the selected {\it stable} stars are then compared to the light curves of the DAWD in the same FoV. The variation around the mean of the {\it stable} star magnitudes was averaged and the average 1-$\sigma$ dispersion was estimated. This dispersion is used as a variability threshold for the systematic observational and instrumental effects
(see Fig.~\ref{fig:wd0554}, ~\ref{fig:sdssj235144} and the figures in the Appendix).

Fig.~\ref{fig:wd0554} shows the single epoch minus the weighted mean instrumental magnitude as a function of the Mid Julian Date (MJD) for WD0554-165 (black filled dots). Averaged magnitudes for a set of {\it stable} stars of comparable instrumental magnitude in the same FoV are also plotted as cyan filled dots.
The selected comparison stars have a $\chi^2$ index less than 1.2, while WD0554-165 has an index of 10.5. 
WD0554-165 shows clear signs of variability, 
with a measurement 1-$\sigma$ dispersion of $\sim$ 0.04 mag, four times larger compared to the {\it stable} star average dispersion of $\sigma \sim$ 0.01 mag (Fig.~\ref{fig:wd0554}).

\begin{figure*}
\begin{center}
\includegraphics[height=0.75\textheight,width=0.55\textwidth, angle=90] {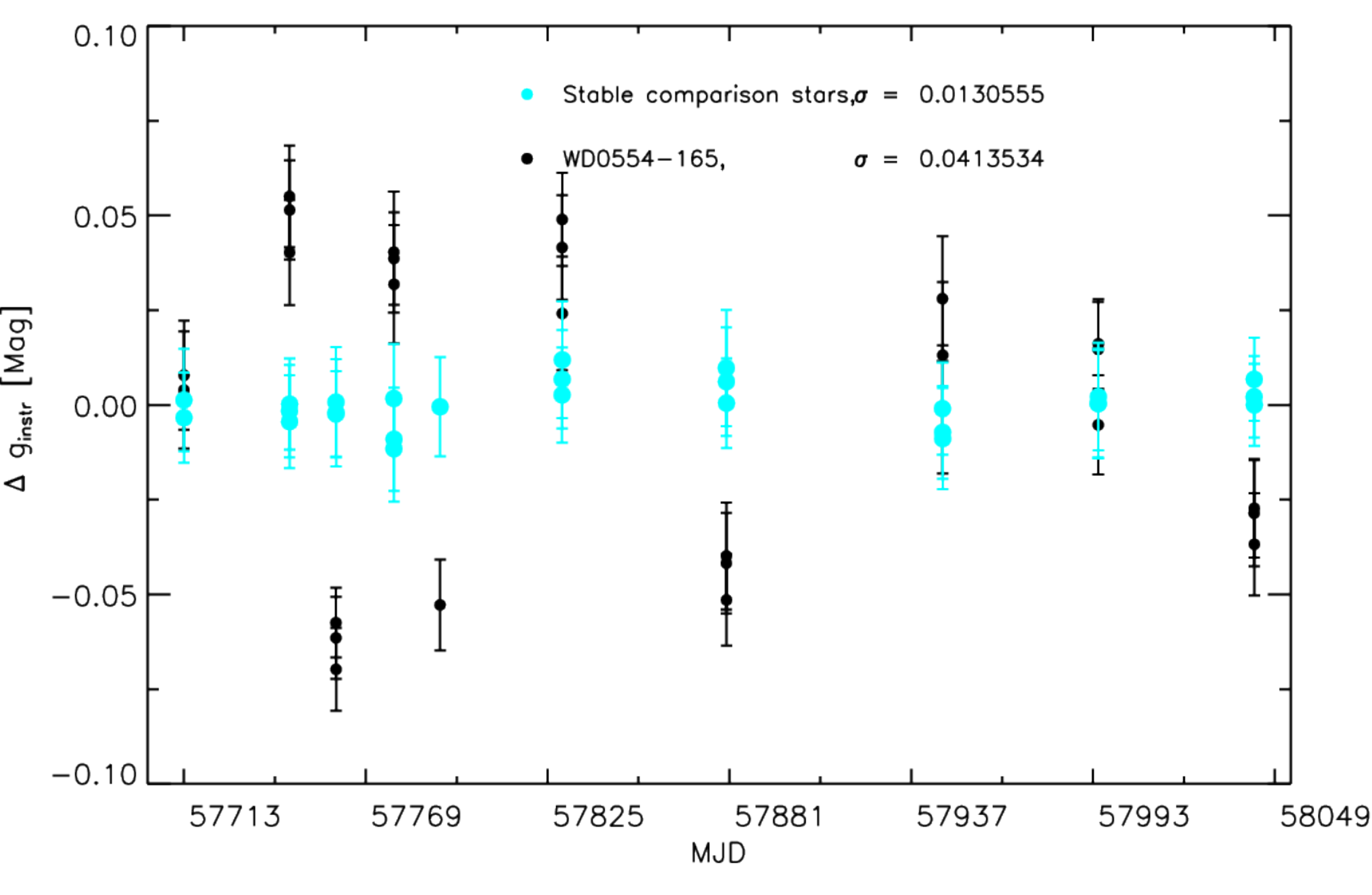}
\caption{Single epoch minus the mean instrumental magnitude measurements for WD0554-165
as a function of observing epoch (black filled dots). Averaged relative magnitudes for a set of {\it stable} stars of comparable instrumental magnitude in the same field of view are overplotted as cyan filled dots. The 1-$\sigma$ dispersion of the measurements of the {\it stable} stars and the DAWD is labeled. Error bars are shown.  \label{fig:wd0554}}
\end{center}
\end{figure*}

\begin{figure*}
\begin{center}
\includegraphics[height=0.75\textheight,width=0.55\textwidth,angle=90] {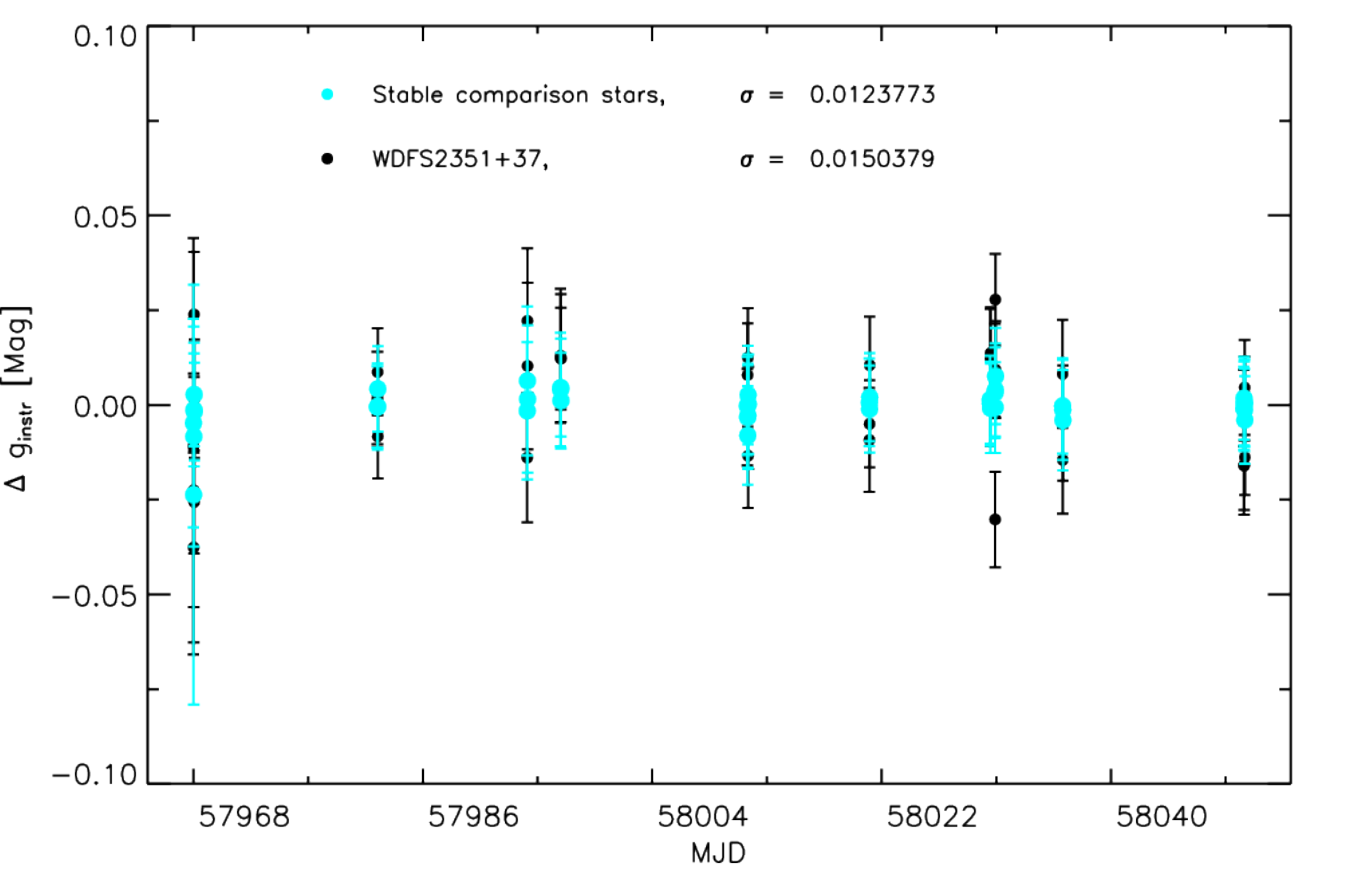}
\caption{Same as Fig.~\ref{fig:wd0554} bur for WDFS2351+37. \label{fig:sdssj235144}}
\end{center}
\end{figure*}

Fig.~\ref{fig:sdssj235144} shows the light curve plot for WDFS2351+37:
its $\chi^2$ index is 1.0 and the dispersion of the measurement is $\sim$ 0.015 mag, of the same order as the measurement dispersion of the {\it stable} stars, $\sigma \sim$ 0.012 mag. This DAWD was considered stable and included in our spectrophotometric standard network.

The light curve for SDSSJ203722.169-051302.964 (see Fig.~\ref{fig:sdssj20372} in the Appendix), a candidate binary system from spectroscopic data, shows hints of variability with a $\chi^2$ index of 3.8 and a dispersion of the measurements of $\sigma \sim$ 0.04 mag, a factor of two larger 
than the comparison star average measurement dispersion, $\sigma \sim$ 0.02 mag.

Stars SDSSJ203722.169-051302.964 and WD0554-165 were excluded from our network of spectrophotometric standard DAWDs due to their variable nature (see also the discussion in CA19 and NA19).

\subsection{Alternative variability indices}\label{sec:4pt1}
To further refine our variability analysis we also used two other variability indices, namely the interquartile range (IQR) and the von Neumann ratio ($\eta$). These pair of indices were proved to be very effective when working with data affected by outliers 
and different kind of variability and periods \citep{sokolovsky2017}, which is our case. 
However, these indices might be less effective when working with a limited sets of measurements as the LCO data we have for our DAWDs.

The two indices are defined as:

\begin{itemize}
\item the IQR is calculated as the difference between the median value of the upper and the lower half of the data points, by excluding the 25\% higher and lower values:

\item and the von Neumann index is calculated as:

\begin{equation}
\eta = \frac{\frac{\sum_{i=1}^{N-1} (m_{i+1}-m_{i})^2}{(n-1)}}{\frac{\sum_{i=1}^{N}(m_{i}-\bar{m})^2}{(n-1)}}
\end{equation}
\end{itemize}

\begin{figure}
\begin{minipage}{0.5\textwidth}
\includegraphics[height=0.4\textheight,width=0.75\textwidth, angle=90]{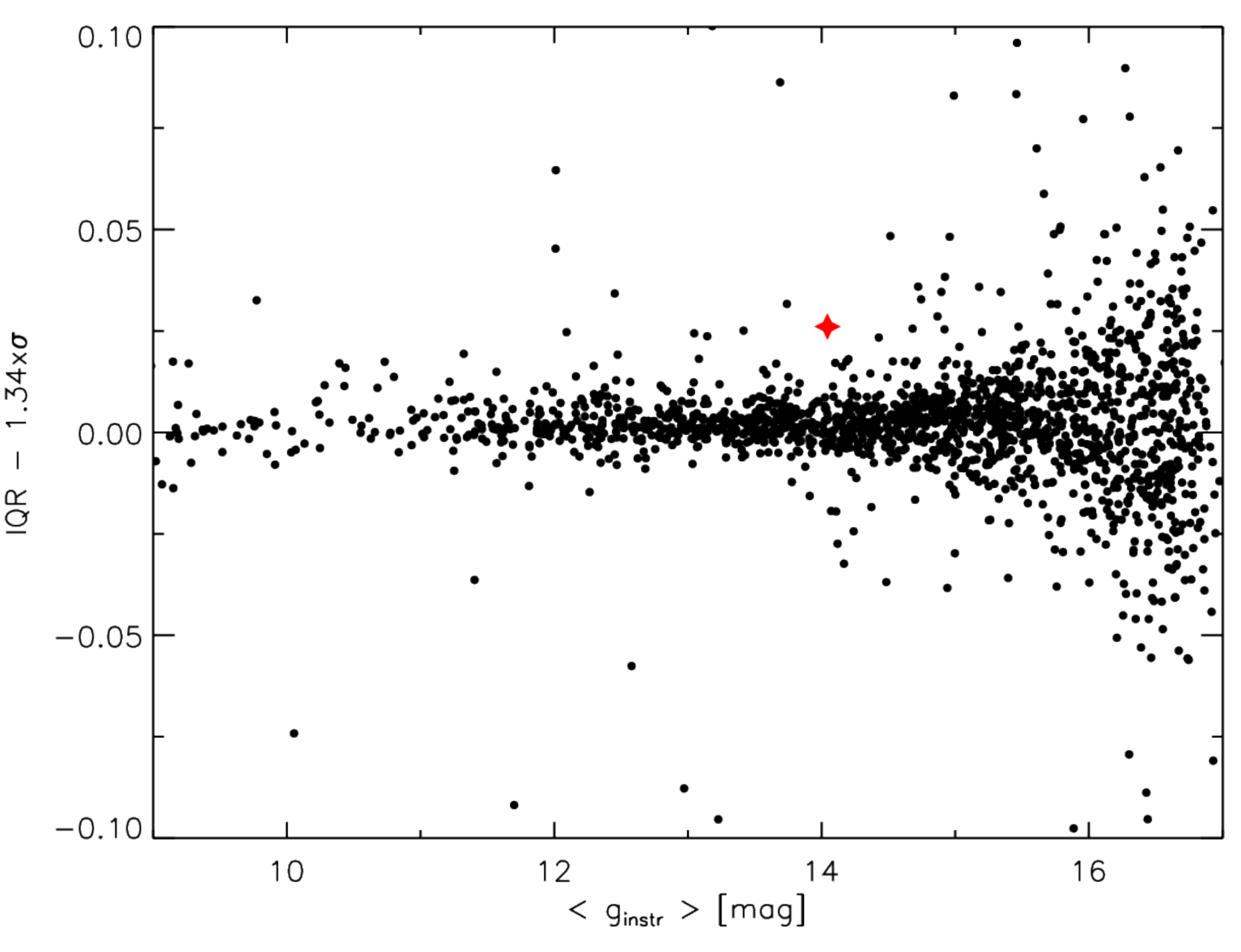} 
\end{minipage}
\begin{minipage}{0.5\textwidth}
\includegraphics[height=0.4\textheight,width=0.75\textwidth,angle=90]{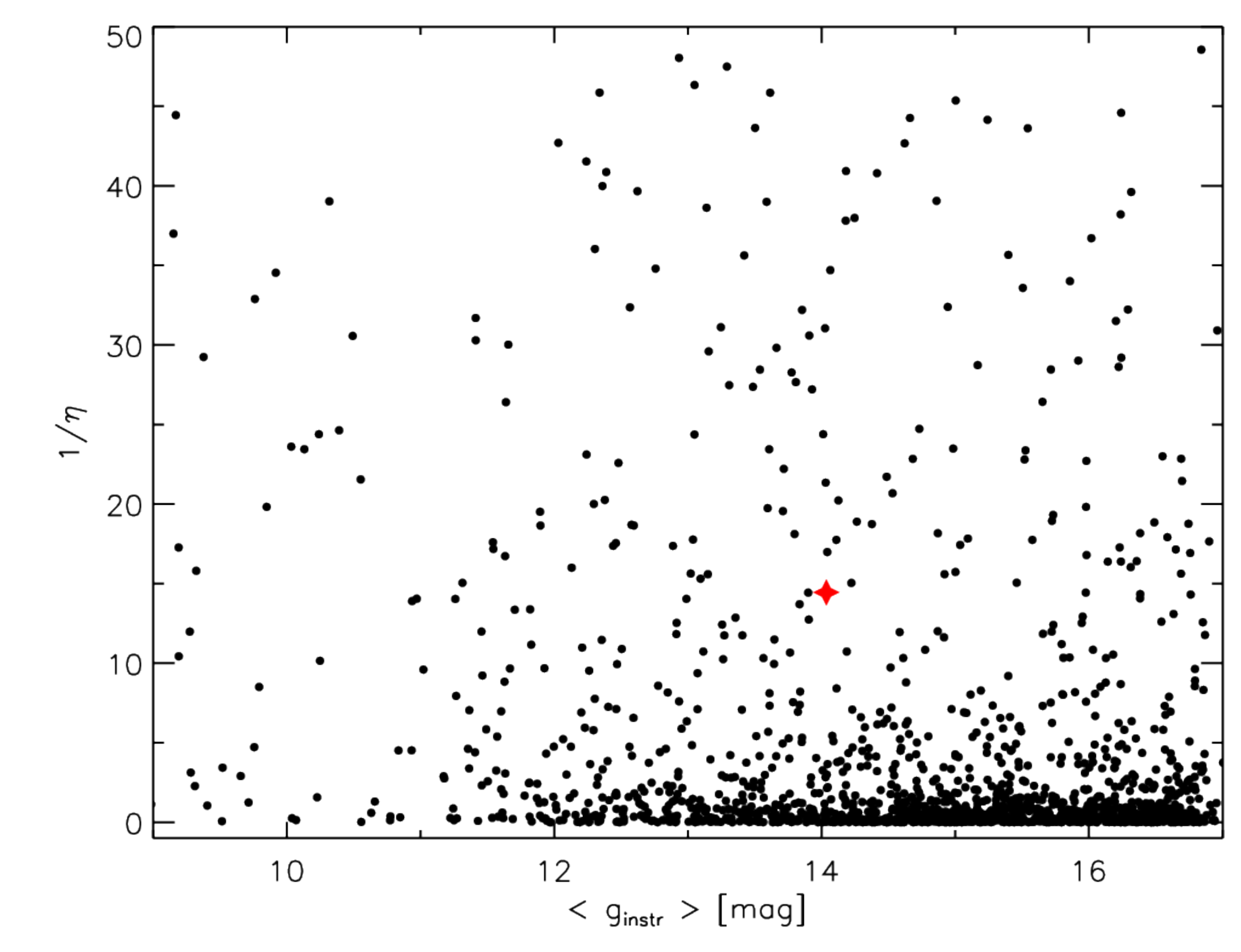} 
\end{minipage}
\vspace{0.5cm}
\caption{IQR index compared to 1.34$\times \sigma$ (top panel) and 1/$eta$ index (bottom) versus the average $g$ instrumental magnitudes for all the stars in the FoV observed 
towards WD0554-165. The DAWD is marked with a red star symbol. \label{fig:wd0554ind2} }
\end{figure}

These variability indices were calculated for all the DAWDs, even those classified as {\it variable} with the $\chi^2$ index and the comparison stars or excluded for other reasons, and for all the stars in the FoV. 
Fig.~\ref{fig:wd0554ind2} shows the difference between the IQR index and 1.34$\times \sigma$ (top panel), where $\sigma$ is the dispersion of the measurements, and the 1/$eta$ index (bottom, the larger 1/$eta$ and the higher is the possibility that a star is variable) plotted versus the average instrumental magnitude ($g$) for the FoV observed towards WD0554-165.
The position of the DAWD on the variability index plots is shown with a red star. The difference between IQR index and 1.34$\times \sigma$ seems consistent with values of most stars in the same FoV, while the 1/$eta$ index of WD0554-165 is $\sim$ 15, much higher compared to the average 1/$eta$ index for the other observed stars in the field, assumed not to be variables, i.e., $\lesssim$ 2.

\begin{deluxetable*}{lccccccccc}
\tablecaption{Spectroscopic and photometric parameters for all the
observed DAWDs. Magnitudes with errors are from Gaia EDR3 ($G$), and
the dispersion of the measurements is from the LCO time-series ($\sigma_{LCO}$). \label{table:4}}
\tablehead{
\colhead{Star}&
\colhead{$N_{exp}$}&
\colhead{$G$}&
\colhead{$\sigma_{LCO}$}&
\colhead{$\chi^2$}&
\colhead{$IQR$}&
\colhead{$1/\eta$}&
\colhead{$IQR_{ATLAS}$}&
\colhead{$1/\eta_{ATLAS}$}&
\colhead{IR excess}\\
\colhead{}&
\colhead{}&
\colhead{mag}&
\colhead{mag}&
\colhead{}&
\colhead{}&
\colhead{}&
\colhead{}&
\colhead{}&
\colhead{}
}
\startdata
\multicolumn{10}{c}{Northern and equatorial DAWDs} \\
\hline
WDFS0103-00\tablenotemark{a} & 39 & 19.301$\pm$0.003 & 0.03 & 2.1 & 0.03 & 0.28 & \ldots& \ldots & \ldots   \\ 
WDFS0228-08\tablenotemark{a} & 40 & 19.975$\pm$0.006 & 0.04 & 2.2 & 0.06 & 0.39 & \ldots& \ldots & \ldots   \\ 
WDFS0248+33		     & 30 & 18.521$\pm$0.002 & 0.03 & 1.5 & 0.03 & 0.11 & \ldots& \ldots & \ldots   \\ 
WDFS0727+32		     & 21 & 18.189$\pm$0.002 & 0.01 & 0.7 & 0.02 & 0.51 & \ldots& \ldots & \ldots   \\ 
WDFS0815+07		     & 24 & 19.932$\pm$0.005 & 0.03 & 1.0 & 0.05 & 0.60 & \ldots& \ldots & \ldots   \\ 
WDFS1024-00		     & 20 & 19.083$\pm$0.003 & 0.02 & 0.7 & 0.03 & 0.60 & \ldots& \ldots & \ldots   \\ 
WDFS1110-17\tablenotemark{a} & 36 & 18.048$\pm$0.001 & 0.03 & 4.3 & 0.05 & 6.78& 0.28  &  1.65  &   No      \\ 
WDFS1111+39		     & 17 & 18.644$\pm$0.002 & 0.02 & 1.6 & 0.02 & 0.20 & \ldots& \ldots & \ldots   \\ 
WDFS1206+02		     & 24 & 18.850$\pm$0.002 & 0.02 & 1.1 & 0.02 & 1.74 & \ldots& \ldots & \ldots   \\ 
WDFS1214+45		     & 6  & 17.979$\pm$0.001 & 0.02 & 1.8 & 0.00 & 0.06 & \ldots& \ldots & \ldots   \\
WDFS1302+10		     & 9  & 17.239$\pm$0.001 & 0.04 & 9.0 & 0.06 & 4.71 & 0.07  &   1.38 &   No     \\ 
WDFS1314-03\tablenotemark{a} & 18 & 19.307$\pm$0.003 & 0.02 & 0.7 & 0.04 & 0.23 & 0.28  &   0.13 & \ldots   \\ 
WDFS1514+00\tablenotemark{a} & 28 & 15.884$\pm$0.001 & 0.03 & 6.0 & 0.04 & 1.60 & 0.05  &  42.95 & Upp. limit  \\ 
WDFS1557+55		     & 20 & 17.691$\pm$0.001 & 0.02 & 1.3 & 0.03 & 4.09 & \ldots& \ldots & \ldots   \\ 
WDFS1638+00\tablenotemark{a} & 12 & 19.025$\pm$0.002 & 0.02 & 0.9 & 0.02 & 23.3 & 0.91  &   2.86 & \ldots   \\ 
WDFS1814+78		     & 18 & 16.745$\pm$0.001 & 0.01 & 1.0 & 0.01 & 0.33 & \ldots& \ldots & \ldots   \\ 
WDFS2101-05\tablenotemark{a} & 17 & 18.827$\pm$0.002 & 0.02 & 2.9 & 0.02 & 0.99 & 0.29  &  77.25 & Upp. limit  \\ 
WDFS2329+00		     & 26 & 18.292$\pm$0.002 & 0.02 & 1.6 & 0.02 & 0.00 & \ldots& \ldots & \ldots  \\ 
WDFS2351+37		     & 40 & 18.235$\pm$0.002 & 0.01 & 1.0 & 0.02 & 0.52 & \ldots& \ldots & \ldots  \\
\hline
\multicolumn{10}{c}{Southern DAWDs} \\
\hline
WDFS0122-30	             & 18 & 18.664$\pm$0.001 & 0.03 &  2.3  & 0.03 &  1.42 & \ldots & \ldots & \ldots	\\
WDFS0238-36\tablenotemark{a} & 17 & 18.236$\pm$0.001 & 0.03 &  6.8  & 0.03 &  8.29 & \ldots & \ldots & \ldots	\\
WDFS0458-56		     & 32 & 17.959$\pm$0.001 & 0.02 &  3.9  & 0.02 &  0.12 & \ldots & \ldots & \ldots	\\
WDFS0541-19\tablenotemark{a} & 27 & 18.433$\pm$0.002 & 0.03 &  3.2  & 0.03 &  0.09 & 0.18   & 33.01  & \ldots	\\
WDFS0639-57		     & 10 & 18.375$\pm$0.002 & 0.02 &  3.2  & 0.03 & 33.00 & \ldots & \ldots & \ldots	 \\
WDFS0956-38		     & 16 & 18.002$\pm$0.001 & 0.02 &  3.1  & 0.04 &  0.14 & \ldots & \ldots & \ldots	 \\
WDFS1055-36\tablenotemark{a} & 14 & 18.196$\pm$0.001 & 0.02 &  3.5  & 0.04 &  1.67 & 0.13   &  1.04  & \ldots	 \\
WDFS1206-27\tablenotemark{a} & 14 & 16.667$\pm$0.001 & 0.02 &  7.7  & 0.02 &  0.73 & 0.07   & 16.68  & No	 \\
WDFS1434-28\tablenotemark{a} & 21 & 18.103$\pm$0.002 & 0.02 &  2.0  & 0.02 &  0.01 & 0.22   &  0.07  & \ldots	 \\
WDFS1535-77\tablenotemark{a} & 24 & 16.765$\pm$0.001 & 0.02 &  1.9  & 0.12 &  1.30 & \ldots & \ldots & No	 \\
WDFS1837-70		     &  9 & 17.91 $\pm$0.001 & 0.02 &  1.3  & 0.01 & 11.72 & \ldots & \ldots & \ldots	 \\  
WDFS1930-52		     & 17 & 17.673$\pm$0.001 & 0.03 &  11.0 & 0.04 &  0.66 & \ldots & \ldots & Upp. limit  \\
WDFS2317-29\tablenotemark{a} & 20 & 18.526$\pm$0.002 & 0.03 &  5.0  & 0.05 &  0.20 & \ldots & \ldots & \ldots	   \\
\hline
\multicolumn{10}{c}{Discarded DAWDs} \\
\hline
SDSSJ041053.632-063027.580   & 38 & 18.990$\pm$0.002 & 0.03 & 2.7 & 0.04 & 0.02 & \ldots& \ldots & \ldots             \\ 
WD0554-165		     & 27 & 17.944$\pm$0.001 & 0.04 & 10.5& 0.08 & 14.45& 0.23  &   0.24 & \ldots            \\ 
SDSSJ172135.97+294016.0	     & 27 & 19.598$\pm$0.003 & 0.04 & 1.4 & 0.06 & 0.13 & \ldots& \ldots & \ldots            \\ 
SDSSJ203722.169-051302.964   & 15 & 19.110$\pm$0.002 & 0.04 & 3.8 & 0.02 & 0.06 & \ldots& \ldots & \ldots            \\ 
WD0418-534		     & 45 & 16.420$\pm$0.001 & 0.03 &  16.4 & 0.04 &  2.98 & \ldots & \ldots & Upp. limit   \\
WD0757-606		     & 11 & 18.953$\pm$0.002 & 0.02 &  1.0  & 0.03 &  0.02 & \ldots & \ldots & \ldots	    \\  
\enddata												   
\tablenotetext{a}{We warn the users that this standard star measurements could be affected by the 	   
presence of close red faint neighbors when observed from the ground.}   				   	      
\end{deluxetable*}

\subsection{ATLAS data \label{atlas}}
\begin{figure}
\begin{minipage}{0.5\textwidth}
\includegraphics[height=0.4\textheight,width=0.75\textwidth, angle=90]{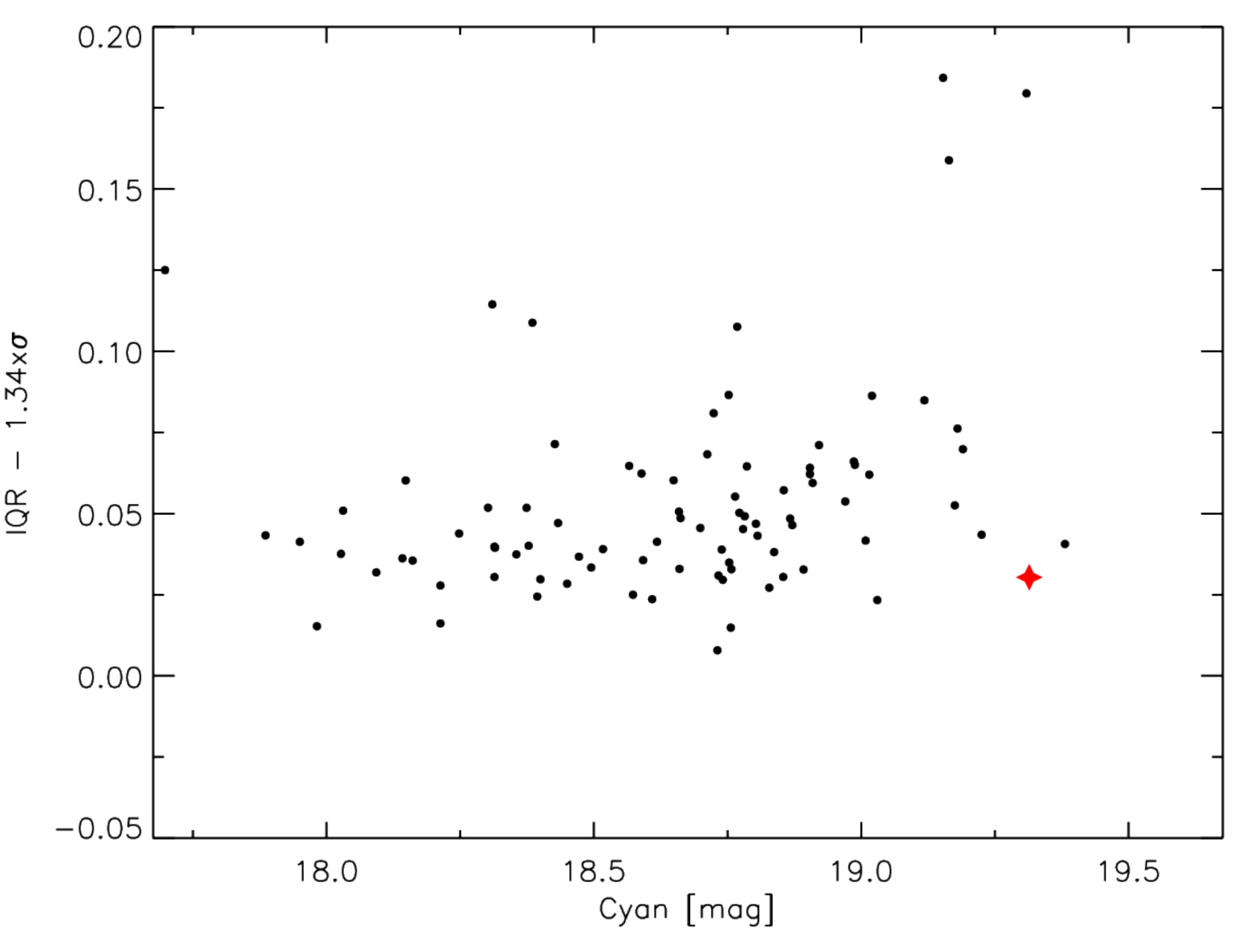} 
\end{minipage}
\begin{minipage}{0.5\textwidth}
\includegraphics[height=0.4\textheight,width=0.75\textwidth,angle=90]{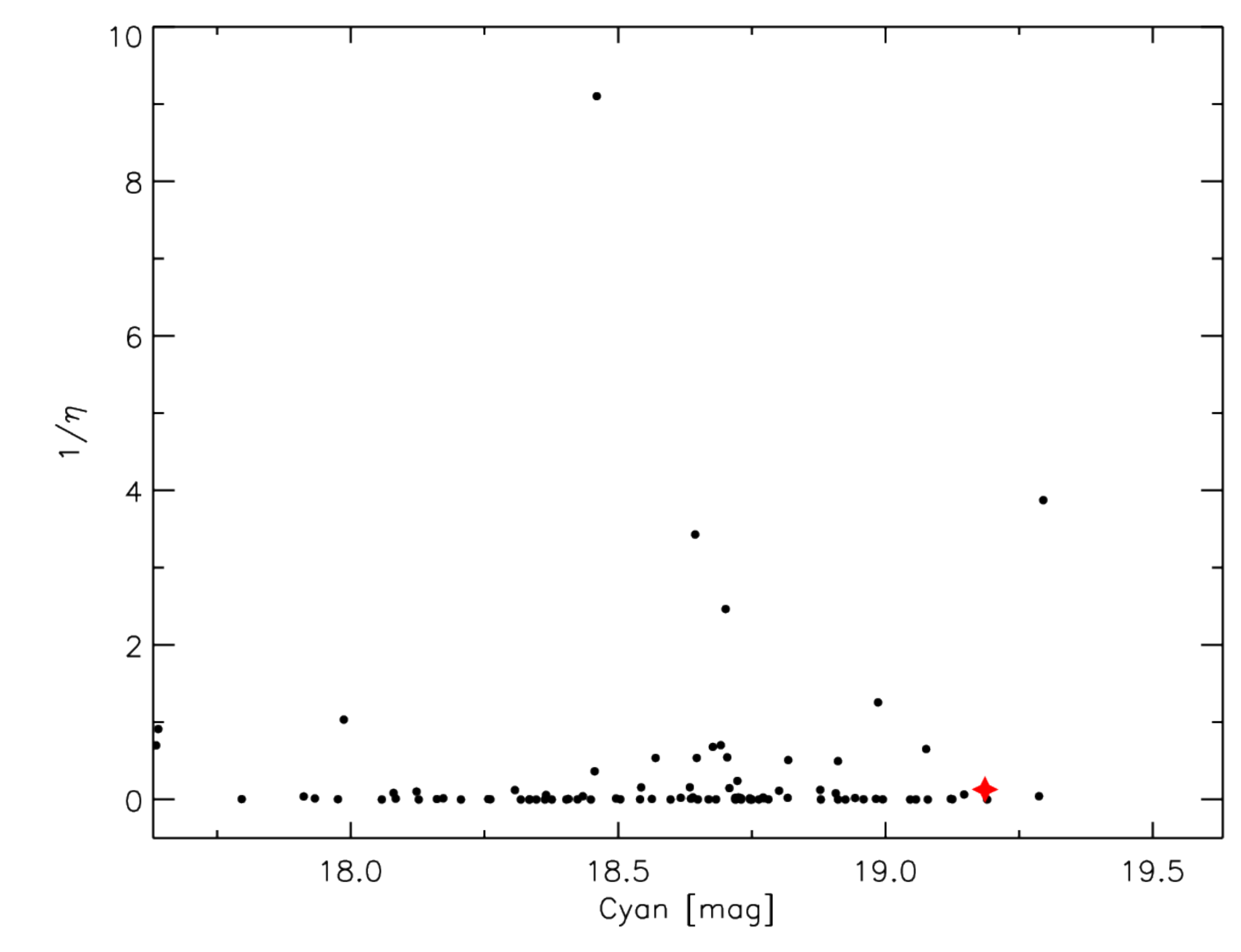}
\end{minipage}
\vspace{0.5cm}
\caption{IQR index compared to 1.34$\times \sigma$ (top panel) and 1/$eta$ index (bottom) versus $Cyan$ average magnitudes 
from the ATLAS survey for WDFS1314-03 (red star symbol) and $\approx$ 100 stars of similar LCO instrumental $g$ magnitude in the same FoV. \label{fig:sdssj131445atlas} }
\end{figure}

For a few DAWDs, the LCO light curves and the different variability indices were still inconclusive to classify them as fully stable stars. Therefore, we downloaded the Asteroid Terrestrial-impact Last Alert System \citep[ATLAS,][]{heinze2018, tonry2018} survey data, when available, for targets with declination north of $-$50\degr. In particular, we downloaded forced photometry in the cyan ($c$) and orange ($o$) filters for 7 DAWDs in the Northern hemisphere and 4 in the Southern one.  We also retrieved ATLAS time-series photometry for a set of up to 100 stars with similar instrumental magnitudes as the DAWD ($-$0.5 $\lesssim m_{DAWD} \lesssim$ $+$0.5) in the LCO FoV. We then calculated the IQR and $\eta$ variability indices for all the observed stars to be compared with the value of the DAWD. The two ATLAS variability indices are also listed in Table~\ref{table:4} when available. 
Fig.~\ref{fig:sdssj131445atlas} shows the IQR compared to 1.34$\times \sigma$ and the $1/\eta$ index versus the ATLAS $cyan$ magnitudes for star WDFS1314-03 and $\approx$ 100 stars of similar LCO instrumental $g$ magnitude. Since we do not have the color information for stars in the LCO FoV, the selected comparison stars are generally brighter (and redder) than the DAWD in the ATLAS filters, as displayed in the figure. The top panel of Fig.~\ref{fig:sdssj131445atlas} shows that the IQR index for WDFS1314-03 is consistent with the dispersion of the measurements and with that of stars of similar LCO $g$ instrumental magnitude, and the $1/\eta$ index (bottom panel) is similar to that of most stars in the FoV. 
WDFS1314-03 is found to be stable and kept in our network of standards. For more details about this DAWD please see Section~\ref{sec:results}.

\subsection{\emph{TESS} data \label{tess}}
We also matched our sample of DAWDs with \emph{TESS} archive and found data for 35 of them (WDFS0228-08, WDFS0815+07, and SDSSJ203722.169-051302.964 do not have available observations). 
All \emph{TESS} data were reduced with the \texttt{TESSreduce} package, which produced flux calibrated difference imaged light curves \citep{tessreduce}. Since calibrating \emph{TESS} data between sectors remains unreliable, for each sector we subtracted the median flux of the sector from the DAWD light curve. Although this process limits the effects of cross sector calibration, it may flatten periods that last for longer than a sector ($\sim27$~days). We checked all the light curves for variability by using a Lomb-Scargle periodigram analysis. In the case of some DAWDs, the light curves contain residual flux from the periodic \emph{TESS} scattered light background, and have issues with image alignment. DAWDs affected by these reduction artifacts contained periodic signals from the drift in image alignment and spikes in the background, substantially decreasing the $S/N$ of the observations. For these targets it is difficult to definitively asses the absence or presence of variability.

\emph{TESS} data for the DAWDs span both the primary and extended missions, which have cadences of 30~minutes and 10~minutes, respectively. For consistency, we binned the light curves to a cadence of 30~minutes, and 1~hour to increase the $S/N$. Analysis of this data does not show any significant sign of variability, or any peculiarities in the DAWD light curves. However, at the wavelength of the \emph{TESS} filter ($\lambda \sim$8,000 \AA) and due to the instrument low spatial resolution (21\arcsec\ per pixel), the light curves of the DAWDs might be contaminated by faint red neighbor stars. For more details on neighbor star contamination see the discussion in Section~\ref{sec:results} and the NIR image cutouts in the Appendix.

\subsection{PS1 and ZTF data \label{ps}}
We matched almost all northern and equatorial DAWDs with Pan-STARRS1 (PS1) data release~2\footnote{\url{https://catalogs.mast.stsci.edu/panstarrs/}} \citep{magnier20} and ZTF data release~8\footnote{We used the SNAD ZTF viewer for ZTF data \citep{malanchev2021} \url{https://ztf.snad.space} } \citep{bellm2019} single epoch detection archives, except for WDFS1557+55.
All ZTF matches have observations in the $g$ and $R$ filters, with most of the single filter light curves of the DAWDs having a modified $\chi^2$ index below $\sim 3.3$. The only exception is the $r$-band light curve for WDFS0815+07, with a $\chi^2$ index value of 5.9 due to a single bright outlier measurement. Most of the PS1 matches have observations in all five filters, $g, r, i, z, y$, while the faint stars WDFS0103-00, WDFS0228-08, and WDFS0815+07, do not have any measurements in $y$. Almost all light curves have modified $\chi^2$ index below $\sim 6.1$. The two exceptions are the $z$-band light curve of SDSSJ041053.632-063027.580, which shows a faint outlier observation making the index 36.7, and the brightest DAWD, WDFS1514+00, with $\chi^2 = 18.3$ for the $g-$ and $7.3$ for the $r$-band, possibly caused by an underestimate of the measurement uncertainties.


\subsection{Near-infrared excess \label{nir}}
\begin{figure*}
\begin{center}
\includegraphics[height=0.75\textheight,width=0.55\textwidth, angle=90]{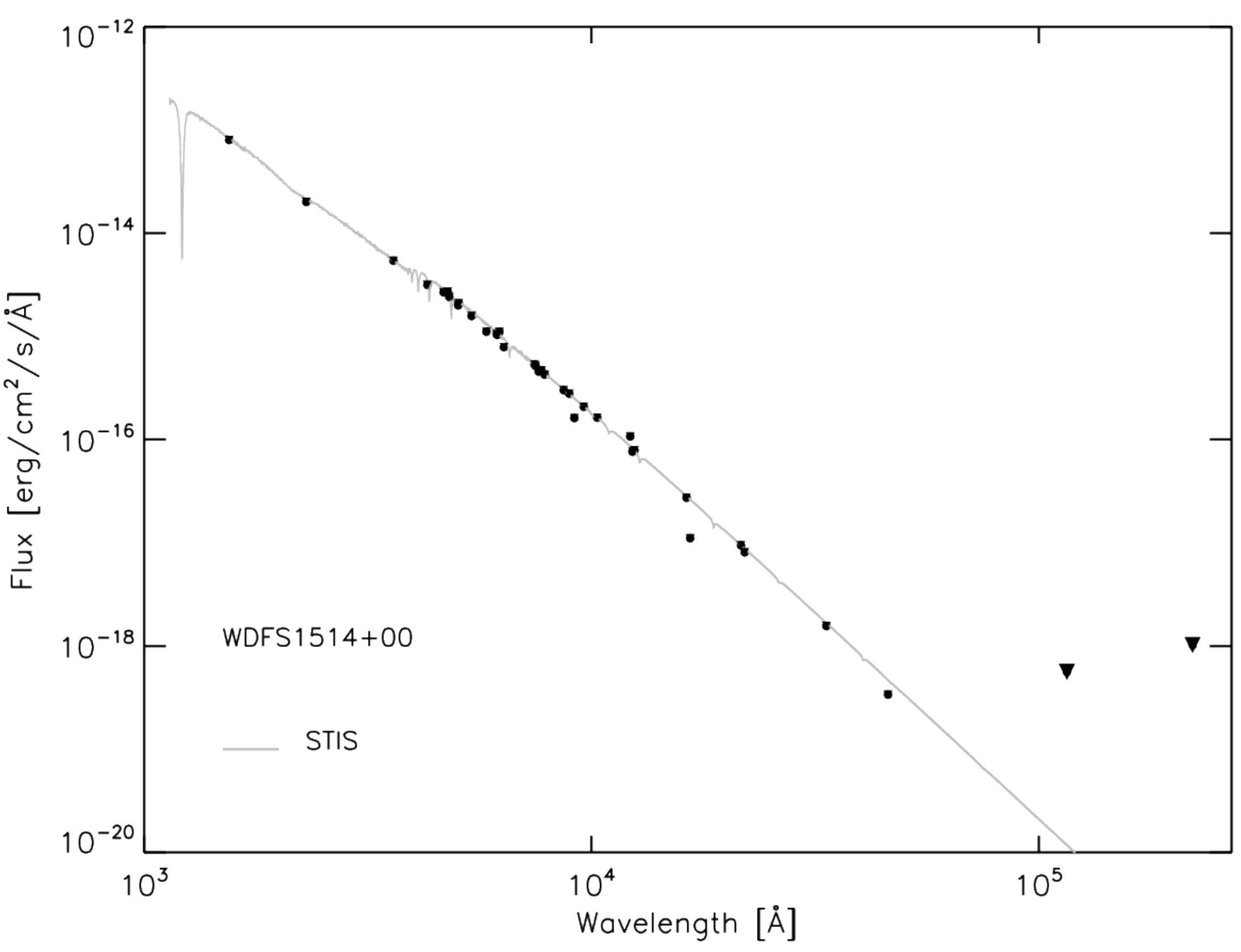}
\caption{CALSPEC STIS spectrum of WDFS1514+00 (grey solid line) with overplotted photometric observations from \emph{GALEX}, SDSS, APASS, \emph{Gaia}, DECam, PS1, DENIS, UKIRT, 2MASS (black filled dots). The WISE upper limit measurements are marked with downward filled triangles. \label{fig:sdssj151421_stis}}
\end{center}
\end{figure*}

For some of the brightest DAWDs that showed hints of variability based on some indices 
near-infrared (NIR) photometry from PS1, UKIRT, DENISE, VISTA and WISE was available and we downloaded it from the VOSA (Virtual Observatory SED Analyzer) database\footnote{http://svo2.cab.inta-csic.es/theory/vosa/index.php} \citep[see Table~\ref{table:4}]{bayo2008}. For four stars, two in the Northern hemisphere (WDFS1110-17, WDFS1302+10) and two in the Southern (WDFS1206-2, WDFS1535-77), no clear NIR excess was identified. However, for other four DAWDs, two in the Northern hemisphere (WDFS1514+00, WDFS2101-05) and two in the Southern (WD0418-534, WDFS1930-52), an upper limit to the NIR excess was identified from WISE photometry.
Fig.~\ref{fig:sdssj151421_stis} shows the SED for star WDFS1514+00 with data from the ultraviolet to the NIR retrieved from VOSA (\emph{GALEX}, SDSS, APASS, \emph{Gaia}, DECam, PS1, DENIS, UKIRT, 2MASS, WISE). The three last points are WISE upper limit measurements. Therefore, it is not possible to confirm the presence of NIR excess for this DAWD neither to fully exclude it. The star is kept as part of our spectrophotometric standard and users should be aware of the possibility of NIR excess that could contaminate its measurements. However, the \emph{HST} SED for this star does not show any indication for a NIR excess (NA19). For more details on neighbor star contamination for this DAWD please see the discussion in Section~\ref{sec:results} and the NIR image cutouts in the Appendix.

Seven of these eight DAWDs were included in our set of spectrophotometric standards with the caveat that photometric measurements from ground-based observatory could be affected by the presence of red faint neighbors. WD0418-534 was excluded due to its potential variability and presence of infrared excess (for more details about these eight DAWDs please see Section~\ref{sec:results} and Table~\ref{table:4}).

\section{Description of findings for each DAWD} \label{sec:results}

In the following we describe results for the 38 DAWDs observed with LCO. All the light curves are shown in the Appendix. In the same section we provide finding charts for all the stars based on WFC3/\emph{HST} $F160W$ images.
Table~\ref{table:4} summarizes the results of the LCO time-spaced data analys and lists parameters for the 32 established spectrophotometric 
standard DAWDs. The discarded DAWDs are listed at the bottom of the same table.
The distribution on the sky of the final network of 32 standards is shown in the Hammer-Aitoff projection of Fig.~\ref{fig:allsky}.

\begin{center}
{\bf Northern and equatorial DAWDs}    
\end{center}

\begin{itemize}
\item {\bf WDFS0103-00}
The light curve of this star does not show signs of variability (see Fig.~\ref{fig:sdssj010322} in the Appendix). The $\chi^2$ index is 2.1 and the dispersion of the LCO photometric measurements is $\sigma \sim$ 0.03 mag (see Table~\ref{table:4}, where the values $\chi^2$ is also listed), compared to the {\it stable} star dispersion of $\sigma \sim$ 0.02 mag.
The WFC3 $F160W$ image for this star (see Fig.~\ref{fig:chart_north1} in the Appendix) shows a neighbor faint red source at $\sim$ 5\arcsec~ that could contaminate the LCO photometry.
WDFS0103-00 is classified as a stable WD by \citet{kleinman2013}.
On the basis of this evidence, we keep this DAWD in our sample of standard stars and warn the users about the close by faint red star when observing from ground-based observatories.

\item {\bf WDFS0228-08} 
The light curve of this star shows some hints of variability (Fig.~\ref{fig:sdssj022817}). The $\chi^2$ index is 2.2, and the dispersion of the LCO measurements is $\sigma \sim$ 0.04 mag,  compared to the {\it stable} star average dispersion of 
$\sigma \sim$ 0.03 mag. The WFC3 $F160W$ image (Fig.~\ref{fig:chart_north1}) shows a faint red source at $\sim$ 1\arcsec~ distance. The source is not visible in the $F775W$ images and in the bluer WFC3-UVIS filters but could contaminate the LCO photometry. 
We do not exclude this DAWD from our sample of standards but we warn the
users when observing this star from ground-based observatories due to the very close by red source.

\item {\bf WDFS0248+33}
The light curve of this star does not show clear signs of variability (Fig.~\ref{fig:sdssj024854}). The $\chi^2$ index is 1.5, with a dispersion
of the measurements of $\sigma \sim$ 0.04 mag, compared to the {\it stable} star dispersion of 
$\sigma \sim$ 0.03 mag. 
This DAWD is included in our sample of standard stars. However, we caution the user about the high estimated reddening towards this stars, i.e., $A_V \sim$ 0.30 mag.

\item {\bf SDSSJ041053.632-063027.580}
The light curve of this star shows a few hints of variability (Fig.~\ref{fig:sdssj041053}). The $\chi^2$ index is 2.7, with a dispersion of the measurements of $\sigma \sim$ 0.03 mag compared to the {\it stable} star average dispersion of 
$\sigma \sim$ 0.02 mag. 
The WFC3 $F160W$ image shows the presence of a few faint red sources at distances in the range
4--6\arcsec, and 4 luminous galaxies at about 10\arcsec distance (Fig.~\ref{fig:chart_north1}). The galaxies are also visible in the WFC3-UVIS $F775W$ images, but not in WFC3-UVIS images collected in bluer filters.
This star was classified as a DA:ME from \citet{kleinman2013}, i.e., it has a
faint M-dwarf companion. We then exclude this DAWD from our sample of standard stars (see also NA19).

\item {\bf WD0554-165}
This star has already been discussed above and the light curve is shown in Fig.~\ref{fig:wd0554}.
We exclude it from our sample of standard stars.

\item {\bf WDFS0727+32}
The light curve of this star does not show any sign of variability, with a $\chi^2$ index of 0.7 and
a dispersion of the measurements of $\sigma \sim$ 0.01 mag compared to the {\it stable} star dispersion of 
$\sigma \sim$ 0.01 mag. This DAWD is included in our network of standard stars (Fig.~\ref{fig:sdssj072752}).

\item {\bf WDFS0815+07}
The light curve of this star does not show signs of variability, with a $\chi^2$ index of 1.0. 
However, the LCO photometry seems quite noisy, with a dispersion of the measurements of 
$\sigma \sim$ 0.03 mag, smaller compared to the average dispersion of the selected {\it stable} stars, $\sigma \sim$ 0.04 mag (Fig.~\ref{fig:sdssj081508}). This DAWD is included in our network of standard stars.

\item {\bf WDFS1024-00}
The light curve of this star does not show signs of variability, with a 
$\chi^2$ index of 0.7 and a dispersion of the measurements of $\sigma \sim$ 0.02 mag, comparable to the average dispersion of the selected {\it stable} stars, $\sigma \sim$ 0.02 mag (Fig.~\ref{fig:sdssjsdss102430}).
We keep this DAWD in our network of standard stars.

\item {\bf WDFS1110-17}
The light curve of this star shows some hints of variability. The $\chi^2$ index is 4.3, with a dispersion of the measurements of $\sigma \sim$ 0.03 mag, three times larger than the {\it stable} star average dispersion of 
$\sigma \sim$ 0.01 mag (Fig.~\ref{fig:sdssj111059}). However, the IQR index from LCO and ATLAS photometry is low, 0.05 and 0.28, respectively, and comparable to the measurement dispersion and to the IQR index of all other stars in the FoV. 
The inverse of the von Neumann index based on LCO photometry is hight, $\sim$ 7, possibly due to the small number of measurements available (36). On the other hand, the inverse of the von Neumann index based on ATLAS photometry is 1.65, comparable to that of the other stars in the field.
Furthermore, \emph{TESS}, PS1 and ZTF do not show this DAWD as variable, and there is no evidence of IR excess. However, the WFC3 $F160W$ image shows the presence of a few faint red sources at 
$\sim$ 2, 3, and 3.5\arcsec~ that could contaminate the LCO photometry (Fig.~\ref{fig:chart_north1}).
Therefore, we do not exclude this DAWD from our sample of standards but we 
warn the users when observing the star from ground-based observatories due to the 
close by red sources.

\item {\bf WDFS1111+39}
The light curve of this star does not show any sign of variability, with a $\chi^2$ index of 1.6 
and a dispersion of the measurements of $\sigma \sim$ 0.02 mag, slightly larger
than the {\it stable} star dispersion $\sigma \sim$ 0.01 mag (Fig.~\ref{fig:sdssj111127}). This DAWD is included in our sample of standard stars.

\item {\bf WDFS1206+02}
The light curve of this star does not show signs of variability with $\chi^2$ index of 1.1 and
a dispersion of the measurements of $\sigma \sim$ 0.02 mag, comparable 
to the {\it stable} star dispersion $\sigma \sim$ 0.02 mag (Fig.~\ref{fig:sdssj120650}). This DAWD is included in our sample of standard stars.

\item {\bf WDFS1214+45}
The light curve of this star does not show any sign of variability, with a $\chi^2$ index of 1.8 and
a dispersion of the measurements of $\sigma \sim$ 0.02 mag, comparable 
to the {\it stable} star dispersion $\sigma \sim$ 0.02 mag (Fig.~\ref{fig:sdssj121405}). This DAWD is included in our sample of standard stars.

\item {\bf WDFS1302+10}
The light curve of this star shows a few hints of variability, with $\chi^2$ = 9.0 and a dispersion of the measurements $\sigma \sim$ 0.04 mag, compared to the {\it stable} star dispersion, $\sigma \sim$ 0.02 mag (Fig.~\ref{fig:sdssj130234}).
The IQR index based on LCO and ATLAS photometry is 0.06 and 0.07, respectively, and comparable to the measurement dispersion. The $1/\eta$ index from the same datasets is $\sim$ 5 and 1.38, respectively. The relatively high von Neumann index based on LCO data might be due to the low (9) number of measurements available. \emph{TESS}, PS1 and ZTF do not show this DAWD as variable, and there is no evidence of IR excess.
Also, the WFC3 $F160W$ image does not show the presence of close by red sources (Fig.~\ref{fig:chart_north1}). Concluding, we keep this DAWD in our network of standard stars.

\item {\bf WDFS1314-03}
The light curve of this star does not show signs of variability, with $\chi^2$ = 0.7 and 
a dispersion of the measurements of $\sigma \sim$ 0.02 mag, comparable to the {\it stable} star dispersion, $\sigma \sim$ 0.02 mag (Fig.~\ref{fig:sdssj131445}). The IQR and the $1/\eta$ index based on LCO and ATLAS data are also small. However, the WFC3 $F160W$ image shows the presence of faint red neighbor at $\sim$ 0.8, 1, and 3\arcsec~ that could contaminate the LCO photometry (Fig.~\ref{fig:chart_north1}).
Therefore, we do not exclude this DAWD from our sample of standards but we 
warn the users when observing this star from ground-based observatories due to the 
close by red sources.

\item {\bf WDFS1514+00}
The light curve of this star shows some level of variability, with a dispersion
of the measurements of $\sigma \sim$ 0.03 mag compared to the {\it stable} star dispersion of 
$\sigma \sim$ 0.01 mag and a $\chi^2$ index of 6.0 (Fig.~\ref{fig:sdssj151421}). On the other hand, the IQR index based on LCO and ATLAS
photometry is low, 0.04 and 0.05, respectively, and comparable to the measurement dispersion. The $1/\eta$ based 
on these two datasets is $\sim$ 1.7 and 43, but it could be due to the low number of measurements available from LCO and contamination by neighbors in the ATLAS survey.
The WFC3 $F160W$ image shows indeed the presence of faint red neighbors at 
$\sim$ 3 and 4\farcs5~ that could contaminate the photometry and there is hint for the presence of some IR excess, possibly due to the contamination (Fig.~\ref{fig:chart_north1}).
\emph{TESS}, PS1 and ZTF do not show this DAWD as variable, and on the basis of 
all the evidence, we do not exclude this star from our network 
of standards. However, we caution observers when using this DAWD from ground-based 
observatories due to the close by red sources.

\item {\bf WDFS1557+55}
The light curve of this star does not show signs of variability, with a $\chi^2$ index of 1.3
and a dispersion of the measurements of $\sigma \sim$ 0.02 mag comparable 
to the {\it stable} star dispersion $\sigma \sim$ 0.02 mag (Fig.~\ref{fig:sdssj155745}). This DAWD is included in our network of standard stars.

\item {\bf WDFS1638+00}
The light curve of this star does not show signs of variability, with a $\chi^2$ index of
0.9 and a dispersion of the measurements of $\sigma \sim$ 0.02 mag as the {\it stable} star 
dispersion of $\sigma \sim$ 0.02 mag (Fig.~\ref{fig:sdssj163800}).
However, the WFC3 $F160W$ image shows the presence of faint red neighbors at 
$\sim$ 3 and 4\arcsec~ that could contaminate LCO photometry (Fig.~\ref{fig:chart_north2}). 
These red sources are also visible in the $F775W$ images, but disappear in the WFC3-UVIS bluer filter images.
This DAWD is not excluded from our network of standards. However, we warn observers when using this star from ground-based observatories due to the close by red sources.

\item {\bf SDSSJ172135.97+294016.0}
The light curve of this star is quite noisy with a dispersion of the measurements of $\sigma \sim$ 0.04 mag, slightly larger to the {\it stable} star dispersion $\sigma \sim$ 0.03 mag, and $\chi^2$ = 1.4 (Fig.~\ref{fig:sdssj172135}).
This DAWD is excluded from our network of standard stars as a result of its lower $T_{eff}$ (for more information see NA19).

\item {\bf WDFS1814+78}
The light curve of this star does not show any sign of variability, with a $\chi^2$ = 1.0,
and a dispersion of the measurements of $\sigma \sim$ 0.01 mag, comparable to the {\it stable} star dispersion $\sigma \sim$ 0.01 mag (Fig.~\ref{fig:sdssj181424}). This DAWD is included in 
our network of standard stars.

\item {\bf SDSSJ203722.169-051302.964}
This star has already been discussed above and the light curve is shown in the Appendix in Fig.~\ref{fig:sdssj20372}.
We exclude this DAWD from our sample of standard stars due to its binary nature.

\item {\bf WDFS2101-05}
The light curve of this star shows some level of variability, and a $\chi^2$ index of 2.9 (Fig.~\ref{fig:sdssj210150}). However, the dispersion of the measurements is $\sigma \sim$ 0.02 mag, slightly larger compared to the  
{\it stable} star dispersion of $\sigma \sim$ 0.01 mag, and the IQR and $1/\eta$ indices have low values comparable to the other stars in the FoV. The $1/\eta$ index calculated from ATLAS data is very high, but this photometry could be contaminated by neighbor stars.
The WFC3 $F160W$ image shows indeed the presence of faint red neighbors at 
$\sim$ 2 and 5\arcsec~ that could contaminate LCO and ATLAS photometry (Fig.~\ref{fig:chart_north2}). One of the sources
is also visible in the WFC3-UVIS $F775W$ images. 
This DAWD is not excluded from our network of standards. However, we caution observers when using this star from ground-based observatories due to the close by red sources.

\item {\bf WDFS2329+00}
The light curve of this star does not show clear signs of variability. The $\chi^2$ index is 1.6
and the dispersion of the measurements of $\sigma \sim$ 0.02 mag, same as the 
{\it stable} star dispersion $\sigma \sim$ 0.02 mag (Fig.~\ref{fig:sdssj232941}). Moreover, the IQR and the $1/\eta$ indices do not highlight any peculiarities for this star.
On the basis of this evidence, we keep this DAWD in our sample of standard stars.

\item {\bf WDFS2351+37}
This DAWD has already been discussed above and the light curve is shown in Fig.~\ref{fig:sdssj235144}, 
and it is included in our sample of standards.

\end{itemize}

\begin{center}
{\bf Southern DAWDs}    
\end{center}

\begin{itemize}
\item {\bf WDFS0122-30}
The light curve of this star shows some level of variability, with a $\chi^2$ index of 2.3 (Fig.~\ref{fig:a020}). However, the dispersion of the measurements of $\sigma \sim$ 0.03 mag, only slightly larger
compared to the {\it stable} star dispersion $\sigma \sim$ 0.02 mag. Moreover, the IQR and the $1/\eta$ indices do not highlight any peculiarities for this star. This DAWD is included in 
our network of standard stars.

\item {\bf WDFS0238-36}
The light curve of this star shows hints of variability, with a $\chi^2$ index of 6.8 and a
dispersion of the measurements of $\sigma \sim$ 0.03 mag compared to the 
{\it stable} star dispersion $\sigma \sim$ 0.01 mag (Fig.~\ref{fig:sssj023824}). 
The IQR index is comparable to the measurement 
uncertainties, but the $1/\eta$ index is high compared to the other stars in the FoV. However, this could be due to the small number of LCO measurements (17). \emph{TESS}, PS1 and ZTF do not show this DAWD as variable. However, the WFC3 $F160W$ images show the presence of a faint red source at $\sim$ 8\arcsec, also
visible in the WFC3 $F775W$ images, that may contaminate the LCO photometry, since
the quality of observing nights for this star were not very good (see also Table~\ref{table:4}).
Therefore, we keep this DAWD in our network of standards, but warn the users 
when observing this star from ground-based observatories due to the close by red sources.

\item {\bf WD0418-534}
The light curve of this star shows signs of variability, with a $\chi^2$ index of 16.4 and a 
dispersion of the measurements of $\sigma \sim$ 0.03 mag compared to the 
{\it stable} star dispersion $\sigma \sim$ 0.01 mag (Fig.~\ref{fig:wd0418}). The IQR index is compatible with the measurement 
uncertainties, but the $1/\eta$ index is slightly higher compared to the other stars in the FoV.
Unfortunately, no ATLAS data are available for this DAWD. \emph{TESS}, PS1 and ZTF do not show this star as variable. However, the stars might show some infrared excess, that could be due to unseen close by red source. In light of this evidence, we decided to exclude this DAWS from our network of standards.

\item {\bf WDFS0458-56}
The light curve of this star shows hints of variability. The $\chi^2$ index is 3.9 but
the dispersion of the measurements is $\sigma \sim$ 0.02 mag, slightly larger than the 
{\it stable} star dispersion $\sigma \sim$ 0.01 mag (Fig.~\ref{fig:sssj045822}). The IQR and $1/\eta$ indices are comparable to those of the other stars in the FoV. \emph{TESS}, PS1 and ZTF do not show this star as variable. 
On the basis of this evidence, we keep this DAWD in our sample of standard stars.

\item {\bf WDFS0541-19}
The light curve of this star shows hints of variability. The $\chi^2$ index is 3.2 but
the dispersion of the measurements is $\sigma \sim$ 0.03 mag, slightly larger than the 
{\it stable} star dispersion $\sigma \sim$ 0.02 mag (Fig.~\ref{fig:sssj054114}).  The IQR and $1/\eta$ indices are comparable to those of the other stars in the FoV, and \emph{TESS}, PS1 and ZTF do not show this star as variable. 
However, $1/\eta$ calculated from ATLAS photometry is relatively high (33) but this could be due to contamination of the photometry from neighbor stars. The WFC3 $F160w$ images shows a very bright galaxy at $\sim$ 8\arcsec and a bright star at $\sim$ 12 \arcsec that could contaminate ground-based photometry (Fig.~\ref{fig:chart_south}).
Therefore, we keep this DAWD in our network of standard stars. However, we warn observers when using this star from ground-based observatories due to the close by bright red sources.

\item {\bf WDFS0639-57}
The light curve of this star shows hints of variability, with a $\chi^2$ index of 3.2 (Fig.~\ref{fig:sssj063941}). However,
the dispersion of the measurements is $\sigma \sim$ 0.02 mag, slightly larger compared to the 
{\it stable} star dispersion $\sigma \sim$ 0.01 mag. The $1/\eta$ index for this star is quite high (33) compared to that of other stars in the FoV, but this is possibly due to the very low number of LCO measurements (10). \emph{TESS}, PS1 and ZTF do not show this star as variable. 
On the basis of this evidence, we keep this DAWD in our network of standards.

\item{\bf WD0757-606}
The light curve of this star does not show any sign of variability, with a $\chi^2$ index of 1.0 and
a dispersion of the measurements of $\sigma \sim$ 0.02 mag, comparable 
to the {\it stable} star dispersion $\sigma \sim$ 0.02 mag (Fig.~\ref{fig:wd0757}). 
However, this DAWD is in a open cluster, NGC~2516, $\approx$ 40\arcsec~ from a very bright $Be$ star (CD-60 1953), $G \approx$ 9 mag, so its flux could be largely contaminated by the neighbor. Also, this star lies very close to the PSF area of an unidentified Rosat PSPC X-ray source \citep{chu2004}.
This DAWD is then excluded from our network of standards.

\item {\bf WDFS0956-38}
The light curve of this star does not show clear signs of variability. The $\chi^2$ index is 3.1 but
the dispersion of the measurements is $\sigma \sim$ 0.02 mag, slightly larger compared to the
{\it stable} star dispersion $\sigma \sim$ 0.01 mag (Fig.~\ref{fig:sssj095657}). Also, IQR and $1/\eta$ indices are comparable to those of the other stars in the FoV, and \emph{TESS}, PS1 and ZTF do not show this star as variable. 
Therefore, we keep this DAWD in our sample of standard stars.

\item {\bf WDFS1055-36}
The light curve of this star shows a few signs of variability, with a $\chi^2$ index of 3.5. 
However, the dispersion of the measurements is $\sigma \sim$ 0.02 mag,
slightly larger compared to the {\it stable} star dispersion $\sigma \sim$ 0.01 mag (Fig.~\ref{fig:sssj105525}).
The IQR and $1/\eta$ indices from LCO and ATLAS photometry are comparable to those of the other stars in the FoV, and \emph{TESS}, PS1 and ZTF do not show this star as variable. 
However, the WFC3 $F160W$ images shows the presence of faint red neighbors
at $\sim$ 2, 3 and 4\arcsec that could contaminate LCO photometry (Fig.~\ref{fig:chart_south}). 
Therefore, we keep this DAWD in our network of standard stars. However, we warn observers when using this star from ground-based observatories due to the close by red sources.

\item{\bf WDFS1206-27}
The light curve of this star shows some level of variability, with a $\chi^2$ index of 7.7 and 
a dispersion of the measurements of $\sigma \sim$ 0.02 mag, more than 3 times larger
than the {\it stable} star dispersion $\sigma \sim$ 0.005 mag (Fig.~\ref{fig:wd1203}). However, IQR and $1/\eta$ indices 
are comparable to those of the other stars in the FoV. The $1/\eta$ calculated from
ATLAS data is high compared to the values of stars in the field, but this could be due to 
neighbors contaminating the DAWD photometry. 
The WFC3 $F160W$ images show indeed the presence of a faint red source at $\sim$ 5\arcsec, also
visible in the WFC3 $F775W$ images, that could contaminate the LCO and ATLAS photometry (Fig.~\ref{fig:chart_south}). However, no infrared excess seems to be present.
\emph{TESS}, PS1 and ZTF do not show this star as variable. On the basis of this evidence, we keep this 
DAWD in our network of standards, and warn the users 
when observing this star from ground-based observatories due to the close by red sources.

\item {\bf WDFS1434-28}
The light curve of this star does not show signs of variability, with a $\chi^2$ index of 2.0, and
a dispersion of the measurements of $\sigma \sim$ 0.02 mag, compared to the 
{\it stable} star dispersion of $\sigma \sim$ 0.01 mag (Fig.~\ref{fig:sssj143459}). The IQR and $1/\eta$ indices 
based on LCO and ATLAS data are also comparable to those of the other stars in the FoV. 
The WFC3 $F160W$ images show the presence of bright galaxies at $\sim$ 5 and 12\arcsec (Fig.~\ref{fig:chart_south}), also
visible in the WFC3 $F775W$ images, that could contaminate ground-based photometry.
Therefore, we include this DAWD in our network of standards and warn the users 
when observing this star from ground-based observatories due to the close by bright sources.

\item{\bf WDFS1535-77}
The light curve of this star does not show signs of variability, with a $\chi^2$ index of 1.9, and a dispersion of the measurements of $\sigma \sim$ 0.01 mag, comparable
to the {\it stable} star dispersion $\sigma \sim$ 0.01 mag (Fig.~\ref{fig:wd1529}). 
The WFC3 $F160W$ images show the presence of a faint red source at about 5\arcsec, also
visible in the WFC3 $F775W$ images, that could contaminate the LCO photometry (Fig.~\ref{fig:chart_south}). However, no infrared excess seems to be present.
Therefore, we keep this DAWD in our network of standards and warn the users 
when observing this star from ground-based observatories due to the close by red sources.

\item {\bf WDFS1837-70}
The light curve of this star does not show any signs of variability, with a $\chi^2$ index of 1.3 and
a dispersion of the measurements of $\sigma \sim$ 0.02 mag comparable 
to the {\it stable} star dispersion $\sigma \sim$ 0.02 mag (Fig.~\ref{fig:sssj183717}). This DAWD is included in our network of standard stars.

\item {\bf WDFS1930-52}
The light curve of this star shows some level of variability, with a $\chi^2$ index of 11.0 and a 
dispersion of the measurements of $\sigma \sim$ 0.03 mag, three times larger than
the {\it stable} star dispersion $\sigma \sim$ 0.01 mag (Fig.~\ref{fig:sssj193018}). However, the IQR and $1/\eta$ indices 
are comparable to those of the other stars in the FoV. Unfortunately, no ATLAS data are available for this stars but \emph{TESS}, PS1 and ZTF do not show variability. There is an upper limit for the infrared excess of this stars, but there are no identified faint red neighbors in the WFC3 $F160W$ image. 
We then keep this DAWD in our network of standards (Fig.~\ref{fig:chart_south}).

\item{\bf WDFS2317-29}
The light curve of this star shows some level of variability, with a $\chi^2$ index of 5.0 and a 
dispersion of the measurements of $\sigma \sim$ 0.03 mag, three times larger compared to
the {\it stable} star dispersion $\sigma \sim$ 0.01 mag (Fig.~\ref{fig:wd2314}). However, the IQR and $1/\eta$ indices 
are comparable to those of the other stars in the FoV. Unfortunately, no ATLAS data are available for this stars but \emph{TESS}, PS1 and ZTF do not show variability.
The WFC3 $F160W$ image does show the presence of very faint red objects at $\sim$ 2, 3 and 5\arcsec that could contaminated LCO photometry (Fig.~\ref{fig:chart_south}).
This DAWD is then included in our network of standard stars with a warning to the users if observing it from the ground.

\end{itemize}


\section{Summary and conclusions}
In this manuscript we presented a photometric analysis to investigate the stability of a set of 
38 DAWDs, out of which 32 were established as spectrophotometric standards.
The summary of the variability analysis and the list of the selected standards is presented in Table~\ref{table:4}.
Their distribution on the sky is shown in Fig.~\ref{fig:allsky}.

All 38 DAWDs in our set have \emph{HST} WFC3 UVIS and IR photometry, ground-based spectroscopy and LCO time-spaced data.
\emph{HST} photometric and spectroscopic data for 23 DAWDs distributed in the Northern hemisphere 
and around the celestial equators, and the reduction methods, were described in CA19. Final averaged and calibrated magnitudes on the \emph{HST} photometric system, and $T_{eff}$ and log~$g$ for 
these 23 Northern DAWDs, were provided in that work.  

Spectra of 48 candidate WDs in the Southern hemisphere were analyzed here and 15 stars were selected as DAWDs, 
while the others were discarded for being DC WDs, DB degenerates, peculiar or magnetized WDs. 
The selected 15 DAWDs were observed with WFC3/\emph{HST} and the photometry for these stars, as well the methods used to derive the $T_{eff}$ and log~$g$ parameters, will be illustrated in Axelrod et al..

Time-spaced observations in the $g$-band were collected with the LCO network of telescopes for all 
38 DAWDs during 6 semesters in between 2016 and 2018. Our cadence implied observing each star
at least 3 times per night for a few nights in a row and then repeating the same observations monthly during the semester.
The observations could not always be collected due to weather and condition restrictions, so cadence
varied according to the target.
Data were pre-reduced with the LCO Banzai pipeline, while PSF photometry was performed by using a custom-made photometric pipeline that integrates Python with DAOPHOTIV/ALLSTAR and ALLFRAME. 
We derived photometric catalogs with average magnitudes and light curves for all stars identified 
in the FoV ($\approx$ 26\arcmin $\times$ 26\arcmin) including the DAWDs. 
Note that no absolute calibration was performed since the goal of this analysis 
was to investigate the sample of candidate spectrophotometric standards for the presence  of variability. 
However, light curves were corrected for differences in the observing conditions due to the usage
of different detectors (the Sinistro cameras), different telescopes, and different observatories.
Also, light curves were corrected for differences due to the observing conditions, such as seeing, cloud 
cover, and due to the different PSF used in the fits.

Photometry was also performed on the same dataset with DoPhot. This analysis allowed us to identify a color effect in the DAWD observations collected with the LCO telescope "lmsc004" on the Cerro Tololo observing site. The measurements obtained from images collected with this telescope were excluded from the light curves of all the DAWDs.

Time-spaced photometry was analyzed by using several different variability indices, 
namely, a reduced $\chi^2$, the von Neumann ($\eta$) index and the IQR index. 
These indices provided very similar results for most stars.

Moreover, for each target and FoV we selected a sample of {\it stable} stars, approximately in the same magnitude range of the DAWD, measured in all the available images, with a good quality PSF fit and {\it shape} parameter, and low $\chi^2$. The average dispersion of the measurements for the {\it stable} stars is used as a threshold to help establishing the variability of the DAWD.

We also downloaded and used ATLAS time-series photometry in the cyan bandpass when available, i.e., for stars with declination larger than -50 degrees. We calculated the IQR and $\eta$ indices for all stars in the LCO FoV based on ATLAS photometry and used the indices for the DAWDs to help determining their stability.

\emph{TESS}, PS1 and ZTF light curves for some of the DAWDs were also analyzed when available, and did not show any of the stars being variable. However, the low spatial resolution of these surveys and the redder filters used in some of the observations, limit the validity of these datasets for this kind of analysis.

We verified for the presence of IR excess by using WISE photometry available from the Virtual Observatory database for a few of the stars, and obtained upper limits for four DAWDs, out of which one star was excluded from our network, mainly on the basis of the variability detected from the LCO light curve.

In summary, two of the Northern DAWDs, namely SDSSJ203722.169-051302.964 and WD0554-1656, were excluded for the variability detected in their LCO light curves and, in the case of SDSSJ203722.169-051302.964 in the spectra, as also described in CA19 and NA19. 

We then excluded SDSSJ172135.97+294016.0, due to its $T_{eff}$ being lower than 20,000 K, which is the low temperature limit established to select our candidate standard DAWDs, and SDSSJ041053.632-063027.580, classified as a binary with a M-dwarf companion by \citet{kleinman2013}.

From the Southern hemisphere sample of DAWDs, WD0418-534 and WD0757-606 were excluded due to detected variability in the LCO light curves, and in the case of WD0757-606 also for the proximity of a very bright Be stars that could contaminate the ground-based and space observations.


The final list of established spectrophotometric standards includes 32 stars
listed 
in Table~\ref{table:4} of this manuscript and their new names (WDFSXXXX-XX), \emph{Gaia} DR3 coordinates and PMs 
are listed in Table~\ref{table:1}.
At the end of Table~\ref{table:4}, the discarded DAWDs are also listed. These were not assigned a new name.
Fig.~\ref{fig:allsky} shows a Hammer-Aitoff projection of the sky with the distribution of the 32 established 
spectrophotometric standards. These are distributed homogeneously on the sky so that at least two of them would 
be available to be observed from any observatory on the ground at any time at airmass less than two.

Finding charts with positions and \emph{Gaia} magnitudes for all the DAWDs are in the Appendix. 


\acknowledgments
{\it Acknowledgments:}
The authors thank the referee, P. Bergeron, for his very useful comments, that improved the content and presentation of the paper.
They thank D. Buckley for his support to the observations, G. Williams for providing some MMT Director's time, and the MMT staff for their typically excellent help.
They also thank the staff at SOAR for support.
This study was supported by NASA through grant GO-15113 from
the Space Telescope Science Institute, which is operated by AURA, Inc., under NASA contract NAS 5-26555 and the Space Telescope Science Institute. 
The analysis was also supported by the DDRF grant D0001.82481.
E. Olszewski was also partially supported by the NSF through grant AST-1815767.
R. Raddi has received funding from the postdoctoral fellowship programme Beatriu de Pin\'os, funded by the Secretary of Universities and Research (Government of Catalonia) and by the Horizon 2020 programme of research and innovation of the European Union under the Maria Sk\l{}odowska-Curie grant agreement No 801370. C. Stubbs is supported by the US DOE through award DE-SC0007881.
This work has made use of data from the European Space Agency (ESA) mission
\emph{Gaia} (\url{https://www.cosmos.esa.int/gaia}), processed by the \emph{Gaia}
Data Processing and Analysis Consortium (DPAC,
\url{https://www.cosmos.esa.int/web/gaia/dpac/consortium}). Funding for the DPAC
has been provided by national institutions, in particular the institutions
participating in the \emph{Gaia} Multilateral Agreement.
This publication makes use of VOSA, developed under the Spanish Virtual Observatory project supported by the Spanish MINECO through grant AyA2017-84089.
VOSA has been partially updated by using funding from the European Union's Horizon 2020 Research and Innovation Programme, under Grant Agreement nº 776403 (EXOPLANETS-A). 
This work includes data from the Asteroid Terrestrial-impact Last Alert System (ATLAS) project. ATLAS is primarily funded to search for near earth asteroids through NASA grants NN12AR55G, 80NSSC18K0284, and 80NSSC18K1575; byproducts of the NEO search include images and catalogs from the survey area. The ATLAS science products have been made possible through the contributions of the University of Hawaii Institute for Astronomy, the Queen's University Belfast, the Space Telescope Science Institute, and the South African Astronomical Observatory.
G. Narayan and K. Malanchev gratefully acknowledge support from NASA under grant 80NSSC20K0453 issued through the NNH18ZDA001N Astrophysics Data Analysis Program (ADAP). 

\facility{
\emph{LCO}, \emph{HST} (WFC3) , \emph{Gaia}, \emph{ATLAS}, \emph{TESS}, \emph{PanSTARRS}, \emph{ZTF}
}


\clearpage
\bibliographystyle{aa}

\bibliography{calamida}

\clearpage

\appendix

\begin{center}
\subsection*{Northern DAWD light curves}
\end{center}

\begin{figure*}
\begin{center}
\includegraphics[height=0.75\textheight,width=0.55\textwidth, angle=90]{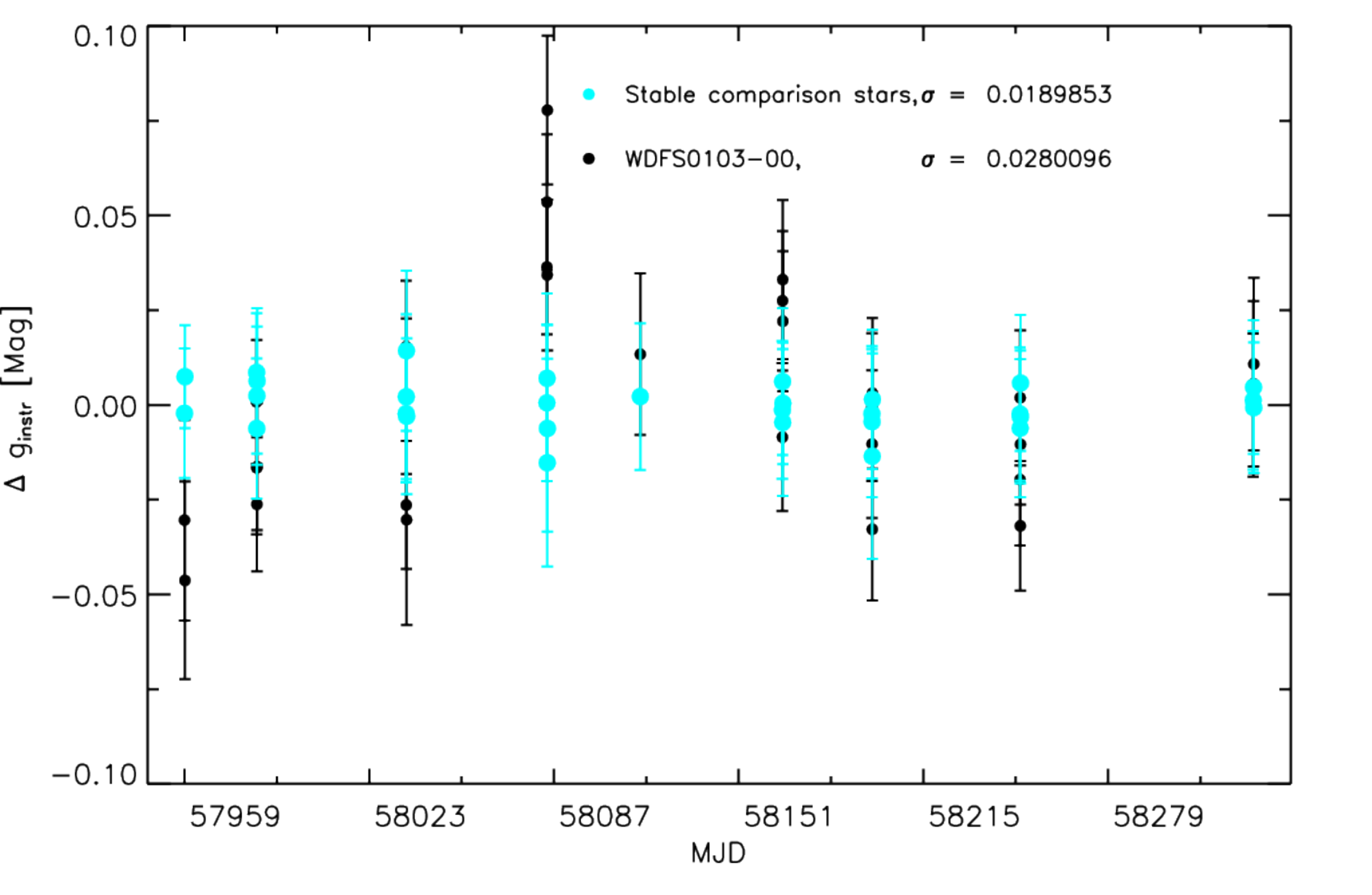} 
\caption{Single epoch minus the mean instrumental magnitude measurements for the Northern DAWD WDFS0103-00 as a function of observing epoch (black crosses). Averaged and binned relative magnitudes for a set of stable stars of comparable
instrumental magnitude in the same FoV are overplotted as a red shaded area. The variability index of the selected stars and the measurement
dispersion are listed. Error bars are shown. \label{fig:sdssj010322}}
\end{center}
\end{figure*}

\begin{figure*}[!h]
\begin{center}
\includegraphics[height=0.75\textheight,width=0.55\textwidth, angle=90]{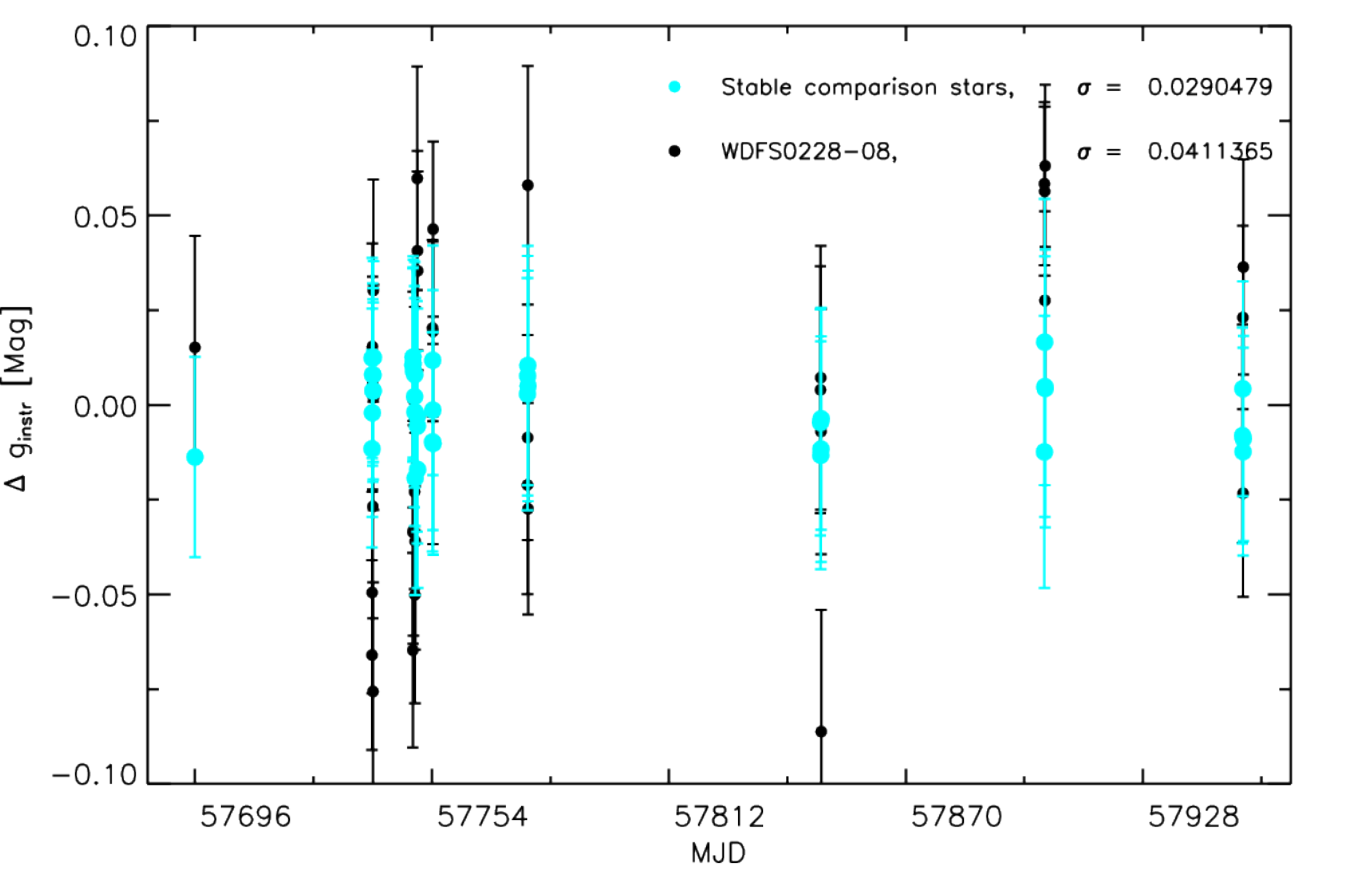} 
\caption{Same as Fig.~\ref{fig:sdssj010322} but for star WDFS0228-08. \label{fig:sdssj022817}}
\end{center}
\end{figure*}

\begin{figure*}[!h]
\begin{center}
\includegraphics[height=0.75\textheight,width=0.55\textwidth, angle=90]{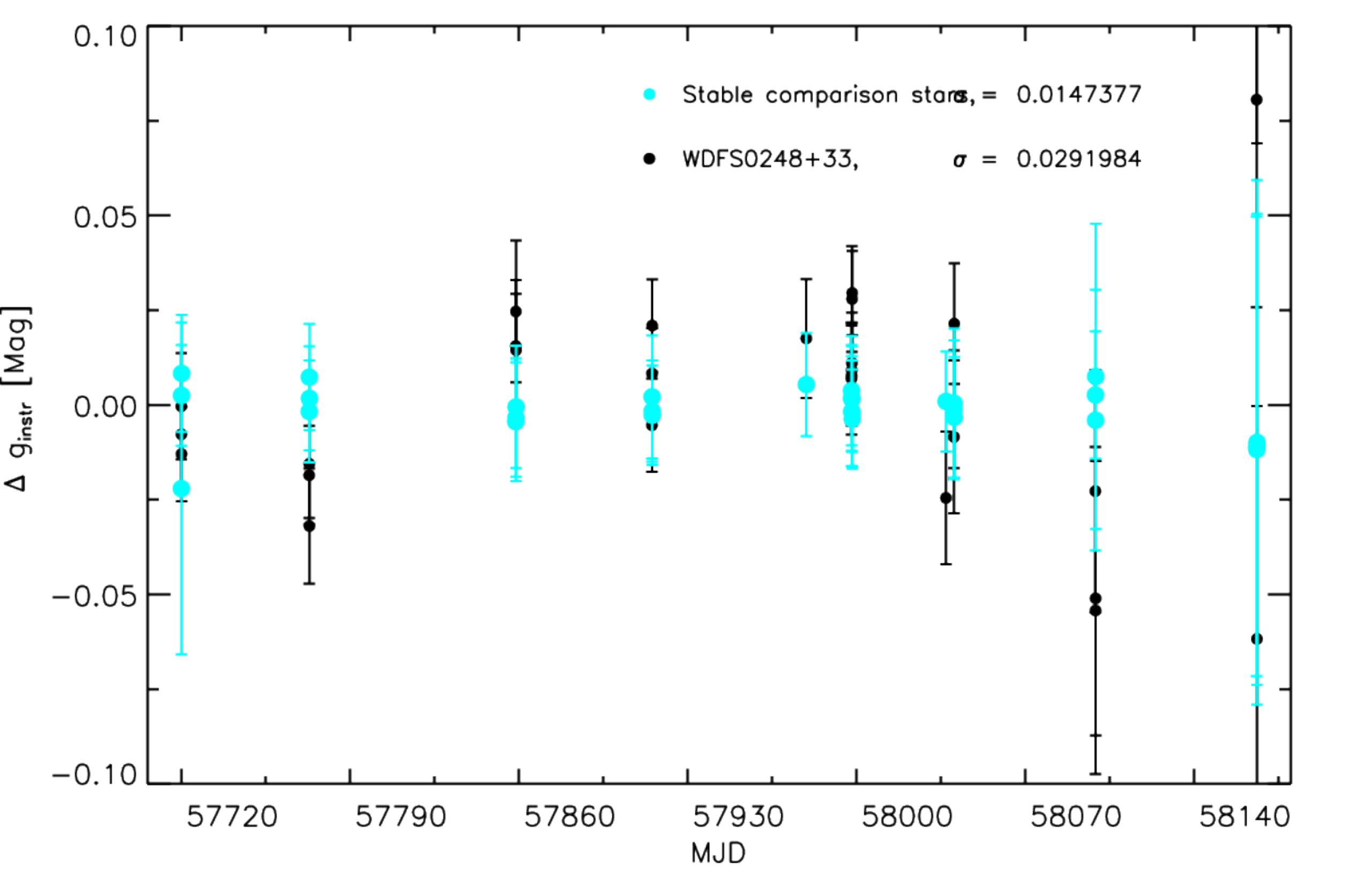} 
\caption{Same as Fig.~\ref{fig:sdssj010322} but for star WDFS0248+33. \label{fig:sdssj024854}}
\end{center}
\end{figure*}

\begin{figure*}[!h]
\begin{center}
\includegraphics[height=0.75\textheight,width=0.55\textwidth, angle=90]{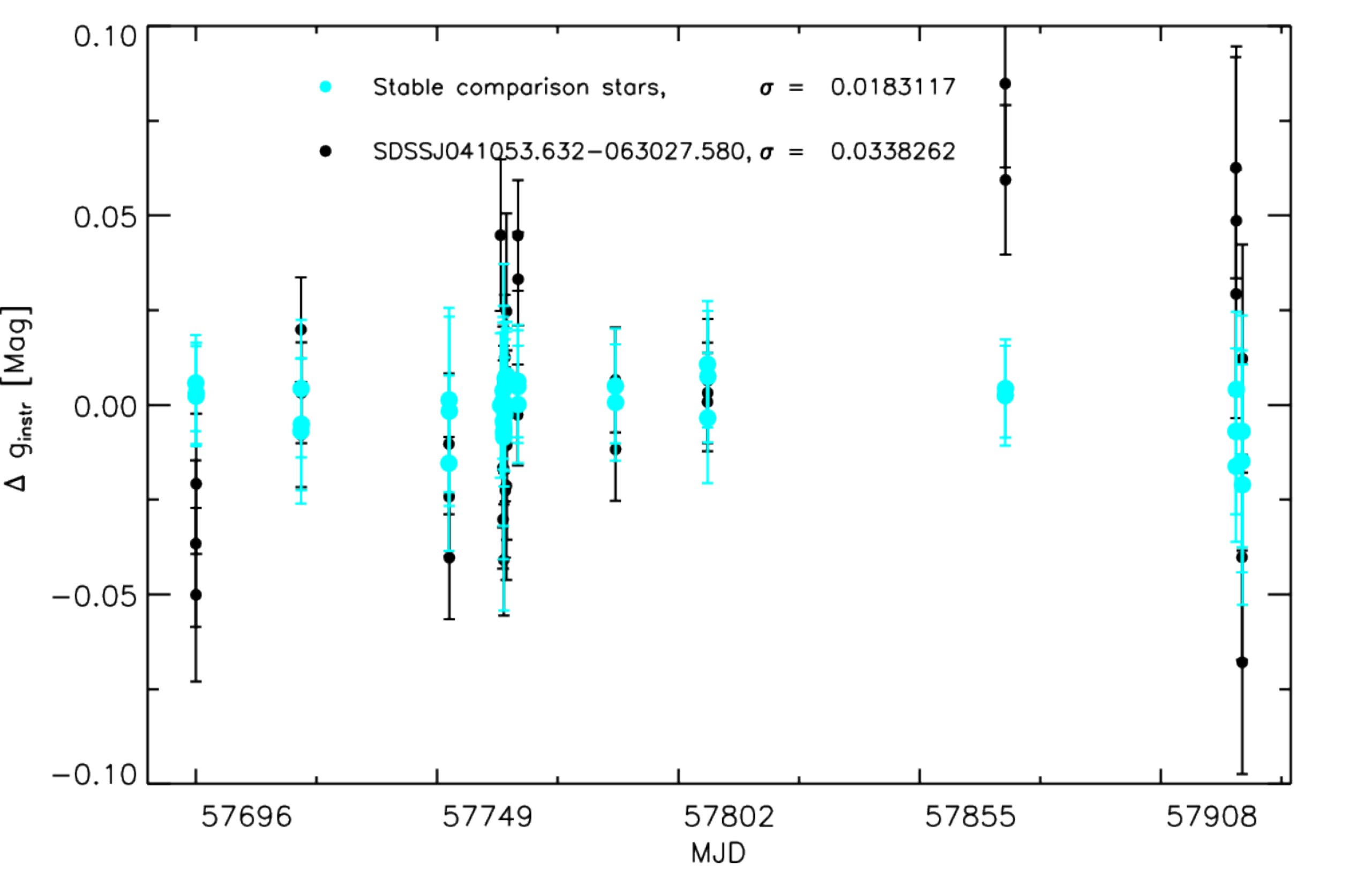} 
\caption{Same as Fig.~\ref{fig:sdssj010322} but for star SDSSJ041053.632-063027.580. \label{fig:sdssj041053}}
\end{center}
\end{figure*}

\begin{figure*}[!h]
\begin{center}
\includegraphics[height=0.75\textheight,width=0.55\textwidth, angle=90]{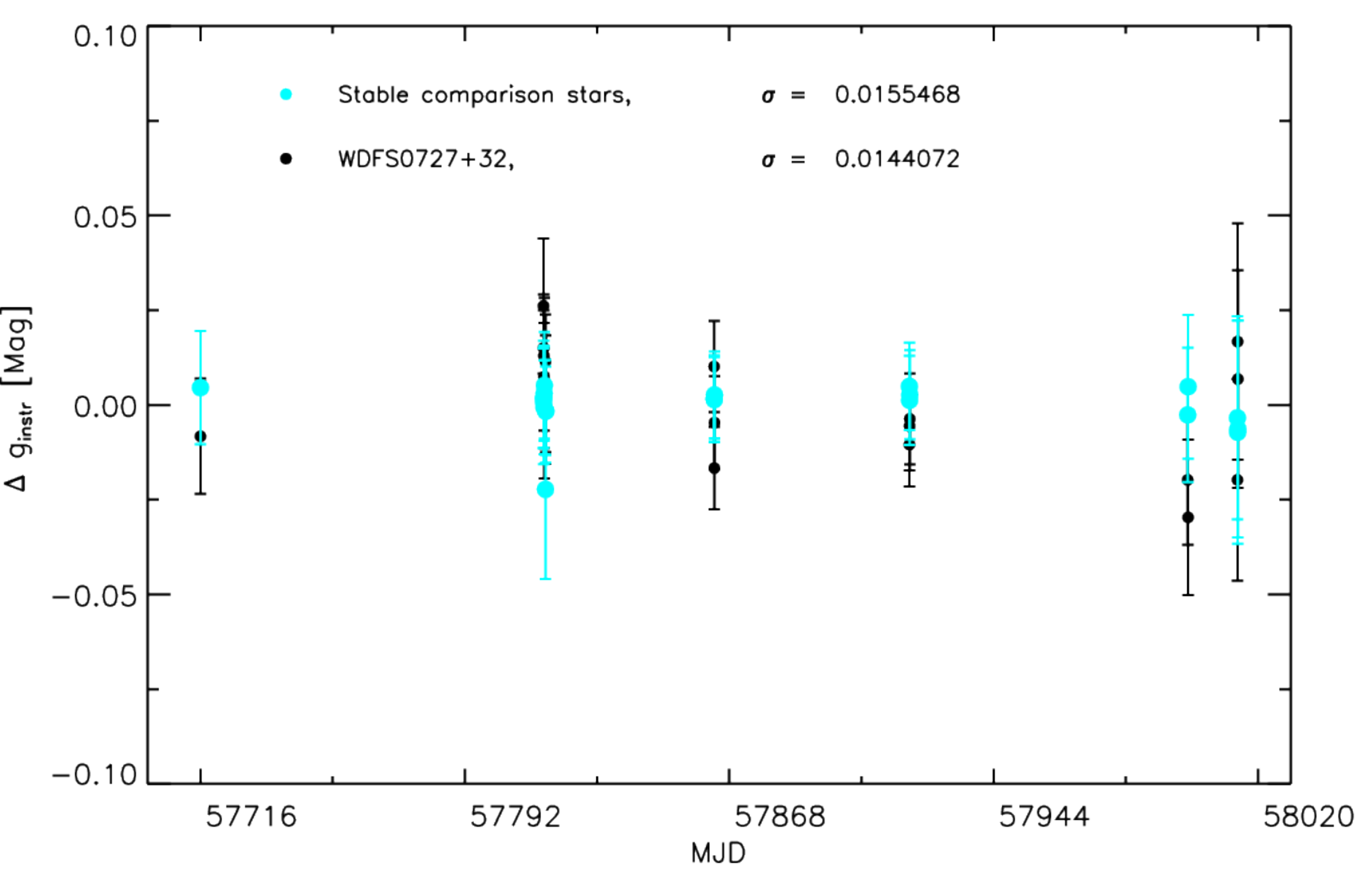} 
\caption{Same as Fig.~\ref{fig:sdssj010322} but for star WDFS0727+32. \label{fig:sdssj072752}}
\end{center}
\end{figure*}

\begin{figure*}[!h]
\begin{center}
\includegraphics[height=0.75\textheight,width=0.55\textwidth, angle=90]{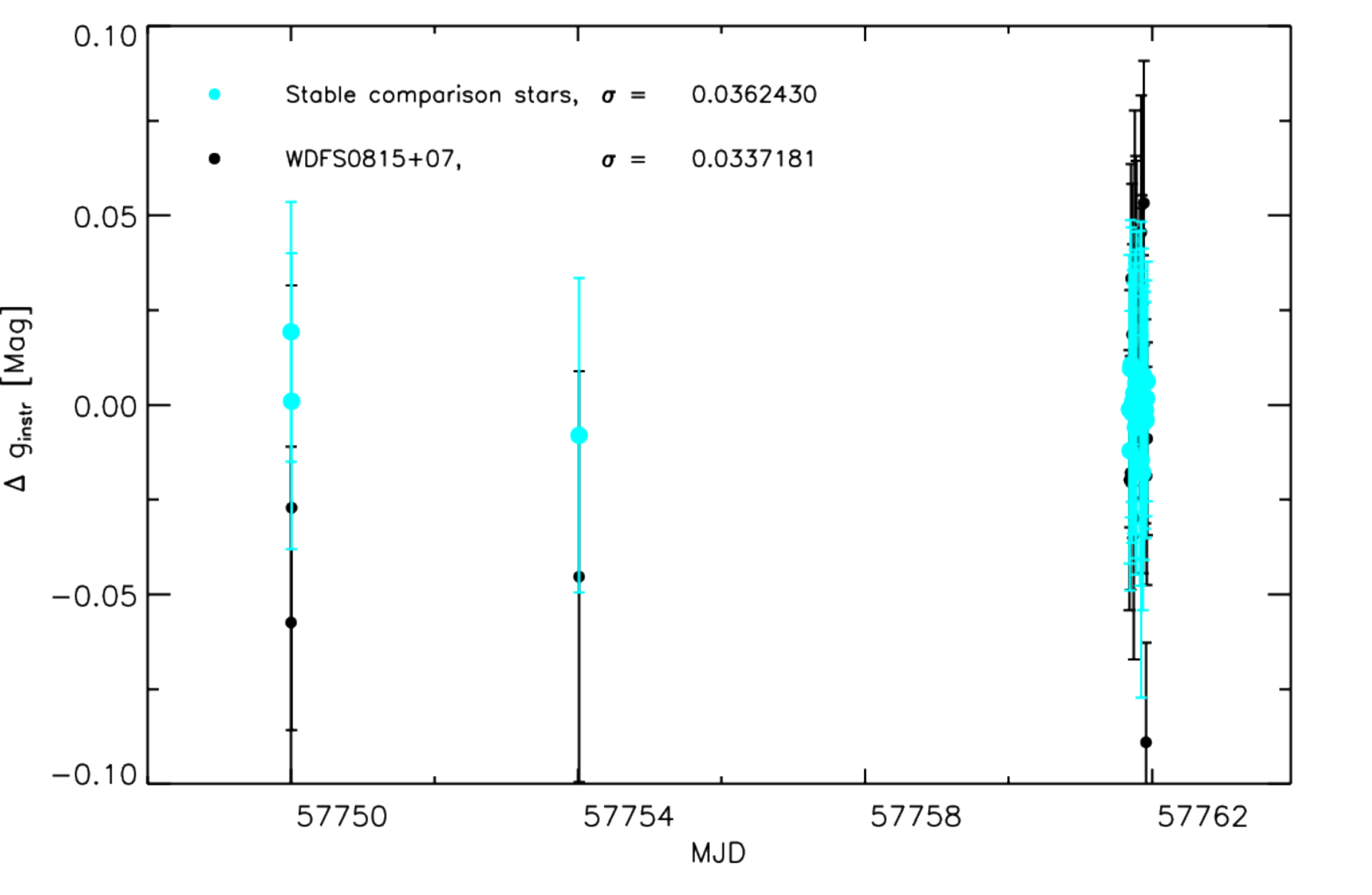} 
\caption{Same as Fig.~\ref{fig:sdssj010322} but for star WDFS0815+07. \label{fig:sdssj081508}}
\end{center}
\end{figure*}

\begin{figure*}[!h]
\begin{center}
\includegraphics[height=0.75\textheight,width=0.55\textwidth, angle=90]{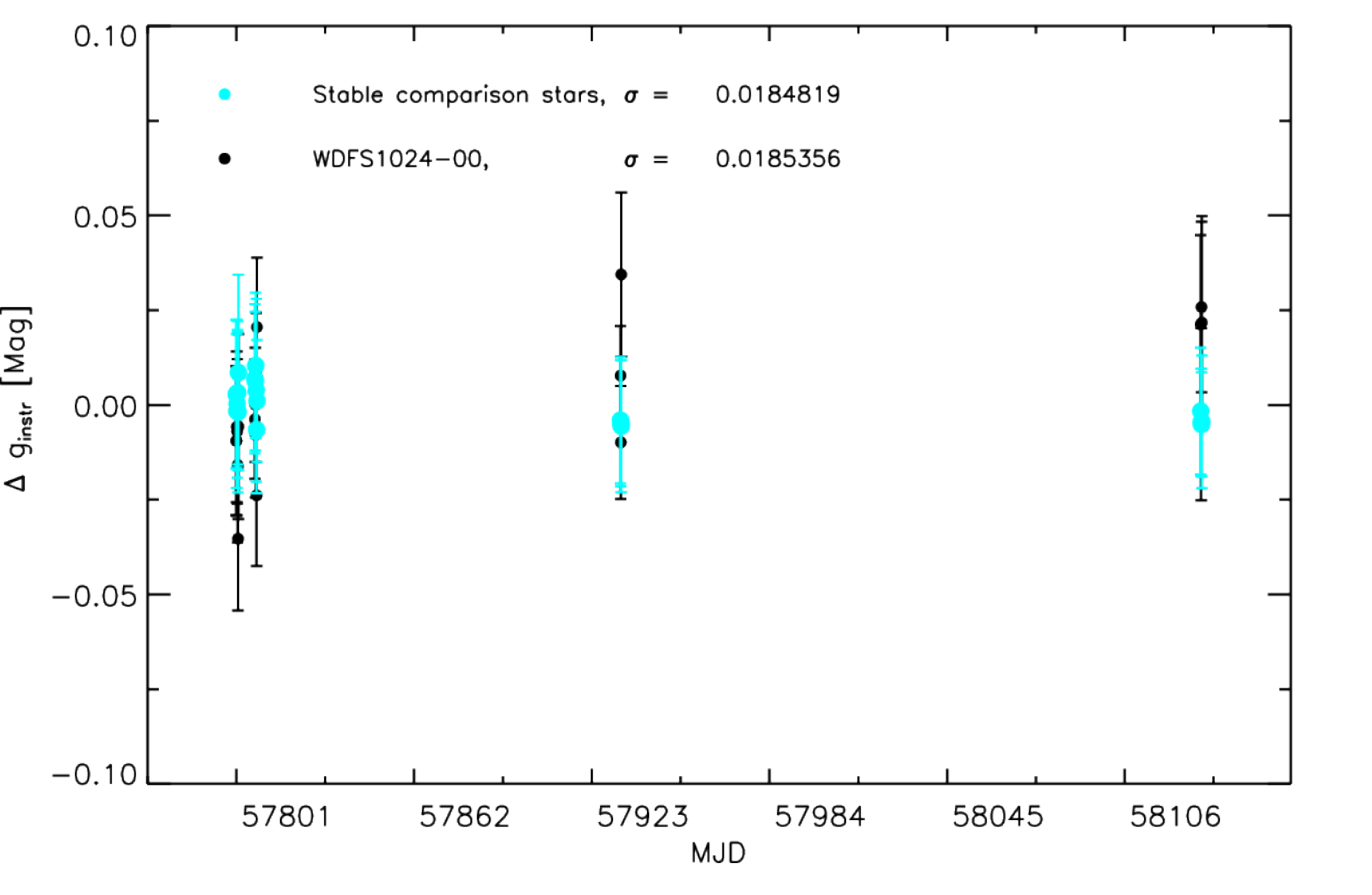} 
\caption{Same as Fig.~\ref{fig:sdssj010322} but for star WDFS1024-00. \label{fig:sdssjsdss102430}}
\end{center}
\end{figure*}

\begin{figure*}[!h]
\begin{center}
\includegraphics[height=0.75\textheight,width=0.55\textwidth, angle=90]{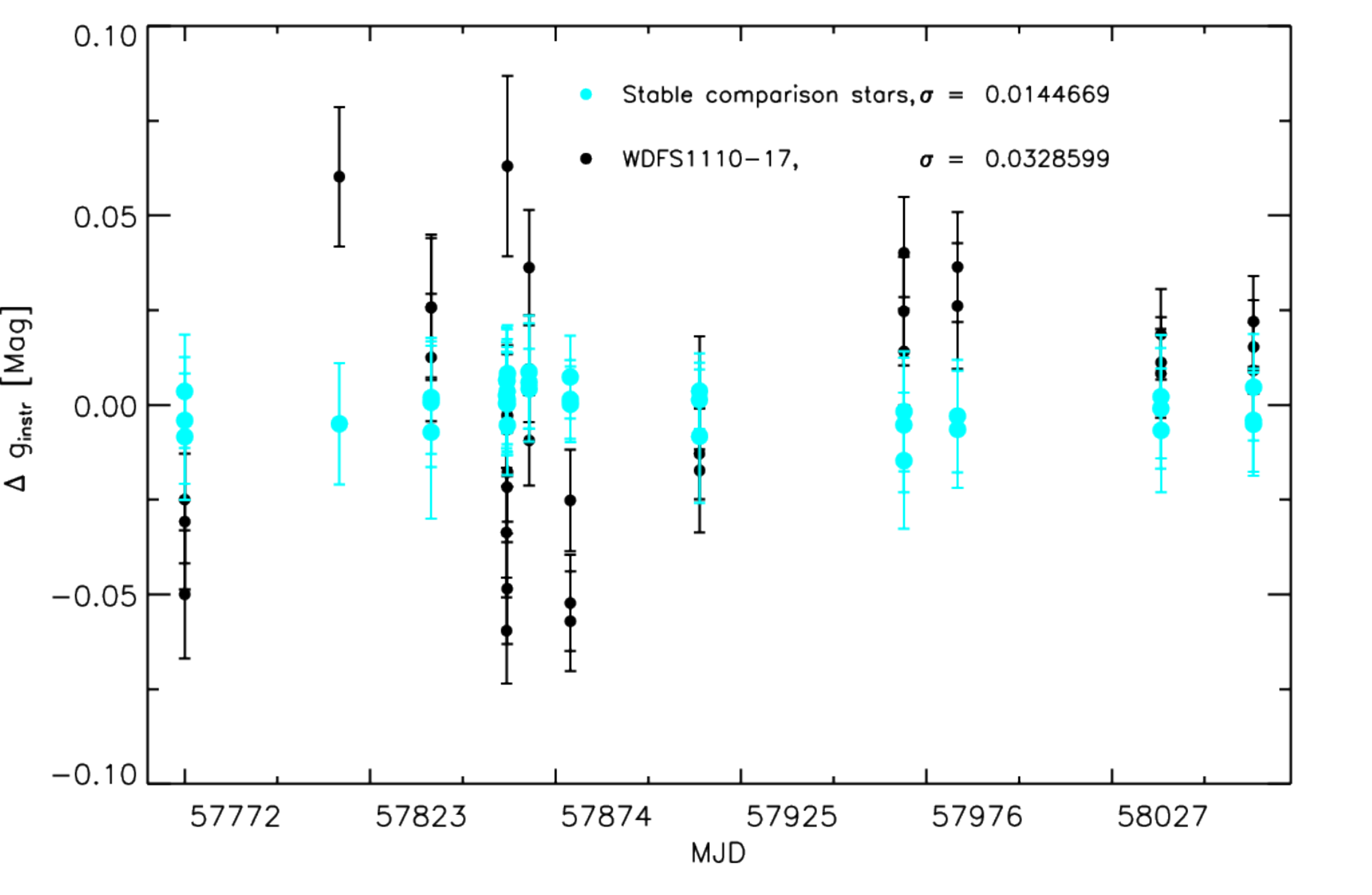} 
\caption{Same as Fig.~\ref{fig:sdssj010322} but for star WDFS1110-17. \label{fig:sdssj111059}}
\end{center}
\end{figure*}

\begin{figure*}[!h]
\begin{center}
\includegraphics[height=0.75\textheight,width=0.55\textwidth, angle=90]{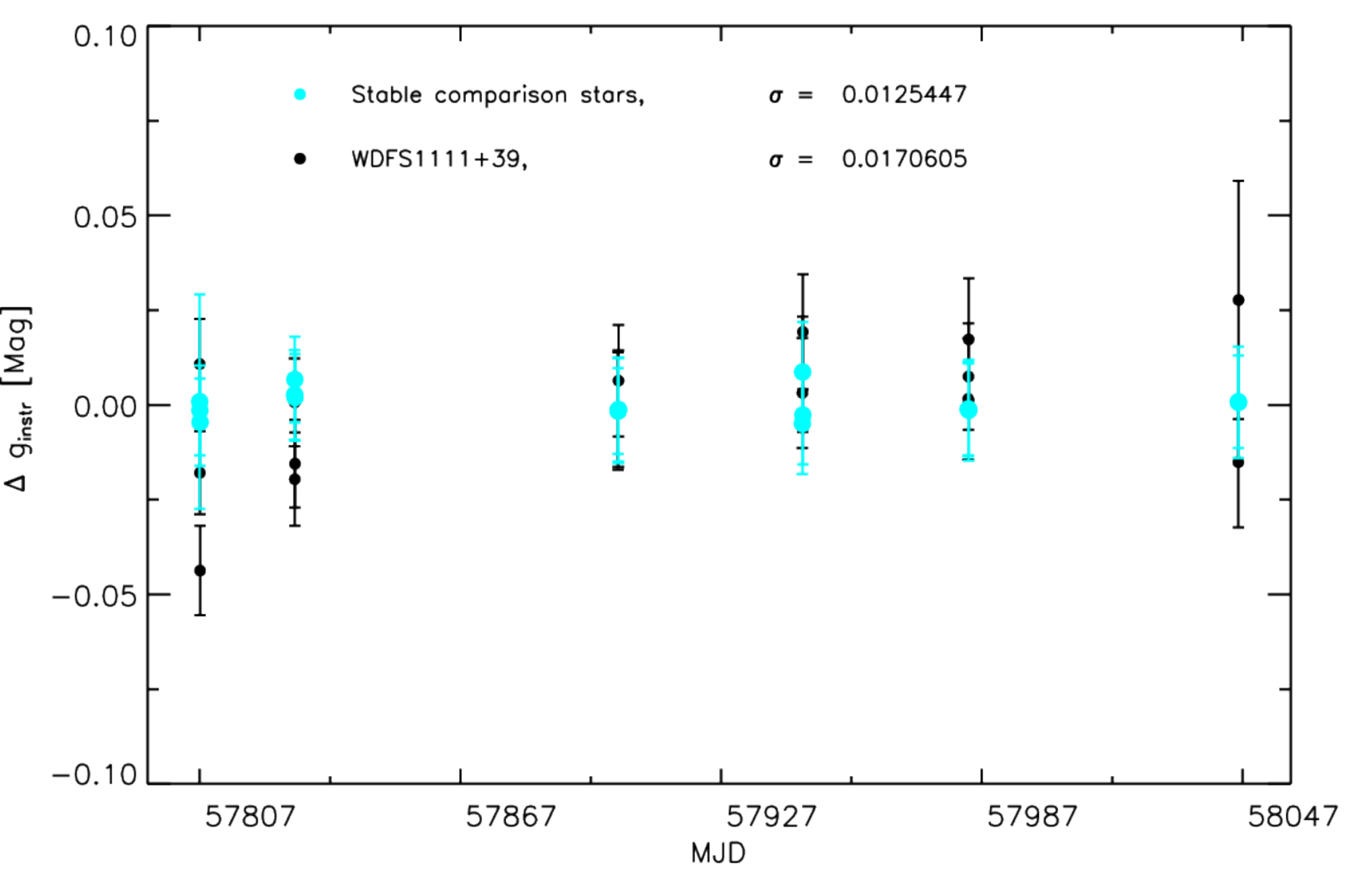} 
\caption{Same as Fig.~\ref{fig:sdssj010322} but for star WDFS1111+39. \label{fig:sdssj111127}}
\end{center}
\end{figure*}

\begin{figure*}[!h]
\begin{center}
\includegraphics[height=0.75\textheight,width=0.55\textwidth, angle=90]{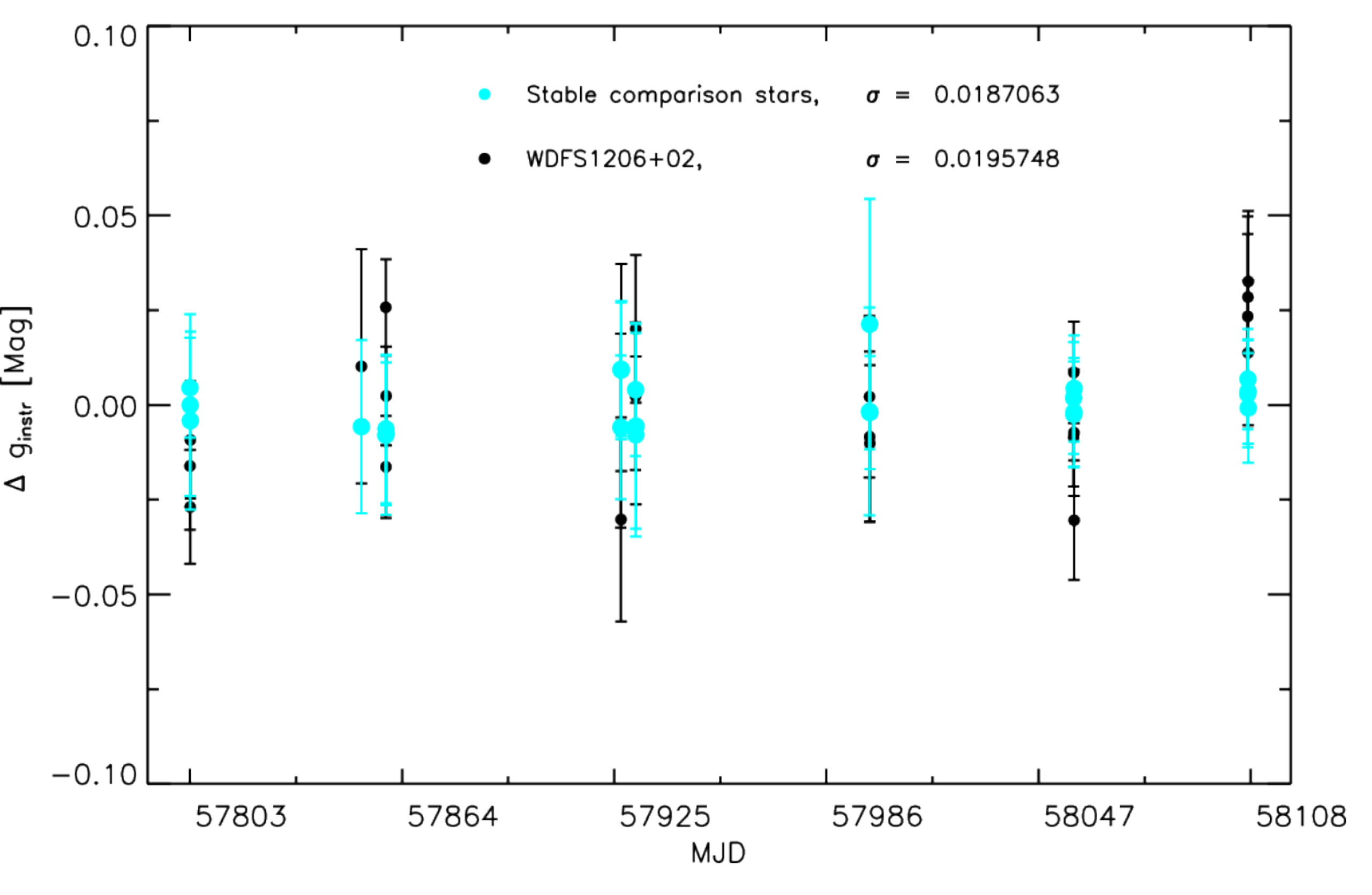} 
\caption{Same as Fig.~\ref{fig:sdssj010322} but for star WDFS1206+02. \label{fig:sdssj120650}}
\end{center}
\end{figure*}

\begin{figure*}[!h]
\begin{center}
\includegraphics[height=0.75\textheight,width=0.55\textwidth, angle=90]{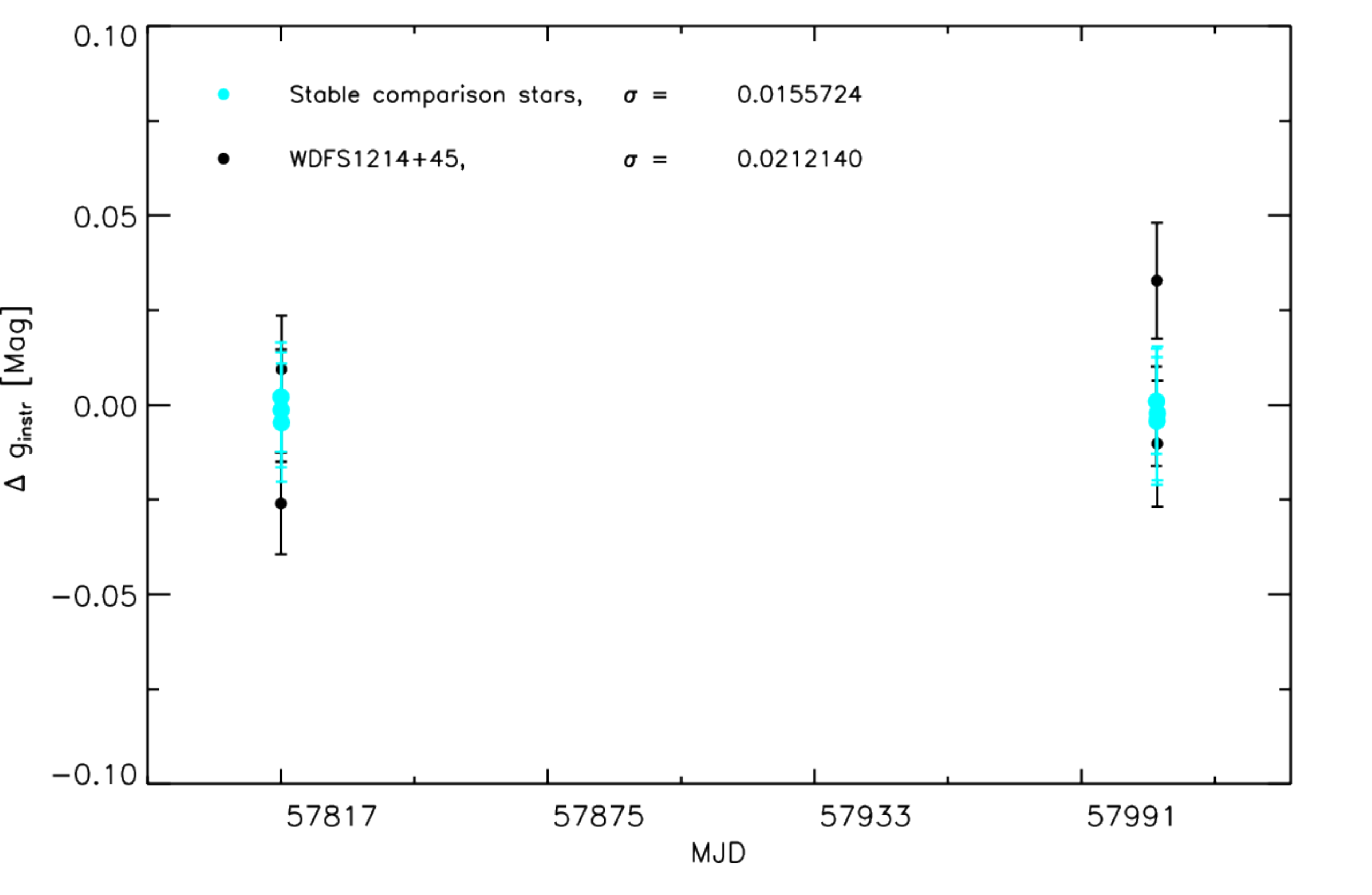} 
\caption{Same as Fig.~\ref{fig:sdssj010322} but for star WDFS1214+45. \label{fig:sdssj121405}}
\end{center}
\end{figure*}

\begin{figure*}[!h]
\begin{center}
\includegraphics[height=0.75\textheight,width=0.55\textwidth, angle=90]{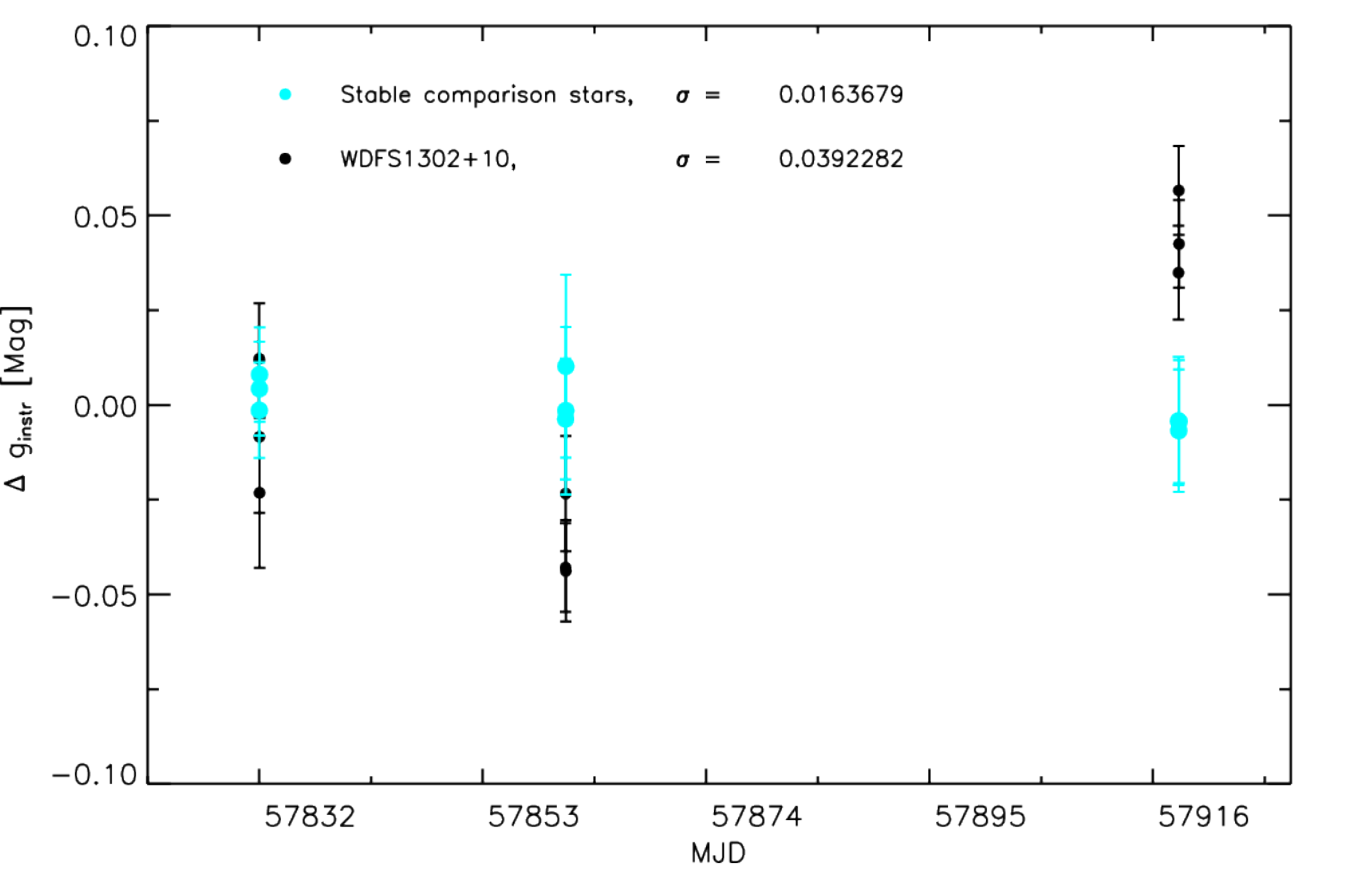} 
\caption{Same as Fig.~\ref{fig:sdssj010322} but for star WDFS1302+10. \label{fig:sdssj130234}}
\end{center}
\end{figure*}

\begin{figure*}[!h]
\begin{center}
\includegraphics[height=0.75\textheight,width=0.55\textwidth, angle=90]{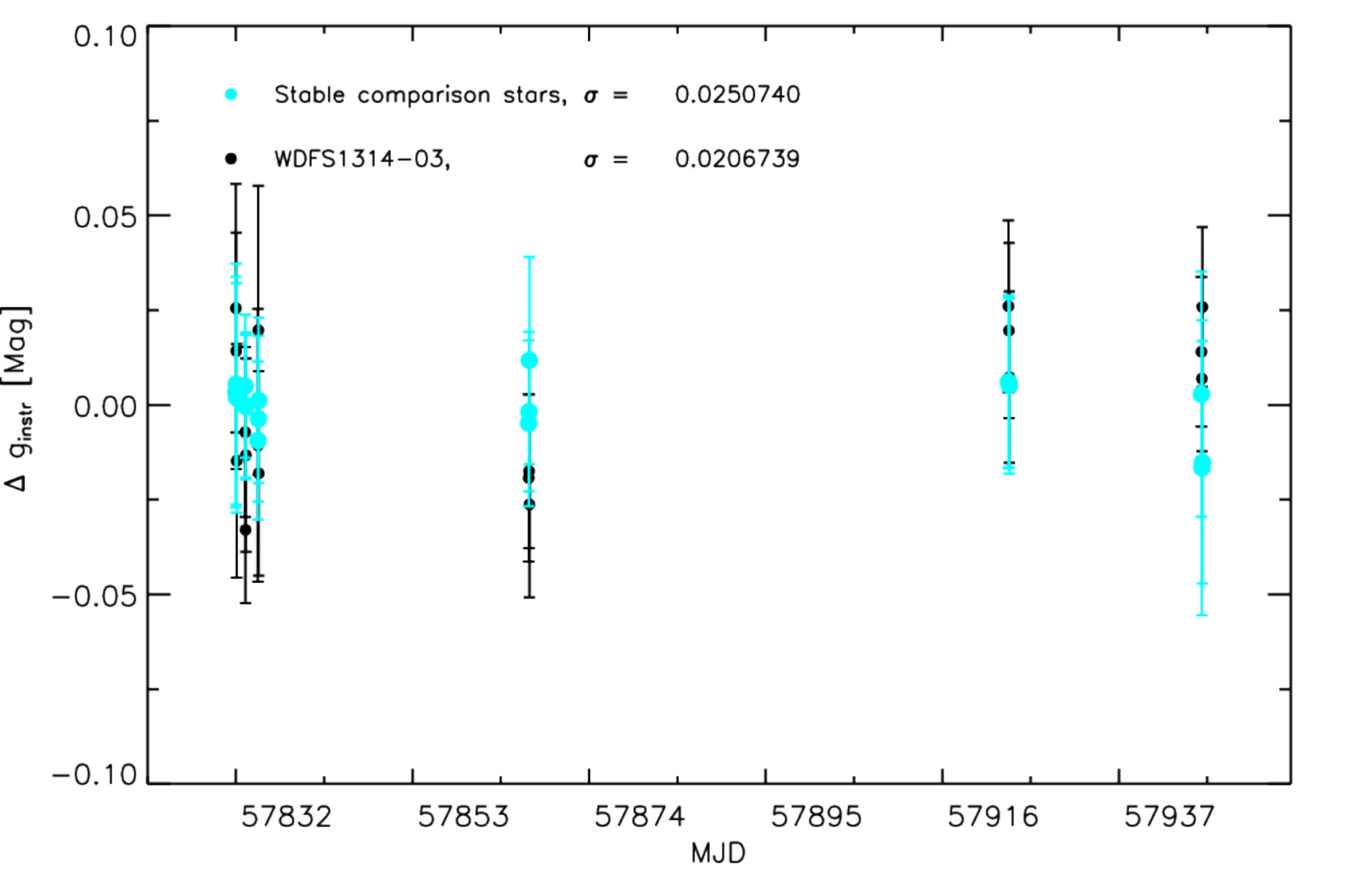} 
\caption{Same as Fig.~\ref{fig:sdssj010322} but for star WDFS1314-03. \label{fig:sdssj131445}}
\end{center}
\end{figure*}

\begin{figure*}[!h]
\begin{center}
\includegraphics[height=0.75\textheight,width=0.55\textwidth, angle=90]{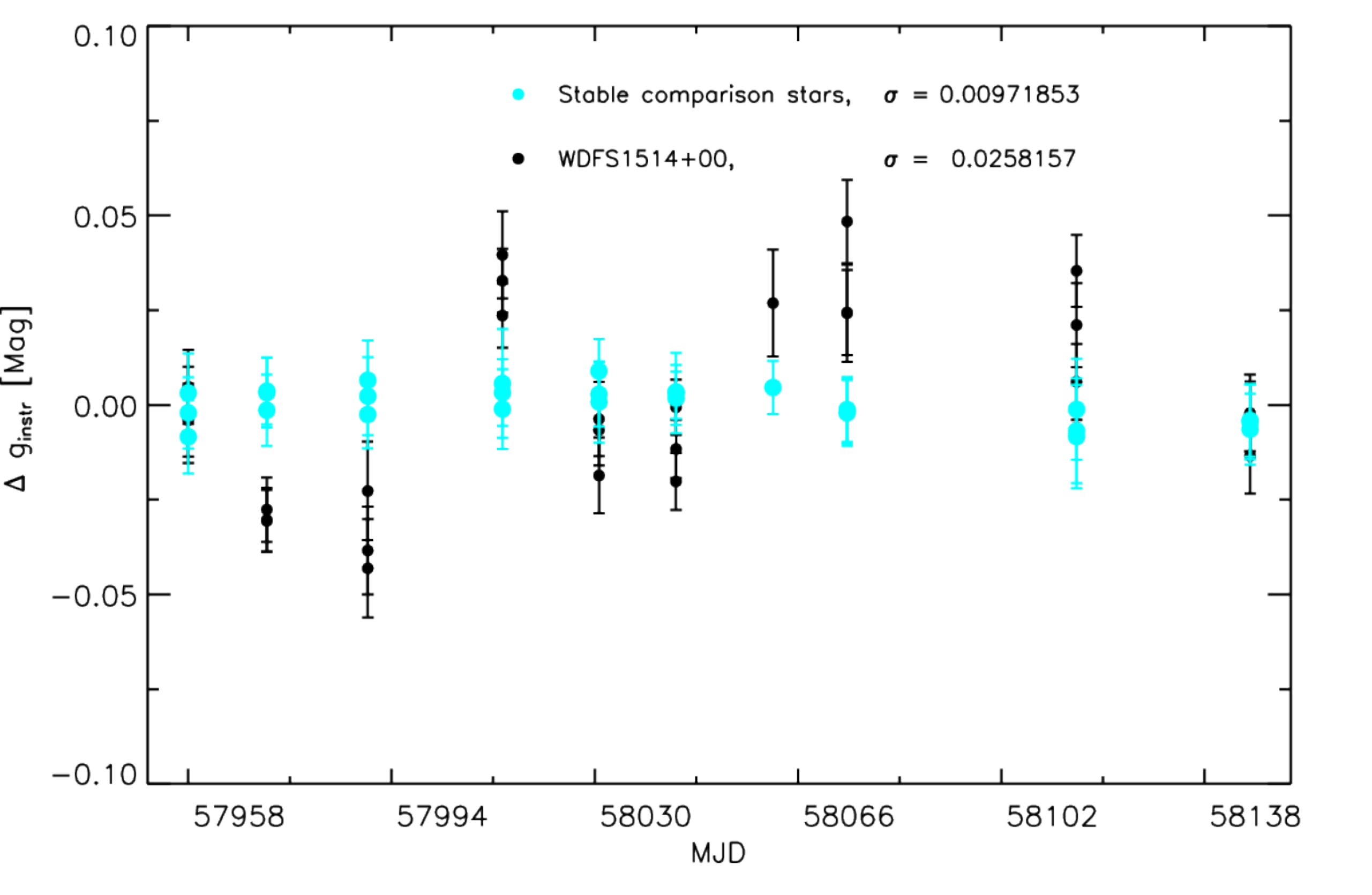} 
\caption{Same as Fig.~\ref{fig:sdssj010322} but for star WDFS1514+00. \label{fig:sdssj151421}}
\end{center}
\end{figure*}

\begin{figure*}[!h]
\begin{center}
\includegraphics[height=0.75\textheight,width=0.55\textwidth, angle=90]{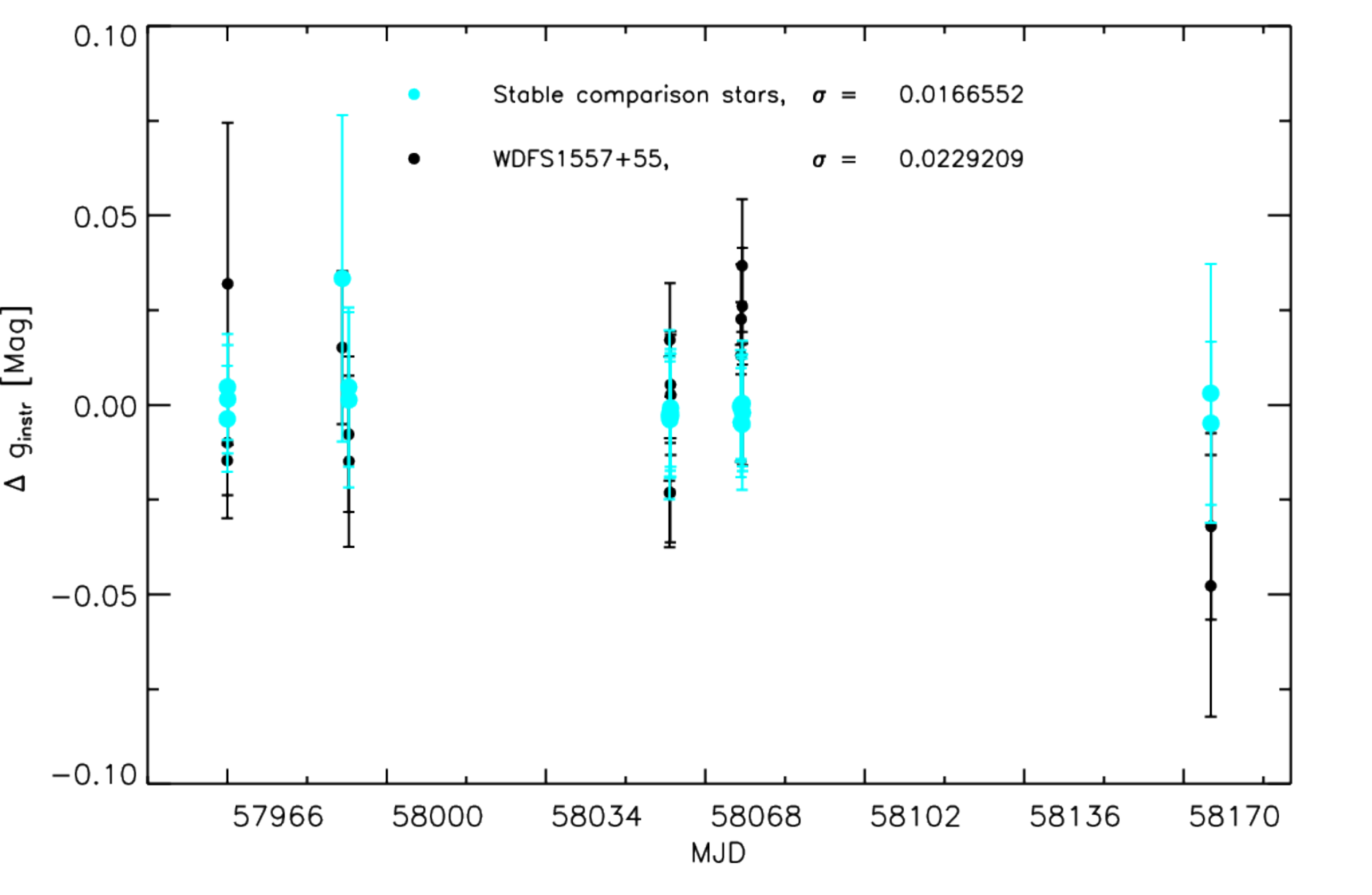} 
\caption{Same as Fig.~\ref{fig:sdssj010322} but for star WDFS1557+55. \label{fig:sdssj155745}}
\end{center}
\end{figure*}

\begin{figure*}[!h]
\begin{center}
\includegraphics[height=0.75\textheight,width=0.55\textwidth, angle=90]{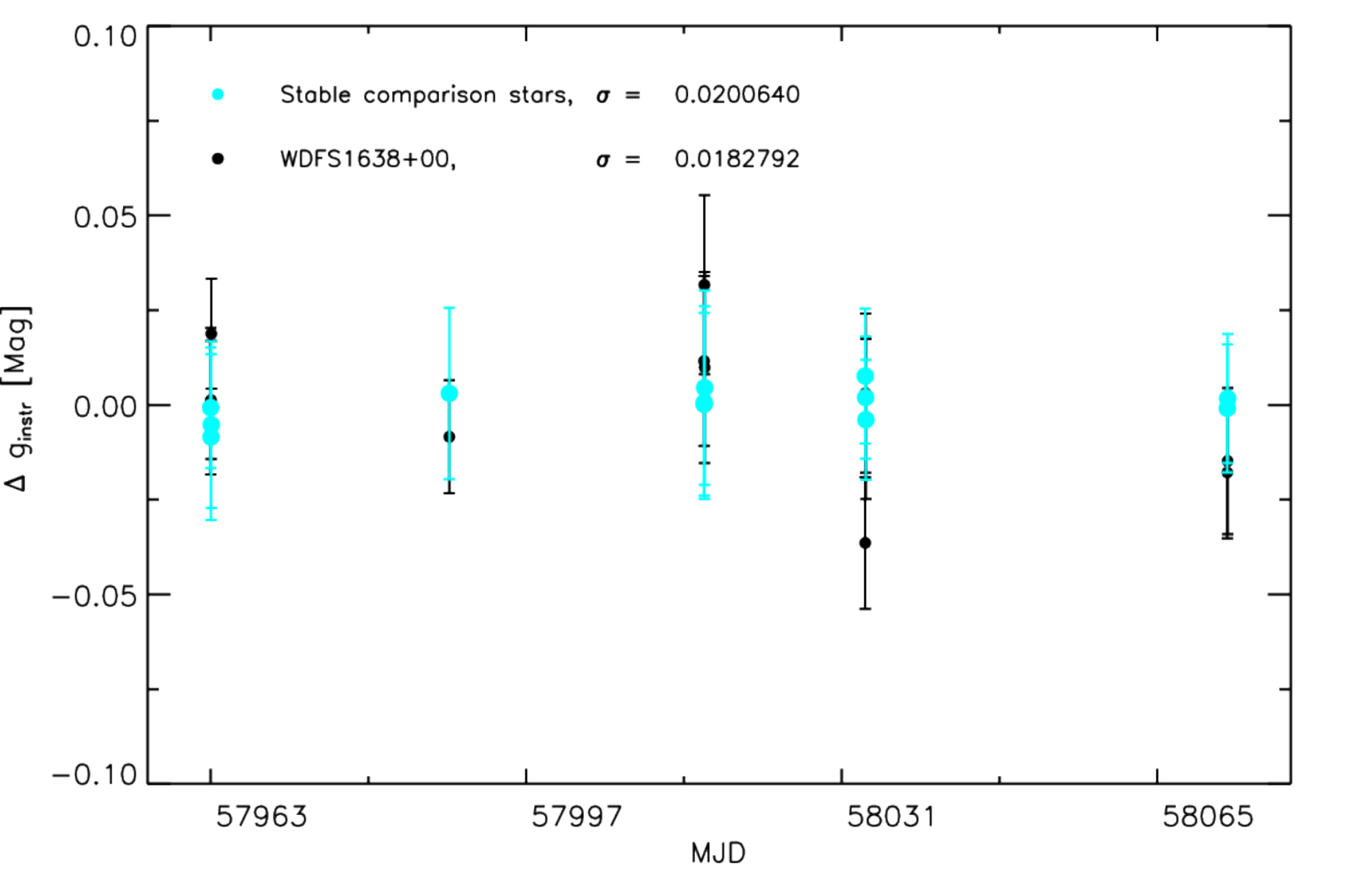} 
\caption{Same as Fig.~\ref{fig:sdssj010322} but for star WDFS1638+00. \label{fig:sdssj163800}}
\end{center}
\end{figure*}

\begin{figure*}[!h]
\begin{center}
\includegraphics[height=0.75\textheight,width=0.55\textwidth, angle=90]{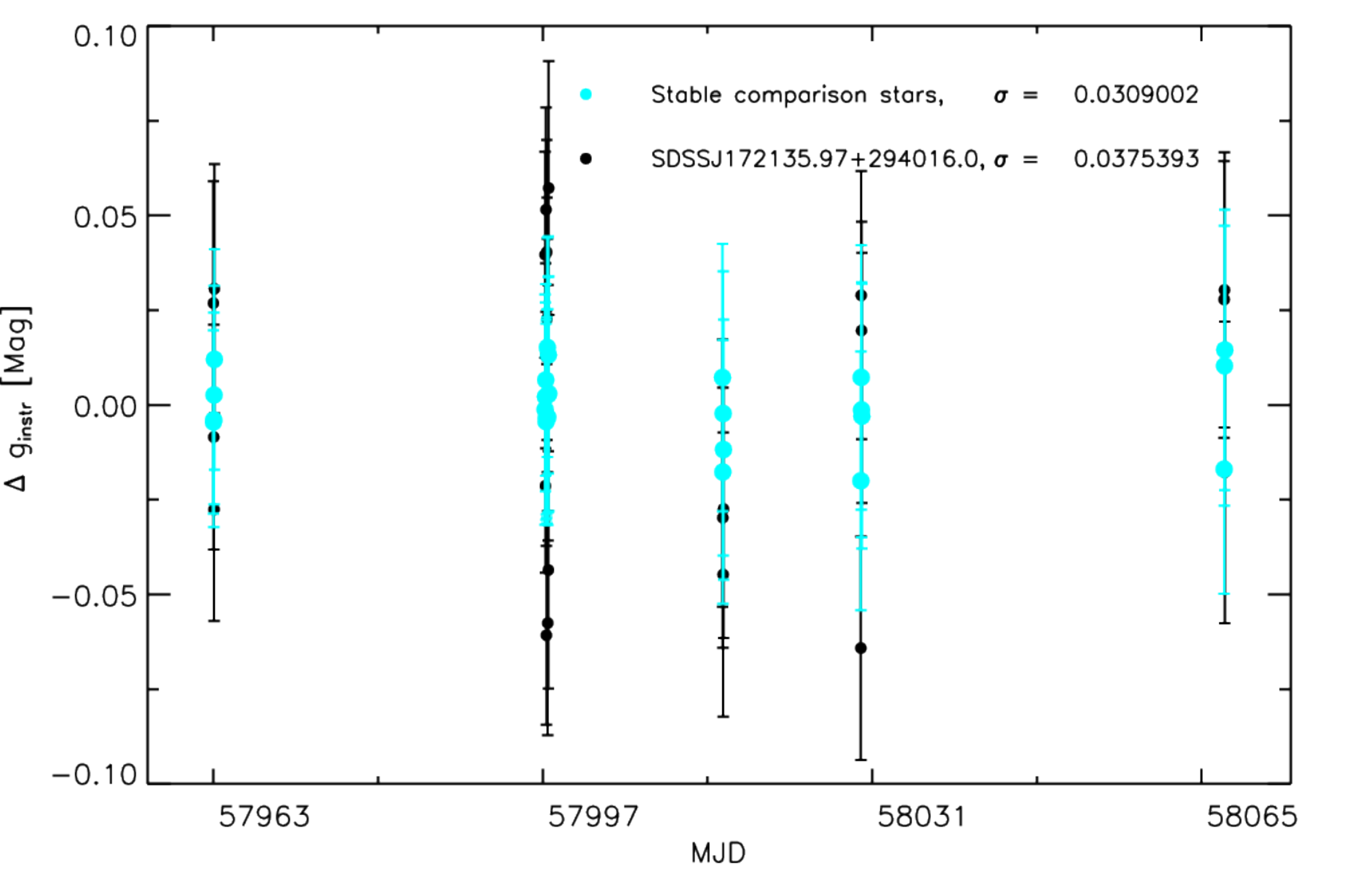} 
\caption{Same as Fig.~\ref{fig:sdssj010322} but for star SDSSJ172135.97+294016.0. \label{fig:sdssj172135}}
\end{center}
\end{figure*}

\begin{figure*}[!h]
\begin{center}
\includegraphics[height=0.75\textheight,width=0.55\textwidth, angle=90]{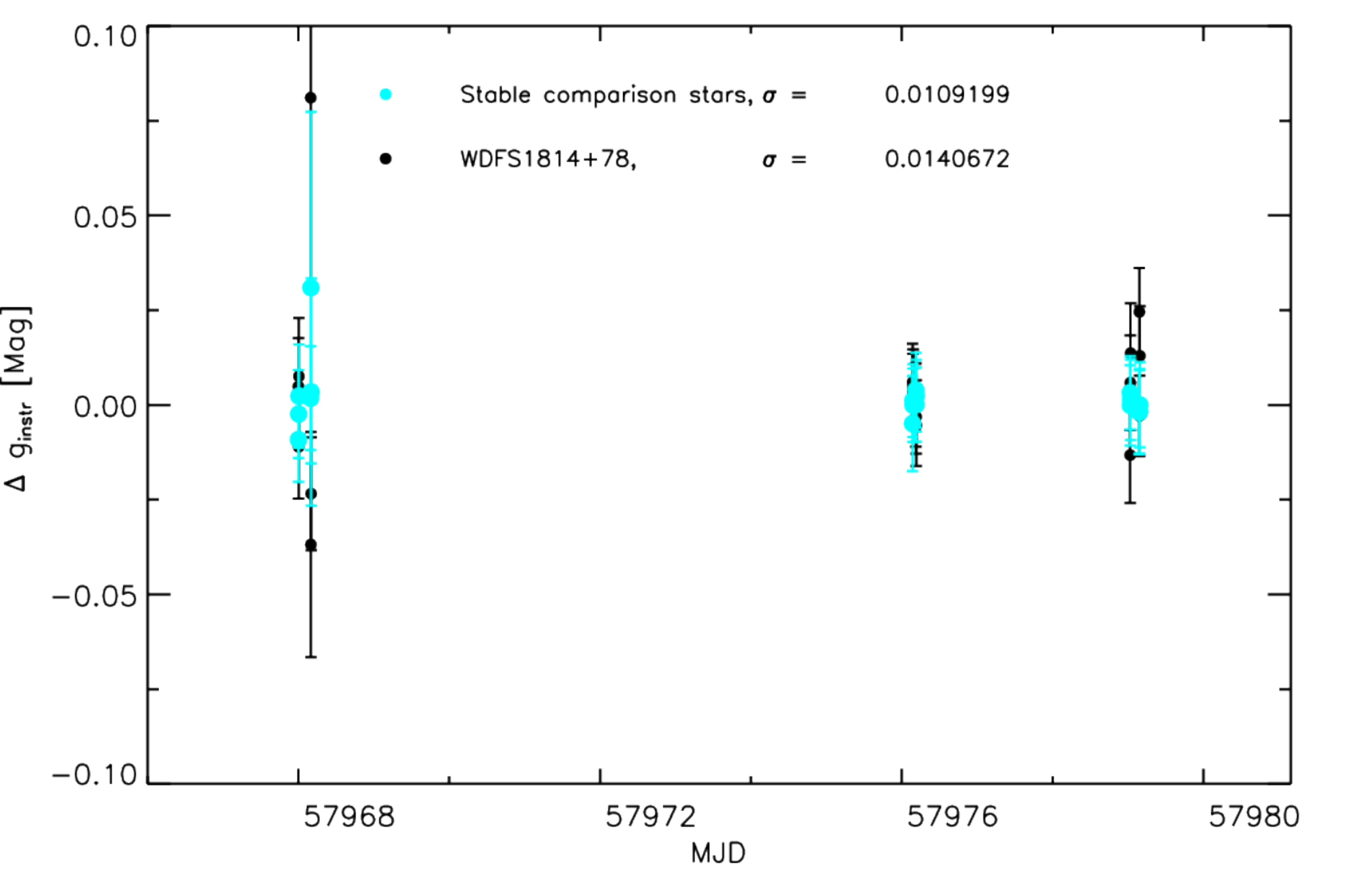} 
\caption{Same as Fig.~\ref{fig:sdssj010322} but for star WDFS1814+78. \label{fig:sdssj181424}}
\end{center}
\end{figure*}

\begin{figure*}[!h]
\begin{center}
\includegraphics[height=0.75\textheight,width=0.55\textwidth, angle=90]{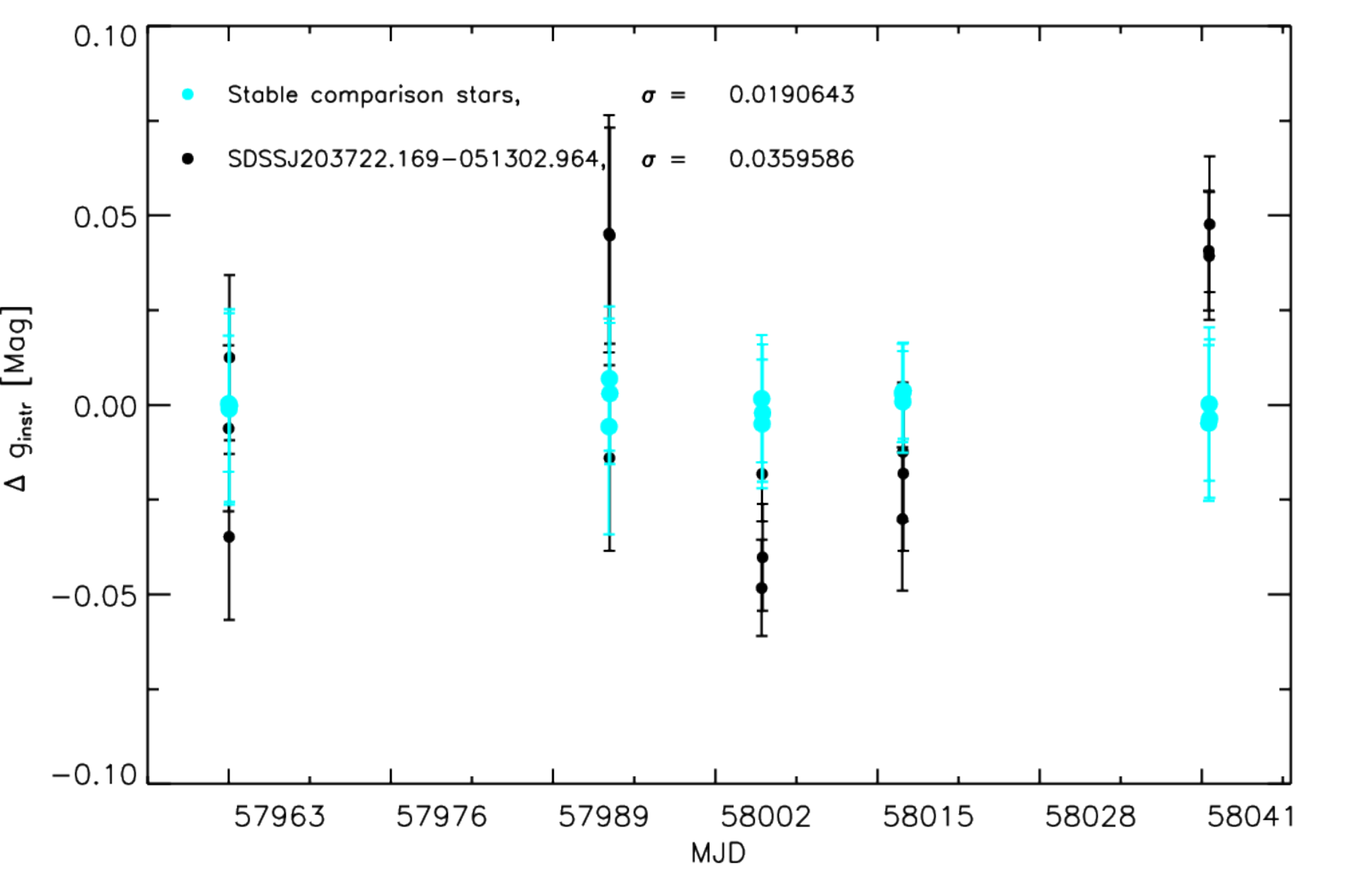} 
\caption{Same as Fig.~\ref{fig:sdssj010322} but for star SDSSJ203722.169-051302.964. \label{fig:sdssj20372}}
\end{center}
\end{figure*}

\begin{figure*}[!h]
\begin{center}
\includegraphics[height=0.75\textheight,width=0.55\textwidth, angle=90]{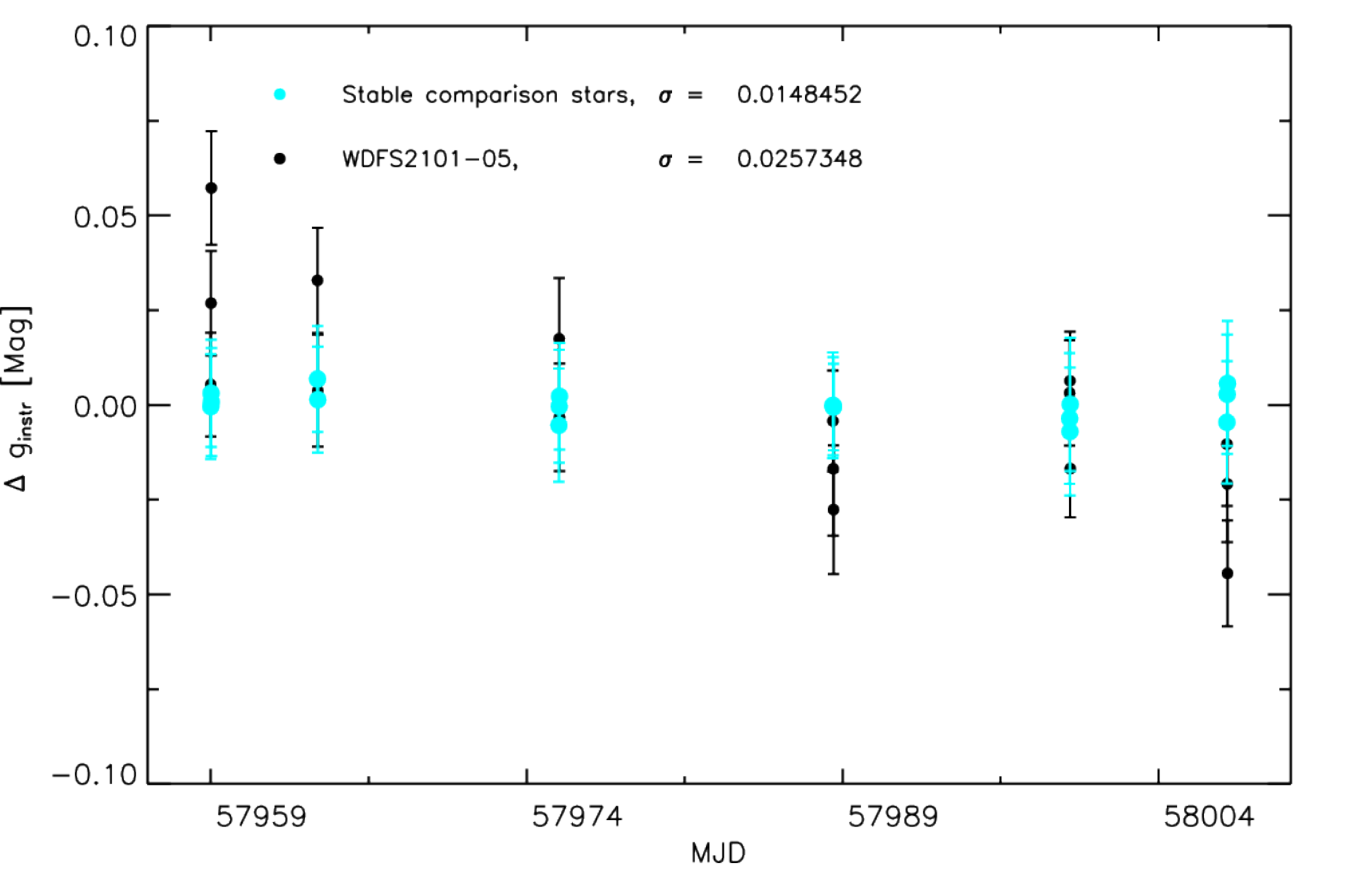} 
\caption{Same as Fig.~\ref{fig:sdssj010322} but for star WDFS2101-05. \label{fig:sdssj210150}}
\end{center}
\end{figure*}

\begin{figure*}[!h]
\begin{center}
\includegraphics[height=0.75\textheight,width=0.55\textwidth, angle=90]{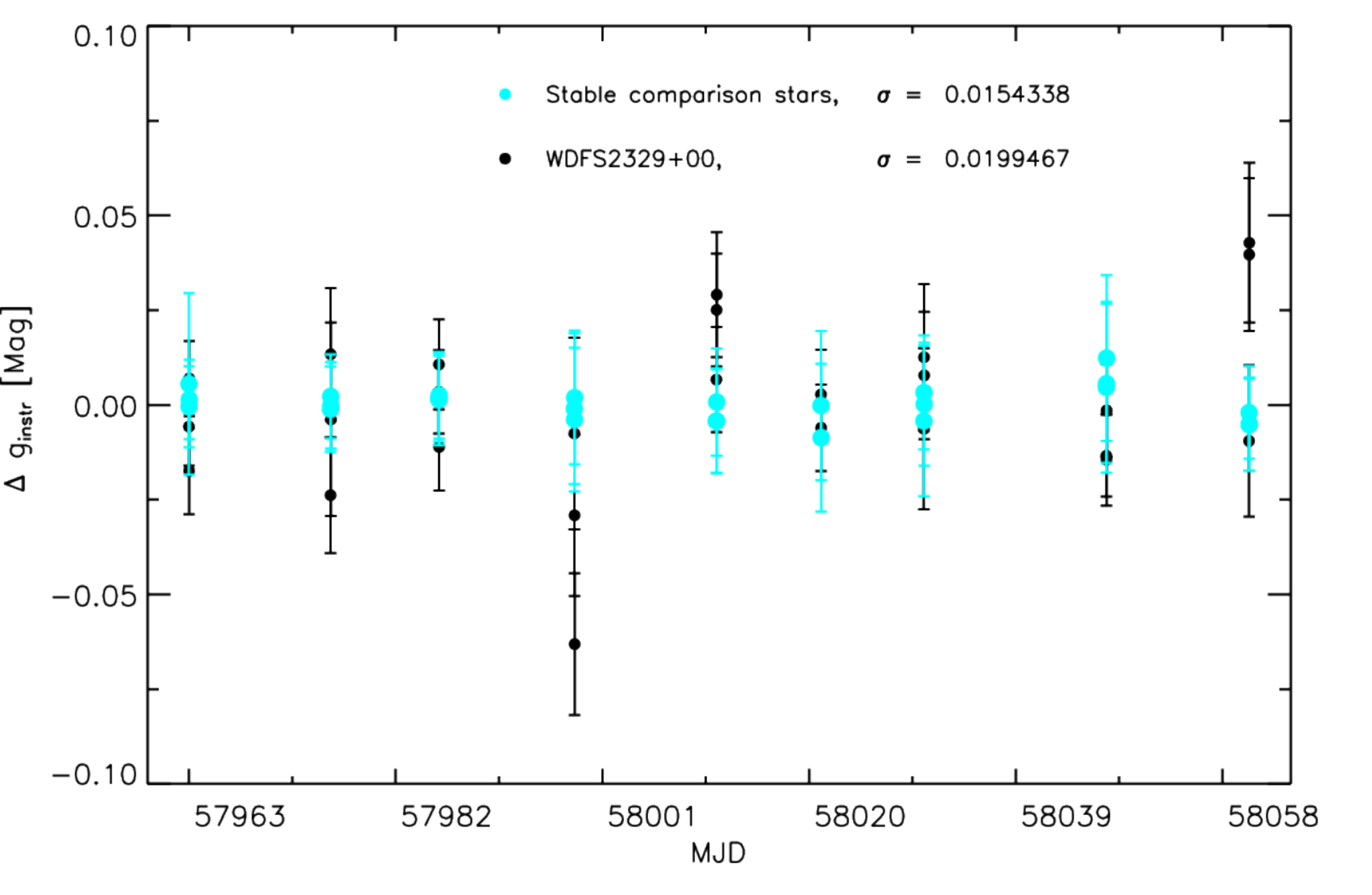} 
\caption{Same as Fig.~\ref{fig:sdssj010322} but for star WDFS2329+00. \label{fig:sdssj232941}}
\end{center}
\end{figure*}

The light curves for all the 23 DAWDs in the Northern hemisphere and around the celestial equators are shown in Fig.~\ref{fig:sdssj010322} to ~\ref{fig:sdssj232941}. 
Plots are listed in order of increasing $RA$.

\clearpage

\begin{center}
\subsection*{Southern DAWD light curves}
\end{center}

\begin{figure*}[!h]
\begin{center}
\includegraphics[height=0.75\textheight,width=0.55\textwidth, angle=90]{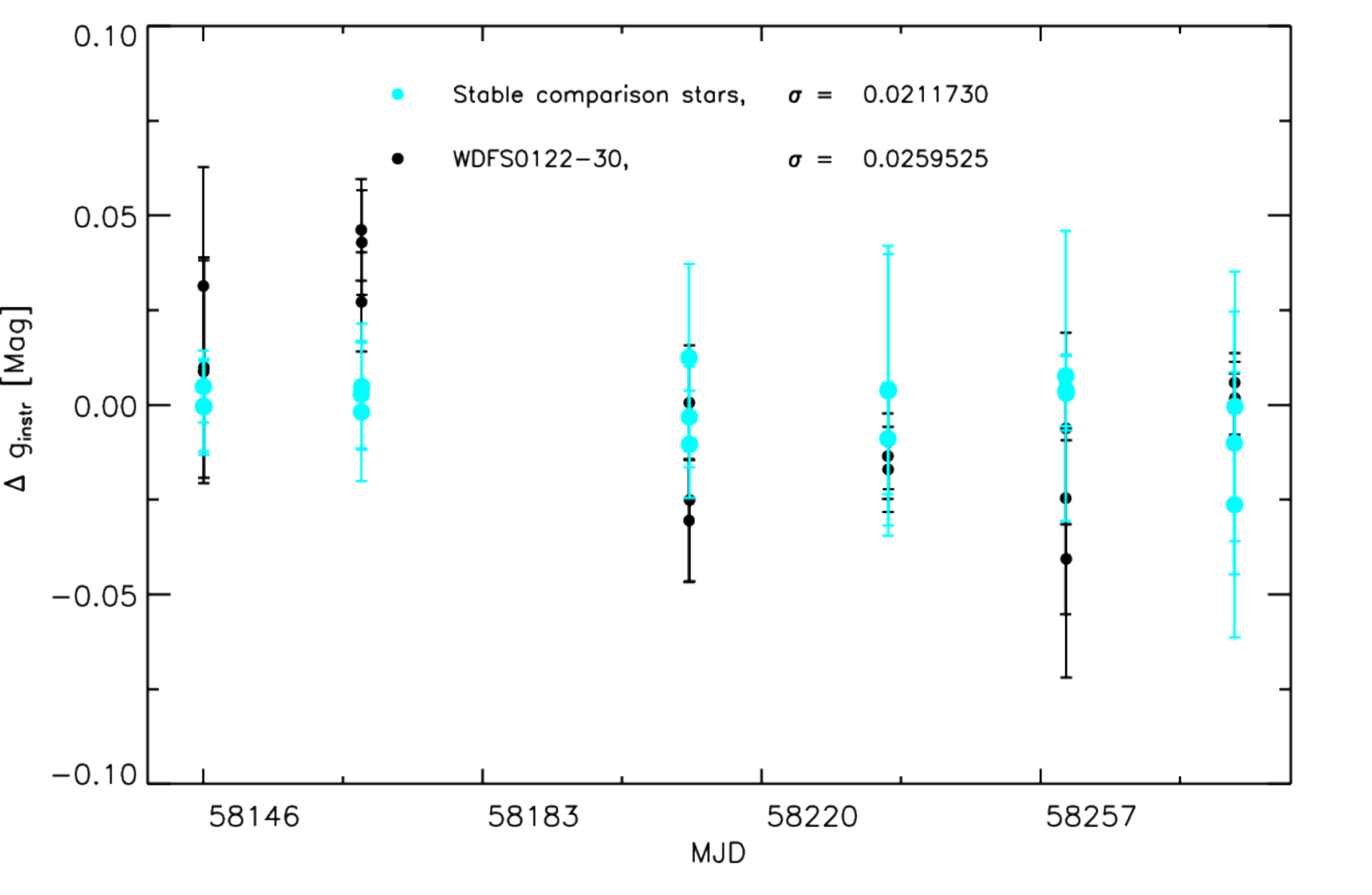} 
\caption{Single epoch minus the mean instrumental magnitude measurements for WDFS0122-30
as a function of observing epoch (black crosses). Averaged and binned relative magnitudes for a set of stable stars of comparable
instrumental magnitude in the same FoV are overplotted as a red shaded area. The variability index of the selected stars and the measurement
dispersion are listed. Error bars are shown. \label{fig:a020}}
\end{center}
\end{figure*}

\begin{figure*}[!h]
\begin{center}
\includegraphics[height=0.75\textheight,width=0.55\textwidth, angle=90]{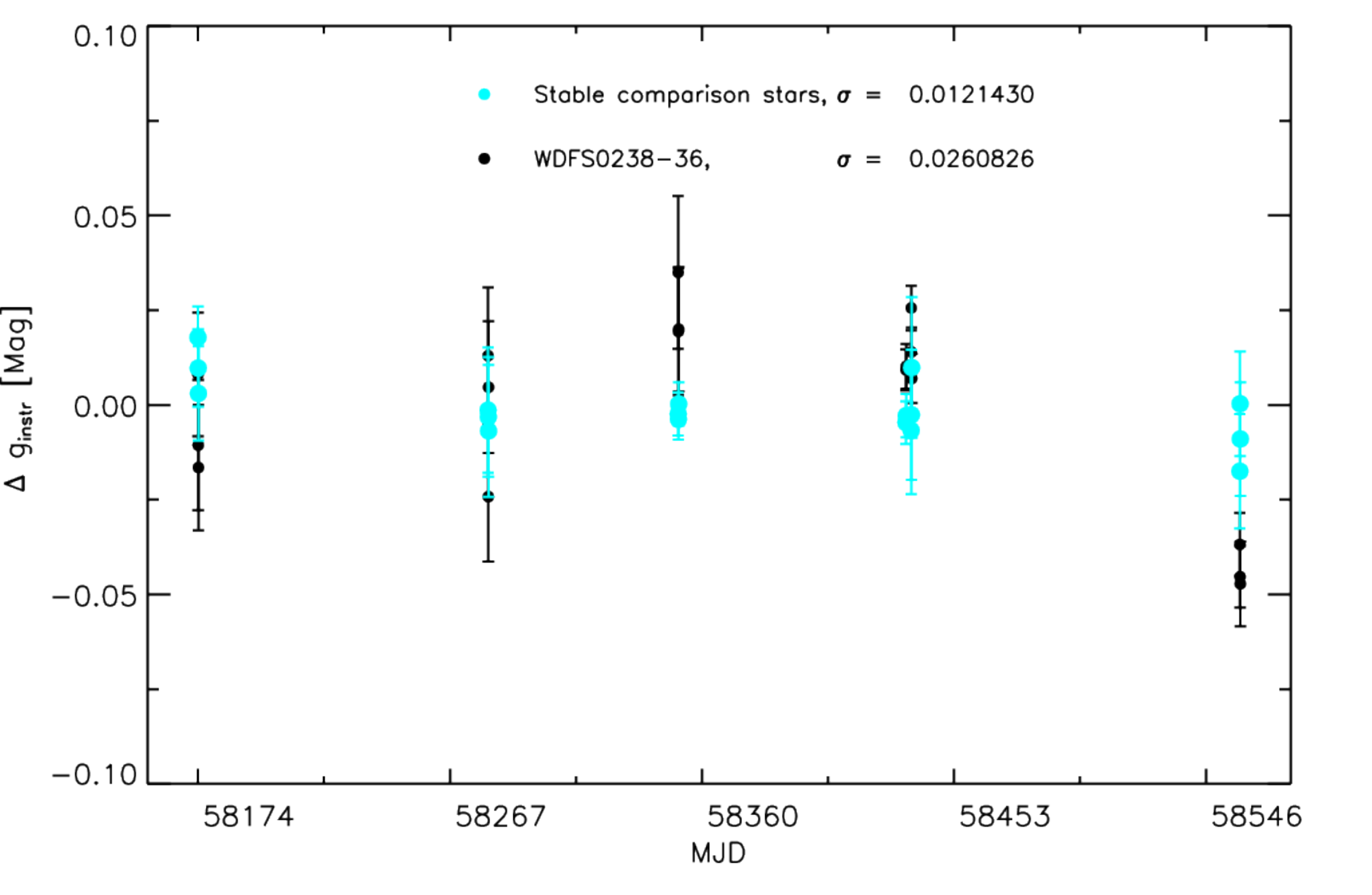} 
\caption{Same as Fig.~\ref{fig:a020} but for star WDFS0238-36. \label{fig:sssj023824}}
\end{center}
\end{figure*}

\begin{figure*}[!h]
\begin{center}
\includegraphics[height=0.75\textheight,width=0.55\textwidth, angle=90]{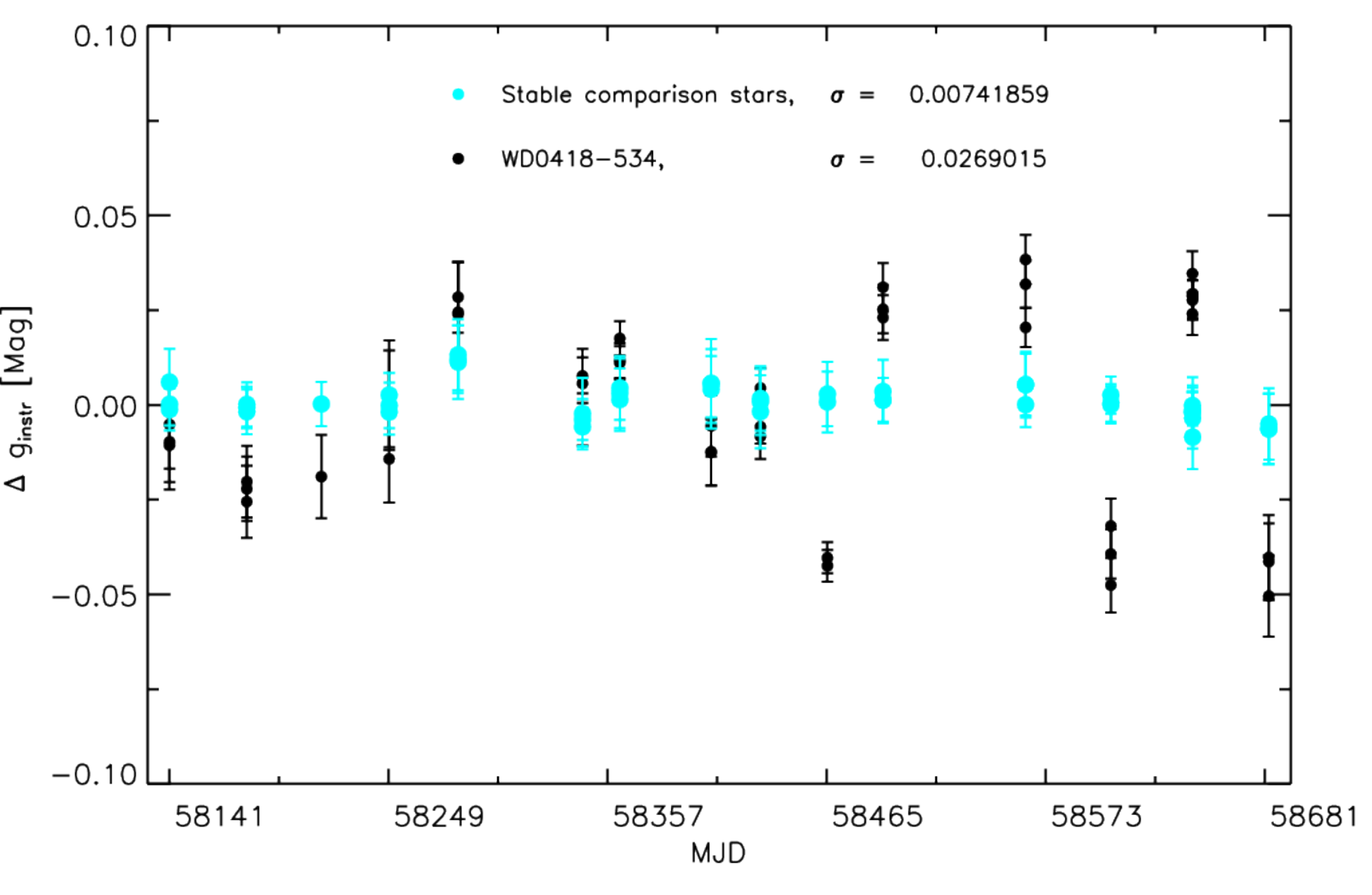} 
\caption{Same as Fig.~\ref{fig:a020} but for star WD0418-534. \label{fig:wd0418}}
\end{center}
\end{figure*}

\begin{figure*}[!h]
\begin{center}
\includegraphics[height=0.75\textheight,width=0.55\textwidth, angle=90]{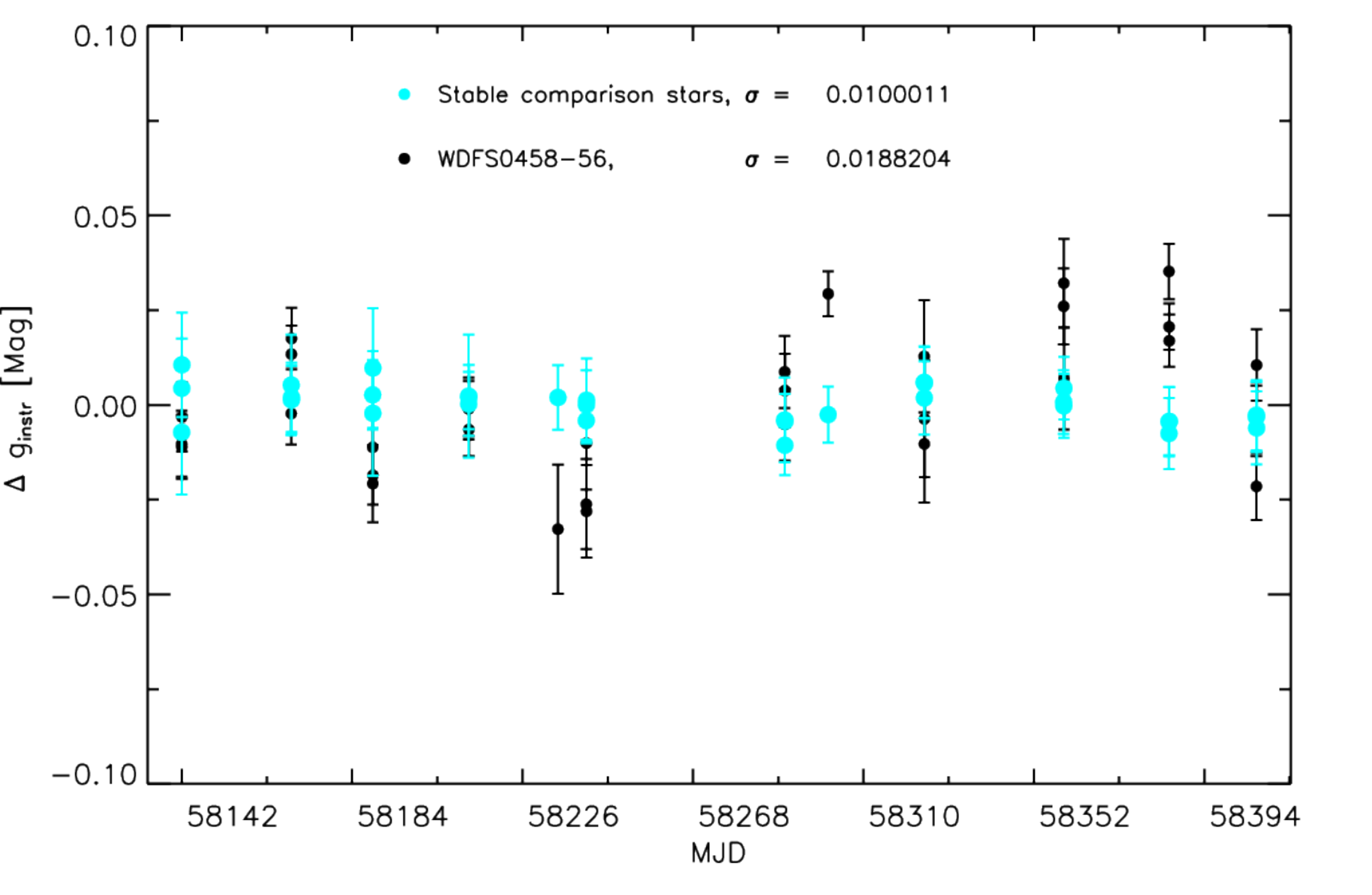} 
\caption{Same as Fig.~\ref{fig:a020} but for star WDFS0458-56. \label{fig:sssj045822}}
\end{center}
\end{figure*}

\begin{figure*}[!h]
\begin{center}
\includegraphics[height=0.75\textheight,width=0.55\textwidth, angle=90]{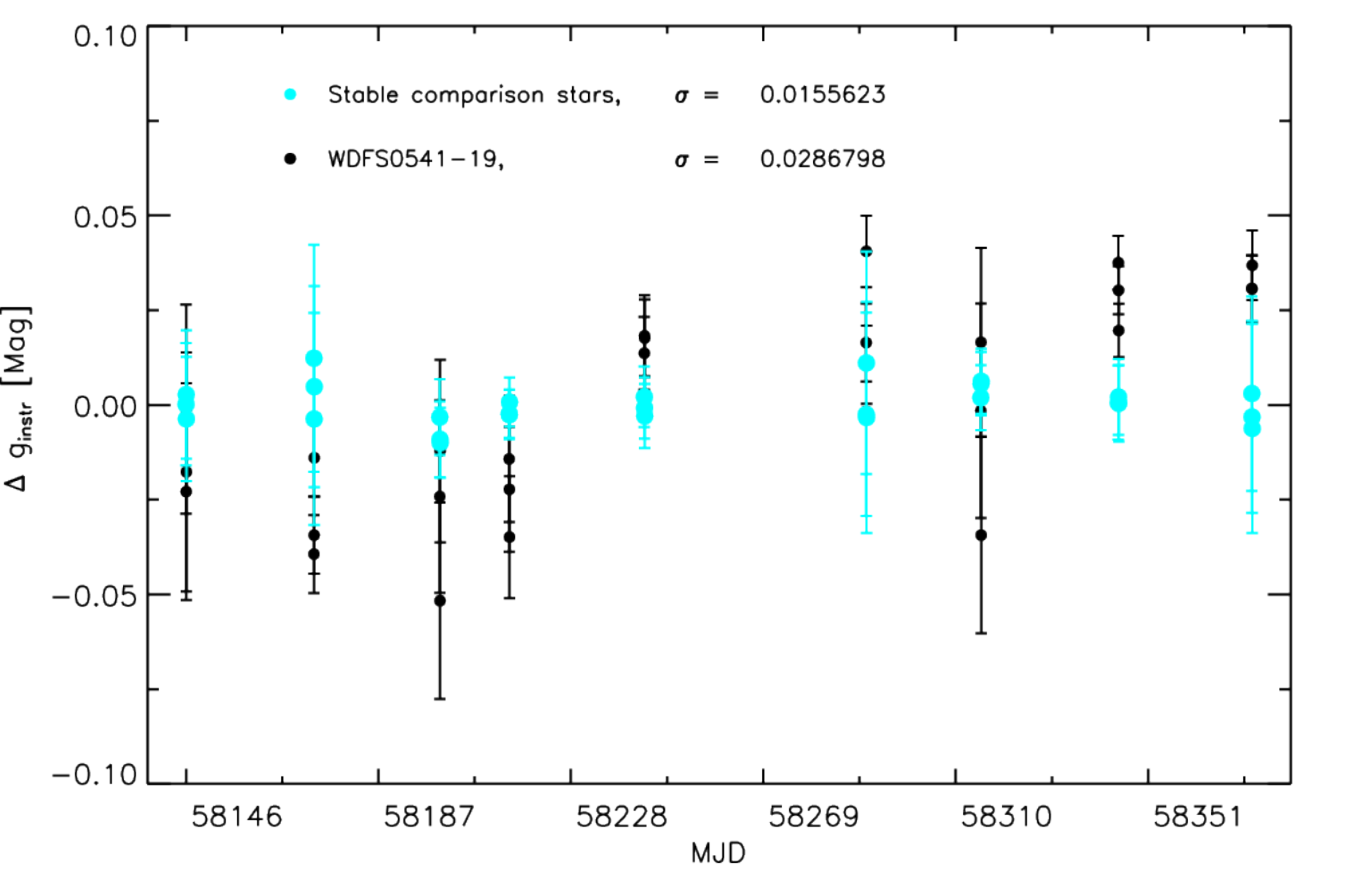} \caption{Same as Fig.~\ref{fig:a020} but for star WDFS0541-19. \label{fig:sssj054114}}
\end{center}
\end{figure*}

\begin{figure*}[!h]
\begin{center}
\includegraphics[height=0.75\textheight,width=0.55\textwidth, angle=90]{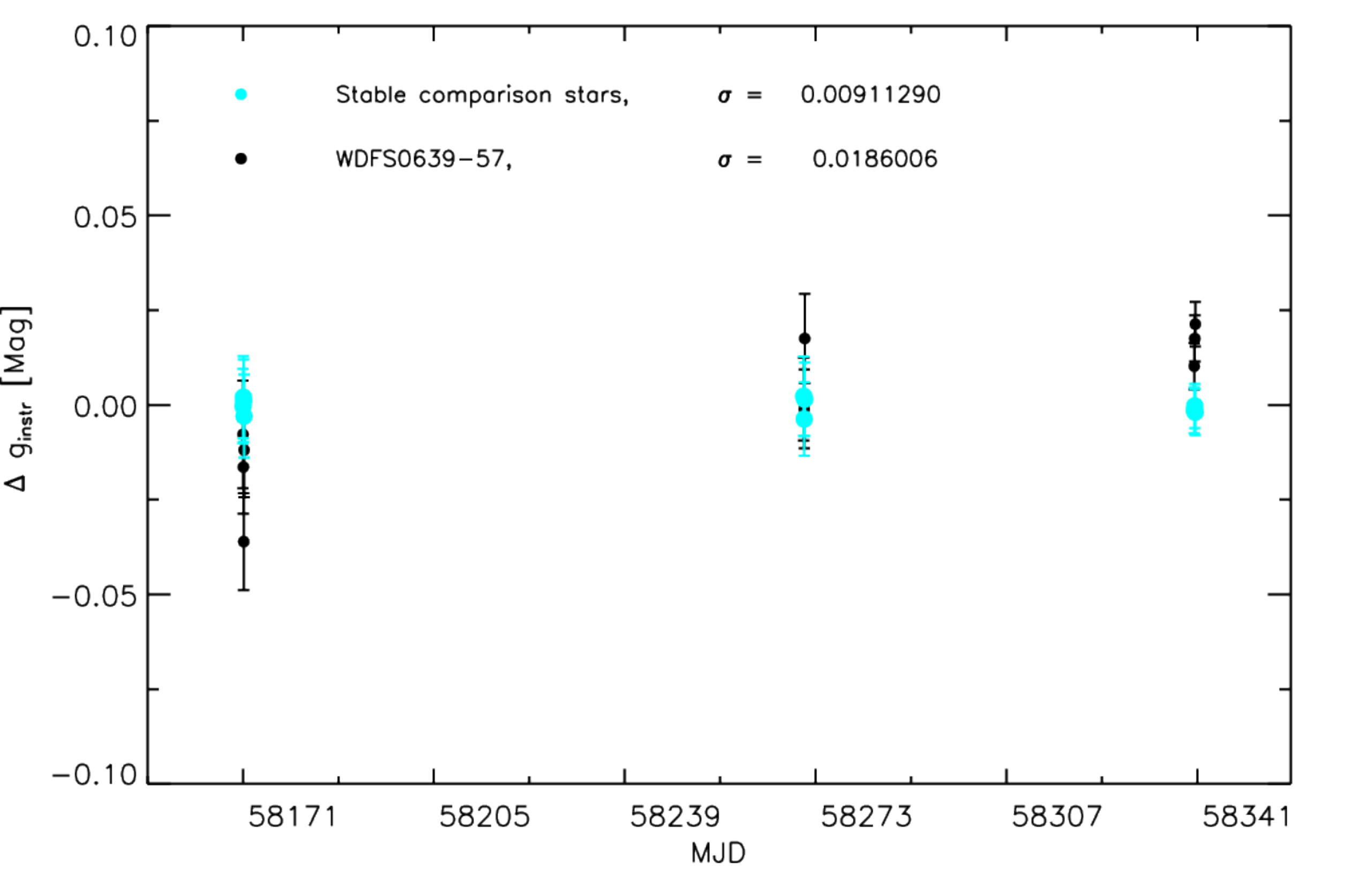} \caption{Same as Fig.~\ref{fig:a020} but for star WDFS0639-57. \label{fig:sssj063941}}
\end{center}
\end{figure*}

\begin{figure*}[!h]
\begin{center}
\includegraphics[height=0.75\textheight,width=0.55\textwidth, angle=90]{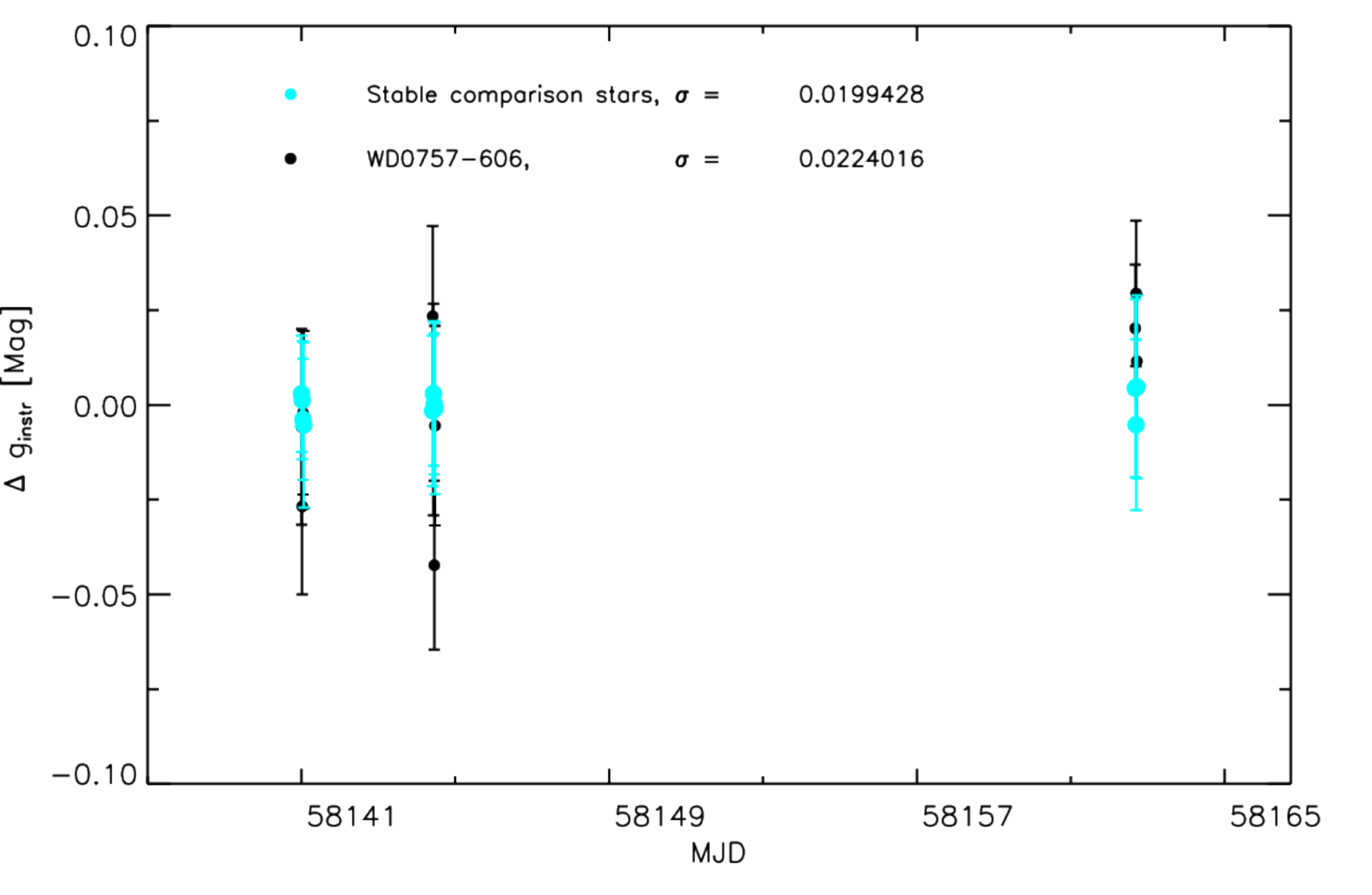} 
\caption{Same as Fig.~\ref{fig:a020} but for star WD0757-606. \label{fig:wd0757}}
\end{center}
\end{figure*}

\begin{figure*}[!h]
\begin{center}
\includegraphics[height=0.75\textheight,width=0.55\textwidth, angle=90]{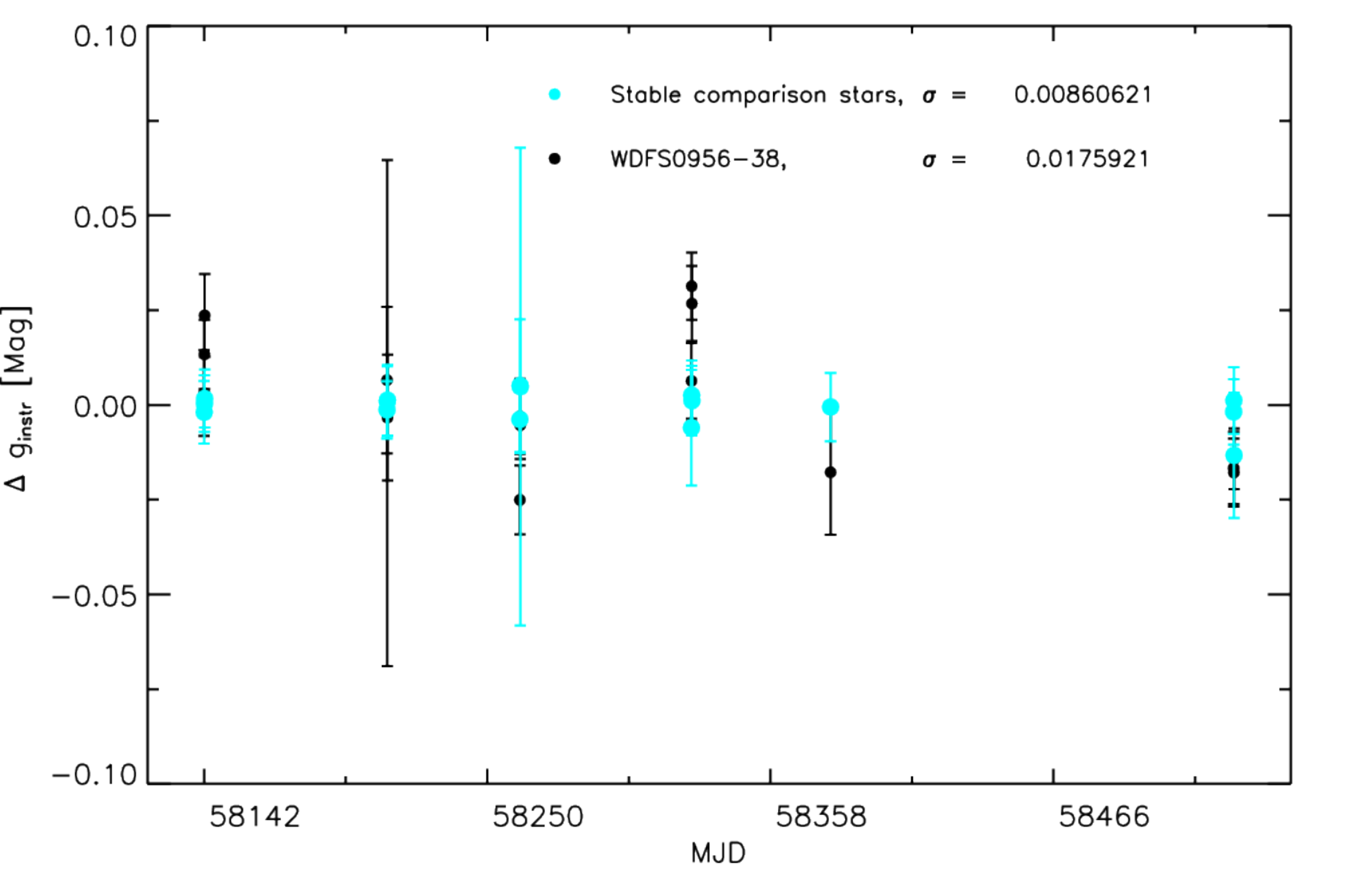} \caption{Same as Fig.~\ref{fig:a020} but for star WDFS0956-38. \label{fig:sssj095657}}
\end{center}
\end{figure*}

\begin{figure*}[!h]
\begin{center}
\includegraphics[height=0.75\textheight,width=0.55\textwidth, angle=90]{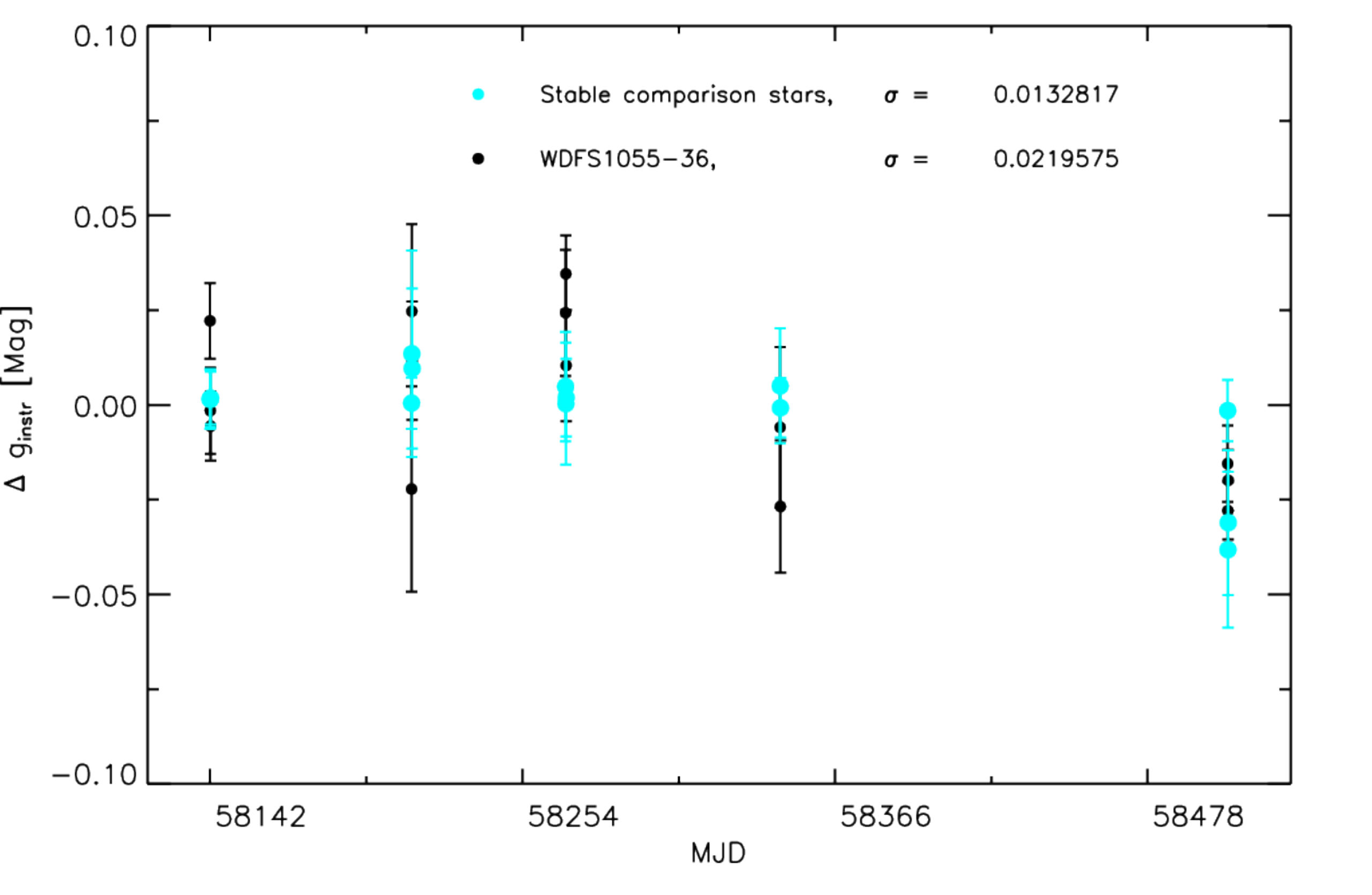} \caption{Same as Fig.~\ref{fig:a020} but for star WDFS1055-36. \label{fig:sssj105525}}
\end{center}
\end{figure*}

\begin{figure*}[!h]
\begin{center}
\includegraphics[height=0.75\textheight,width=0.55\textwidth, angle=90]{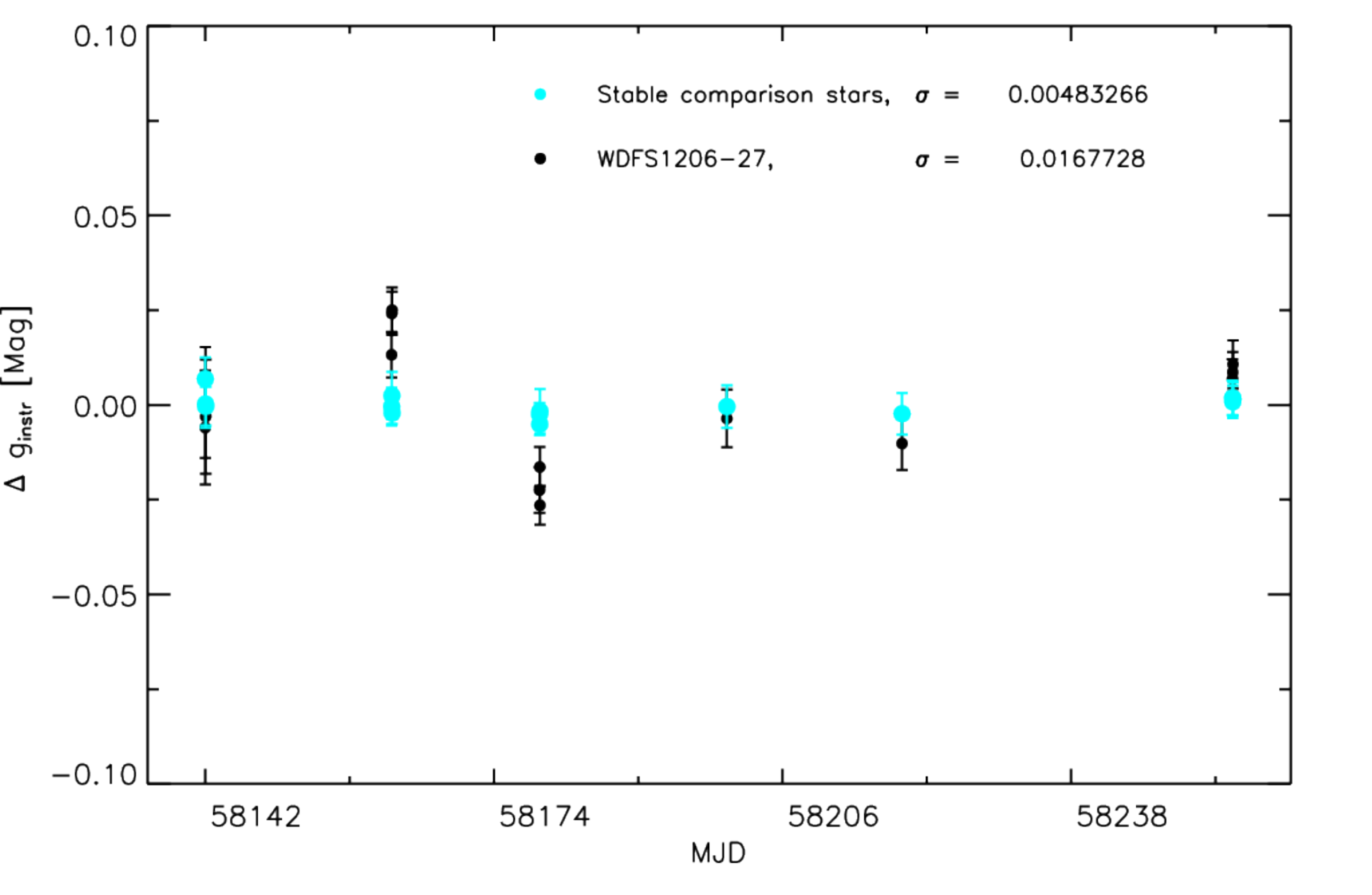} 
\caption{Same as Fig.~\ref{fig:a020} but for star WDFS1206-27. \label{fig:wd1203}}
\end{center}
\end{figure*}

\begin{figure*}[!h]
\begin{center}
\includegraphics[height=0.75\textheight,width=0.55\textwidth, angle=90]{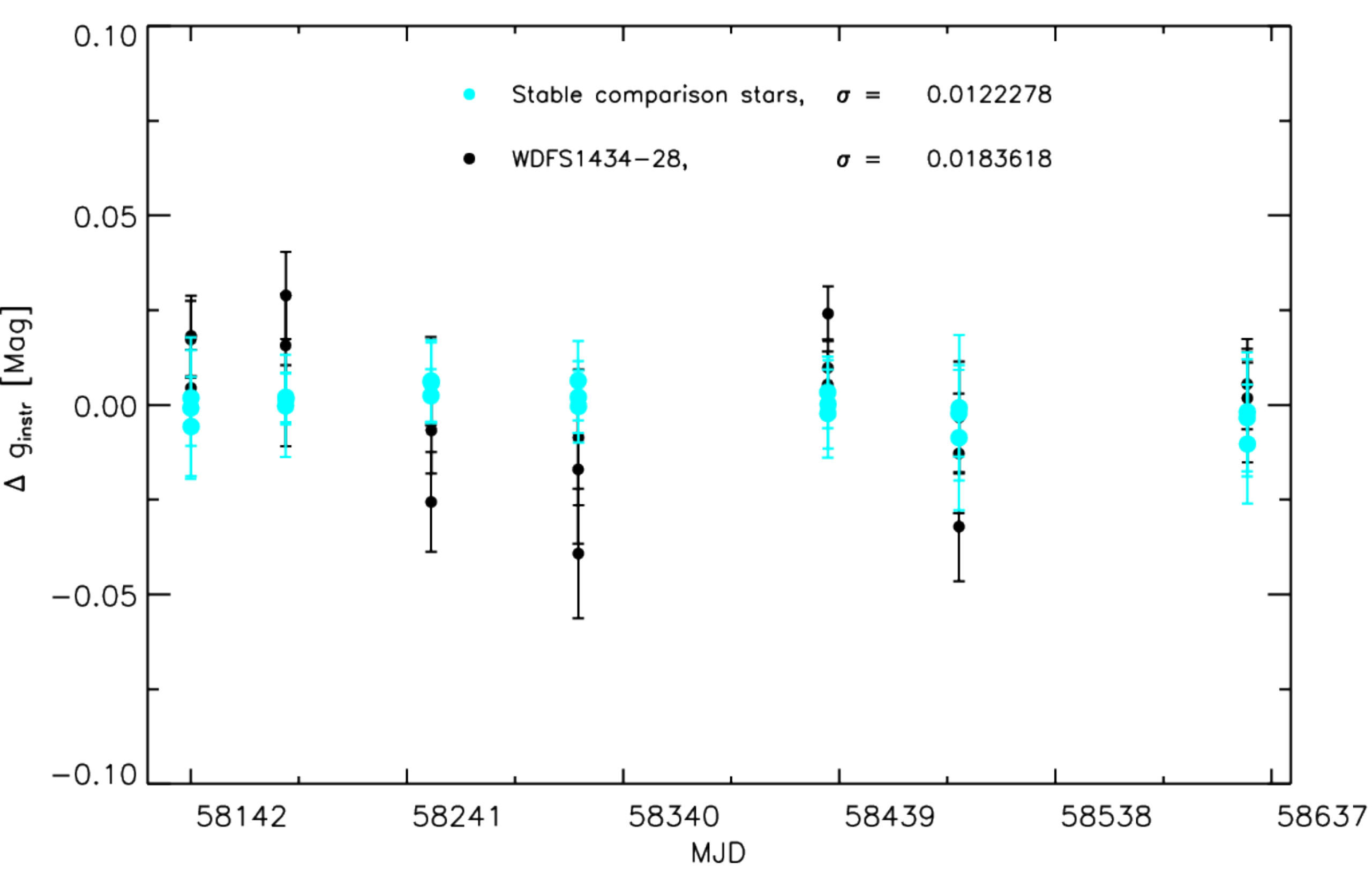} \caption{Same as Fig.~\ref{fig:a020} but for star WDFS1434-28. \label{fig:sssj143459}}
\end{center}
\end{figure*}

\begin{figure*}[!h]
\begin{center}
\includegraphics[height=0.75\textheight,width=0.55\textwidth, angle=90]{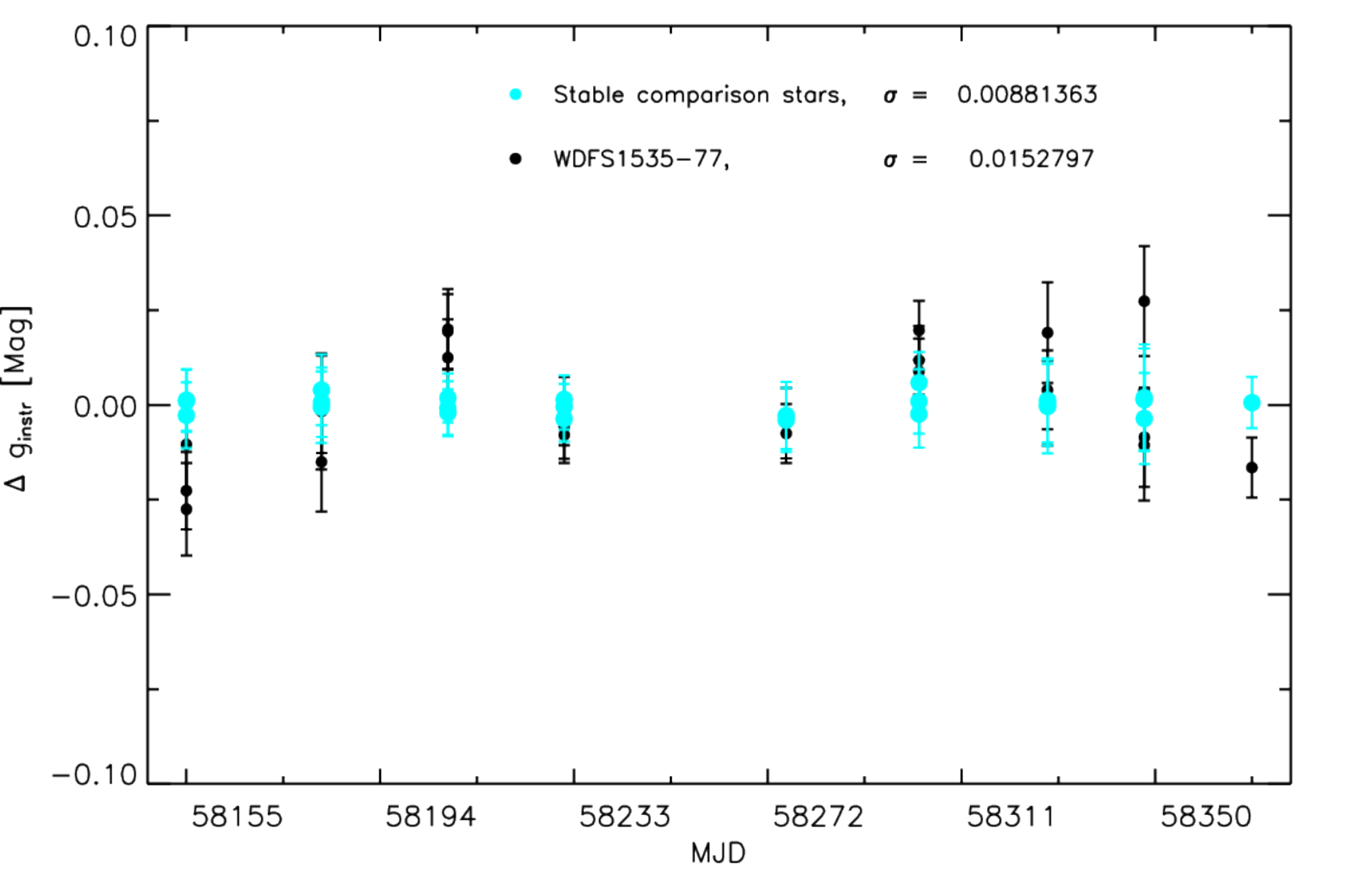} 
\caption{Same as Fig.~\ref{fig:a020} but for star WDFS1535-77. \label{fig:wd1529}}
\end{center}
\end{figure*}

\begin{figure*}[!h]
\begin{center}
\includegraphics[height=0.75\textheight,width=0.55\textwidth, angle=90]{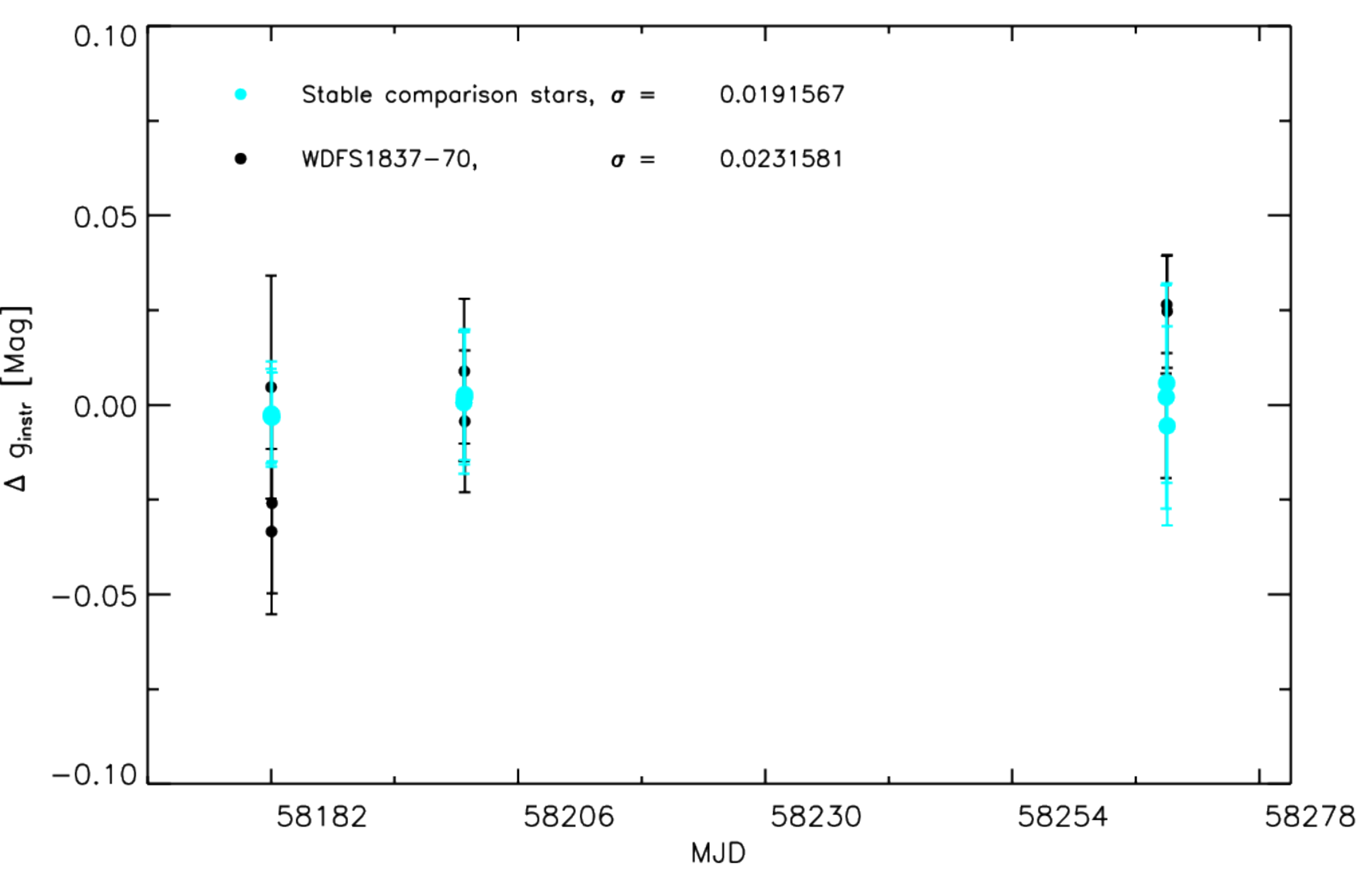} \caption{Same as Fig.~\ref{fig:a020} but for star WDFS1837-70. \label{fig:sssj183717}}
\end{center}
\end{figure*}

\begin{figure*}[!h]
\begin{center}
\includegraphics[height=0.75\textheight,width=0.55\textwidth, angle=90]{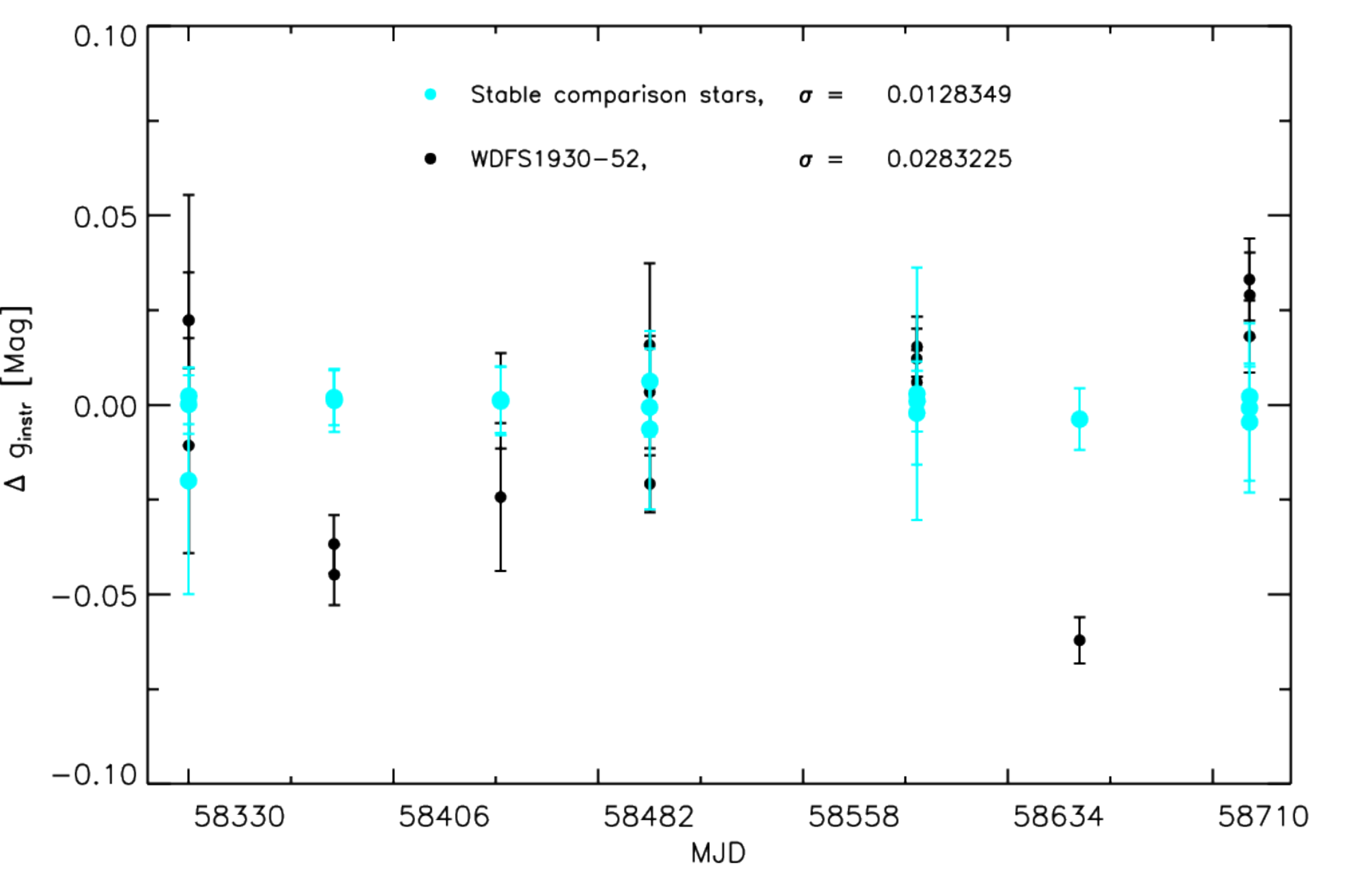} \caption{Same as Fig.~\ref{fig:a020} but for star WDFS1930-52. \label{fig:sssj193018}}
\end{center}
\end{figure*}

\begin{figure*}[!h]
\begin{center}
\includegraphics[height=0.75\textheight,width=0.55\textwidth, angle=90]{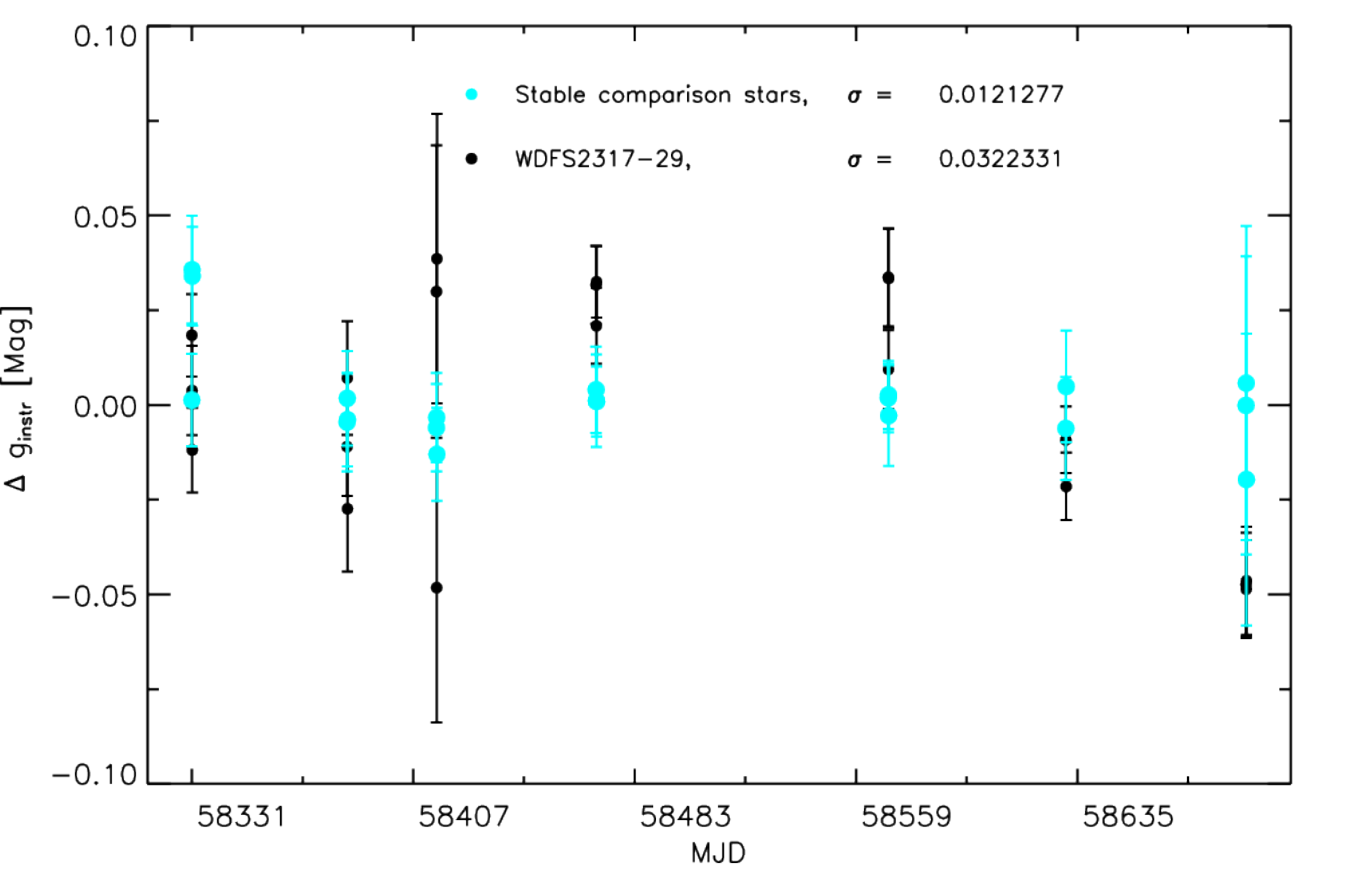} 
\caption{Same as Fig.~\ref{fig:a020} but for star WDFS2317-29. \label{fig:wd2314}}
\end{center}
\end{figure*}

The light curves for all the 15 DAWDs in the Southern hemisphere are shown in Fig.~\ref{fig:a020} to ~\ref{fig:wd2314}. Plots are listed in order of increasing $RA$.

\clearpage

\begin{center}
\subsection*{Finding charts}
\end{center}

We provide here finding charts for the 23 Northern and the 15 Southern DAWDs. These are based on {\it drizzled} WFC3/HST images collected in the $F160W$ filter and cover a FoV of 26\arcsec$\times$26\arcsec. Some background objects and faint red dwarfs are visible in the NIR but are absent in the bluer WFC3/HST filter images and in the LCO g-band images. In particular, due to the larger pixel scale ($\approx$ 0\farcs39 vs 0\farcs13/pixel) and to the seeing, these faint red objects, that could be as close as $\lesssim$ 0\farcs5 - 2\arcsec~ to the DAWDs, might contaminate the LCO light curves. These contaminants are difficult or impossible to separate and identify on the LCO $g$-band images. Thus, we provide the following higher resolution NIR finding charts. In addition, we warn observers using ground-based facilities to be aware of potential NIR contamination for some of the DAWDs selected as standards. For more details see the discussion in Section~\ref{sec:results} and Table~\ref{table:4}.


\begin{figure*}[!h]
\begin{minipage}{0.24\textwidth}
\includegraphics[height=0.2\textheight,width=1.15\textwidth]{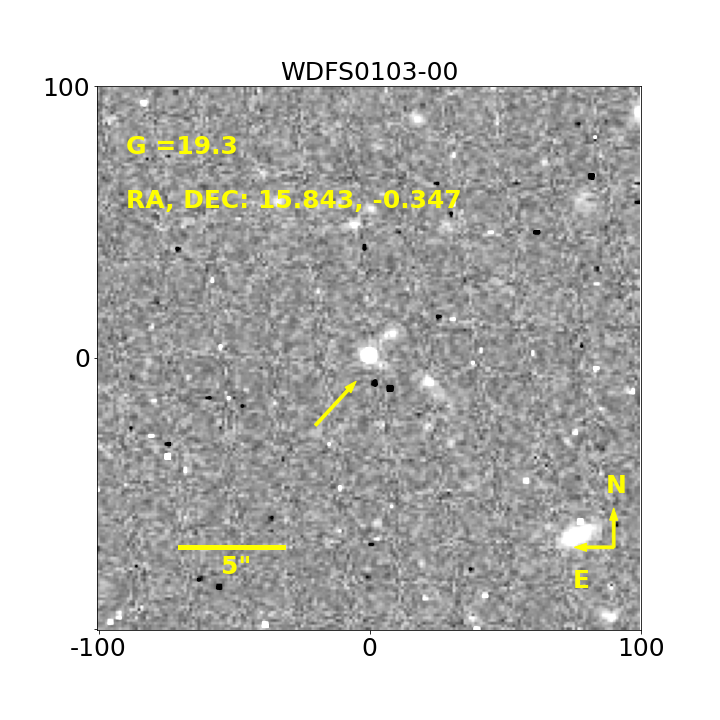} 
\end{minipage}
\begin{minipage}{0.24\textwidth}
\includegraphics[height=0.2\textheight,width=1.15\textwidth]{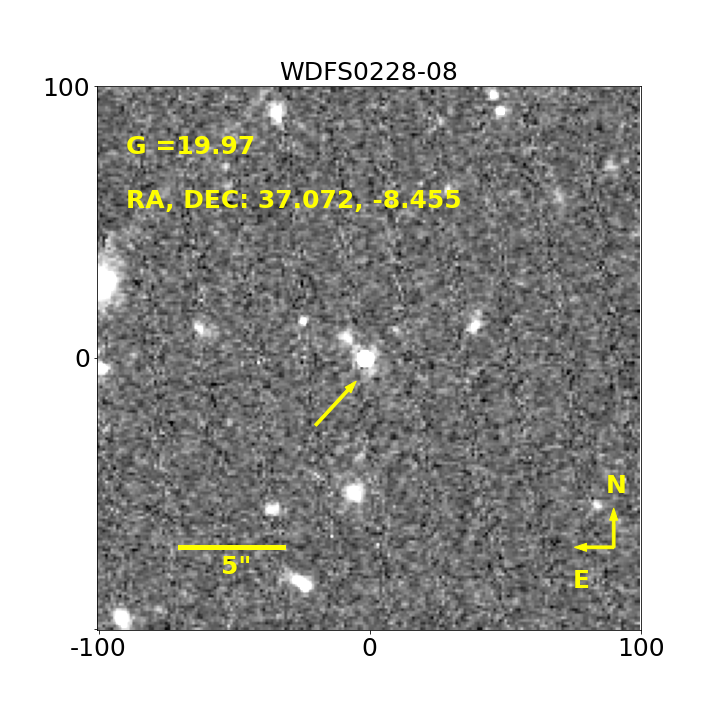} 
\end{minipage}
\begin{minipage}{0.24\textwidth}
\includegraphics[height=0.2\textheight,width=1.15\textwidth]{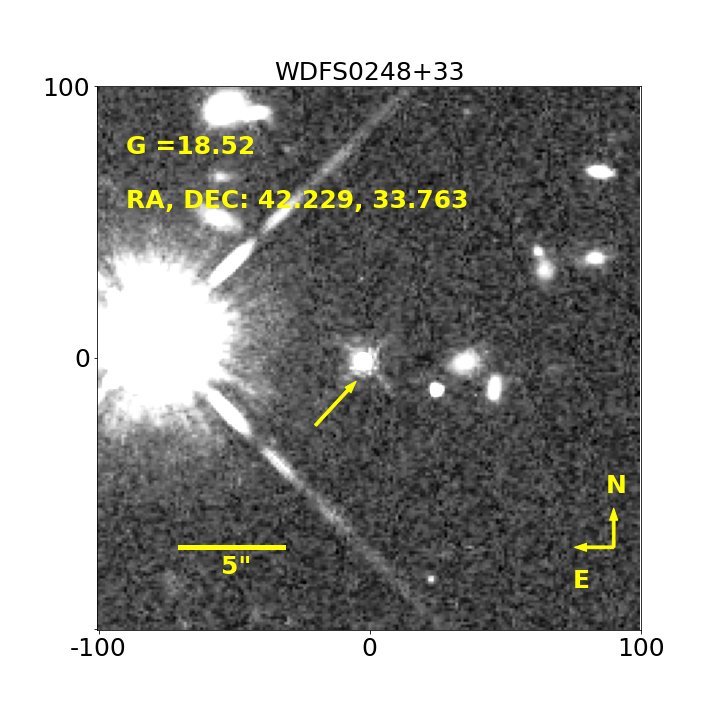} 
\end{minipage}
\begin{minipage}{0.24\textwidth}
\includegraphics[height=0.2\textheight,width=1.15\textwidth]{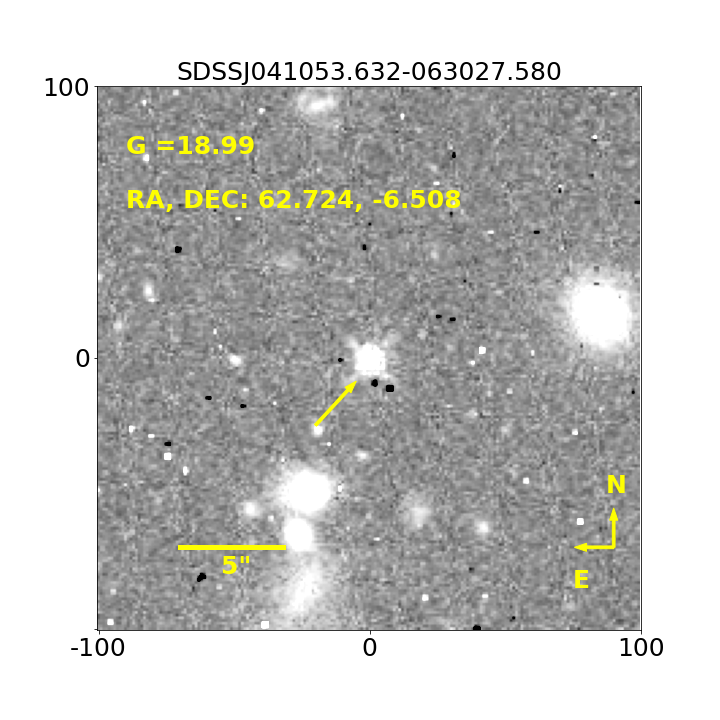} 
\end{minipage}
\begin{minipage}{0.24\textwidth}
\includegraphics[height=0.2\textheight,width=1.15\textwidth]{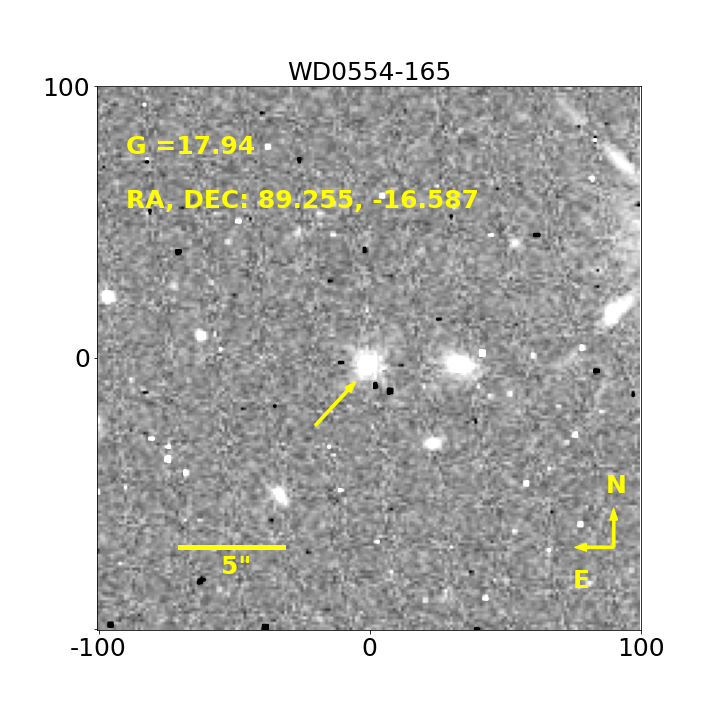} 
\end{minipage}
\begin{minipage}{0.24\textwidth}
\includegraphics[height=0.2\textheight,width=1.15\textwidth]{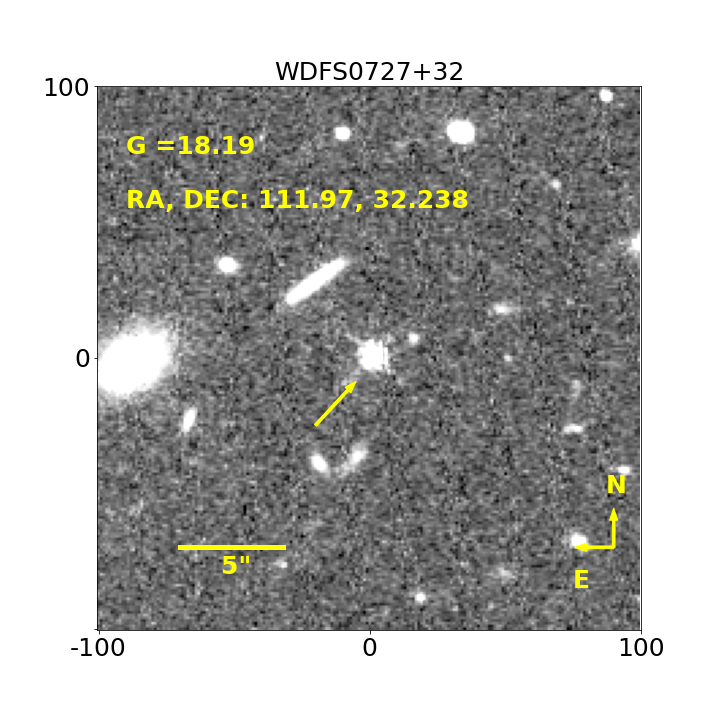} 
\end{minipage}
\begin{minipage}{0.24\textwidth}
\includegraphics[height=0.2\textheight,width=1.15\textwidth]{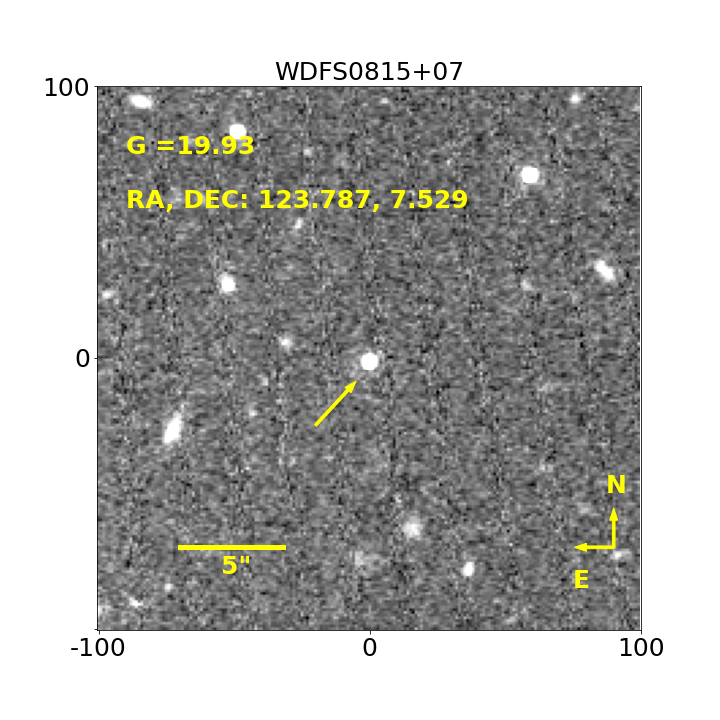} 
\end{minipage}
\begin{minipage}{0.24\textwidth}
\includegraphics[height=0.2\textheight,width=1.15\textwidth]{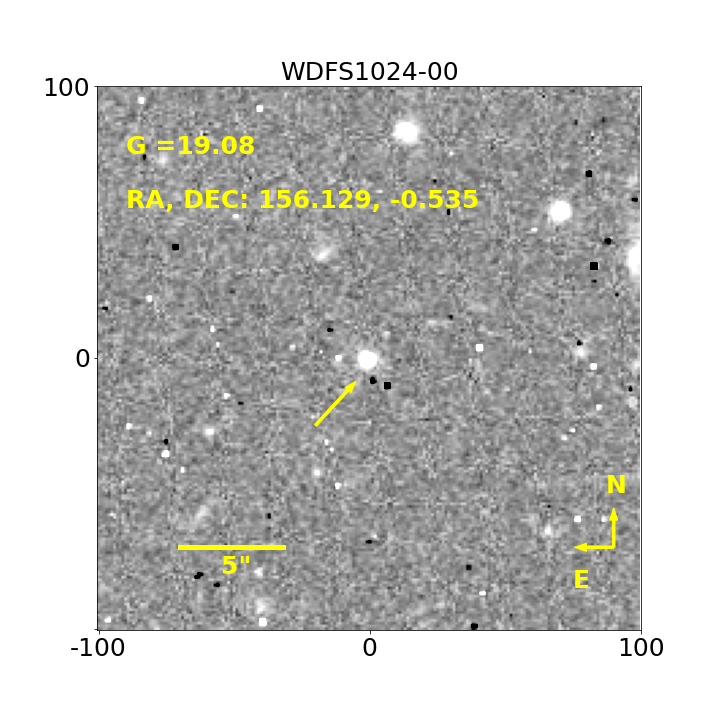} 
\end{minipage}
\begin{minipage}{0.24\textwidth}
\includegraphics[height=0.2\textheight,width=1.15\textwidth]{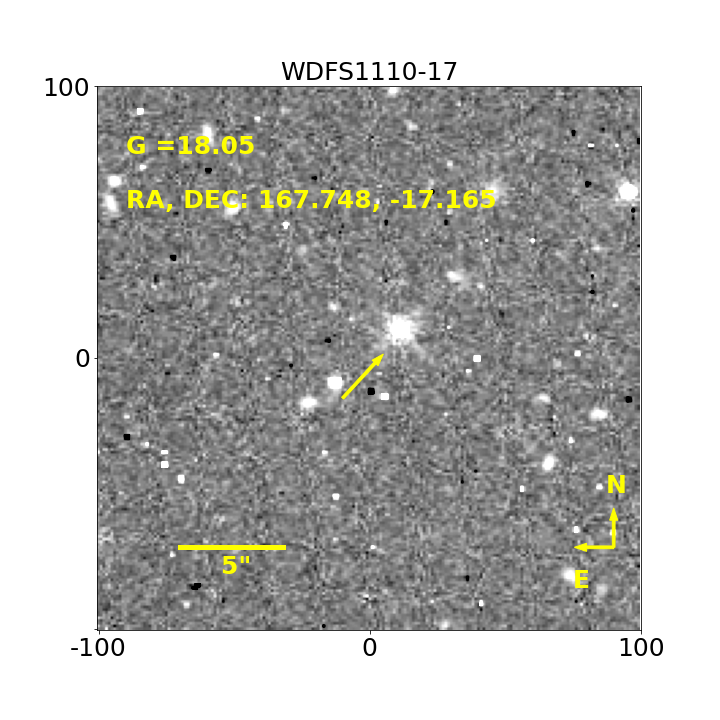} 
\end{minipage}
\begin{minipage}{0.24\textwidth}
\includegraphics[height=0.2\textheight,width=1.15\textwidth]{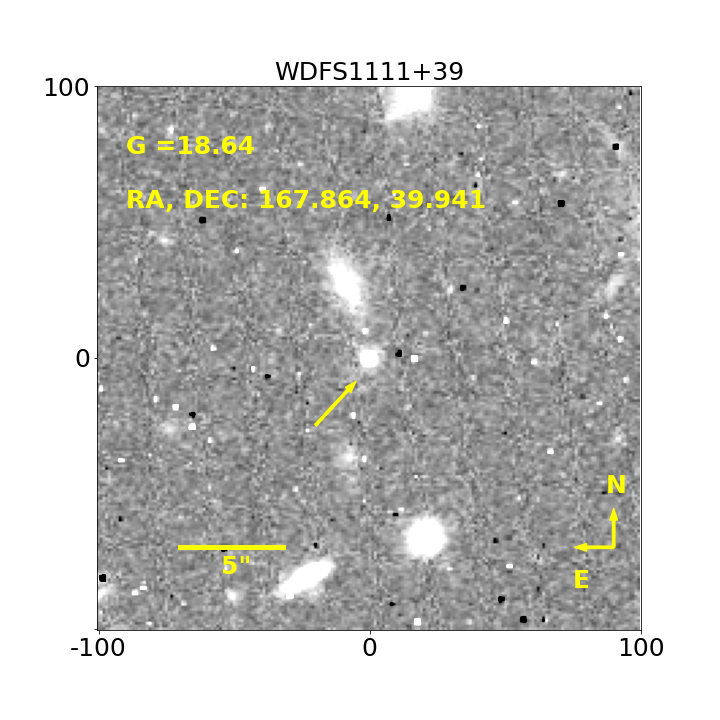} 
\end{minipage}
\begin{minipage}{0.24\textwidth}
\includegraphics[height=0.2\textheight,width=1.15\textwidth]{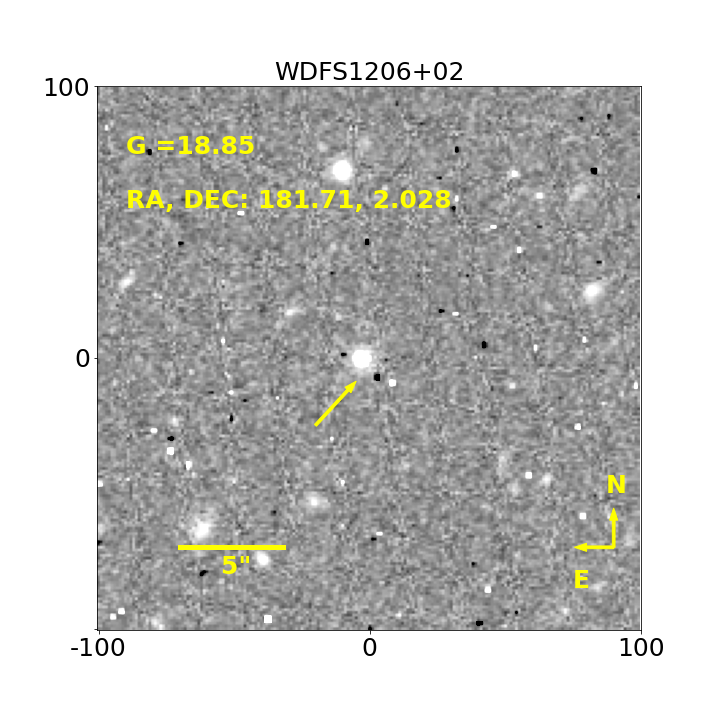} 
\end{minipage}
\begin{minipage}{0.24\textwidth}
\includegraphics[height=0.2\textheight,width=1.15\textwidth]{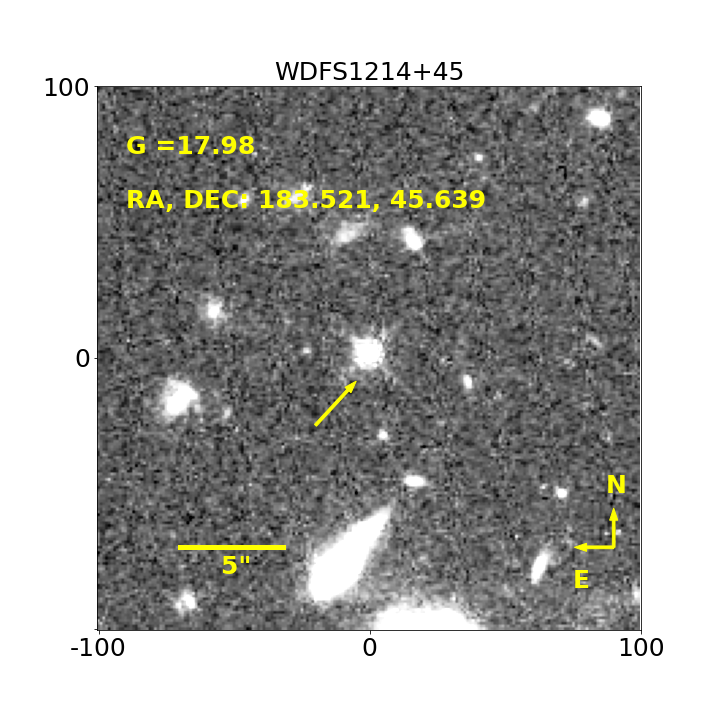} 
\end{minipage}
\begin{minipage}{0.24\textwidth}
\includegraphics[height=0.2\textheight,width=1.15\textwidth]{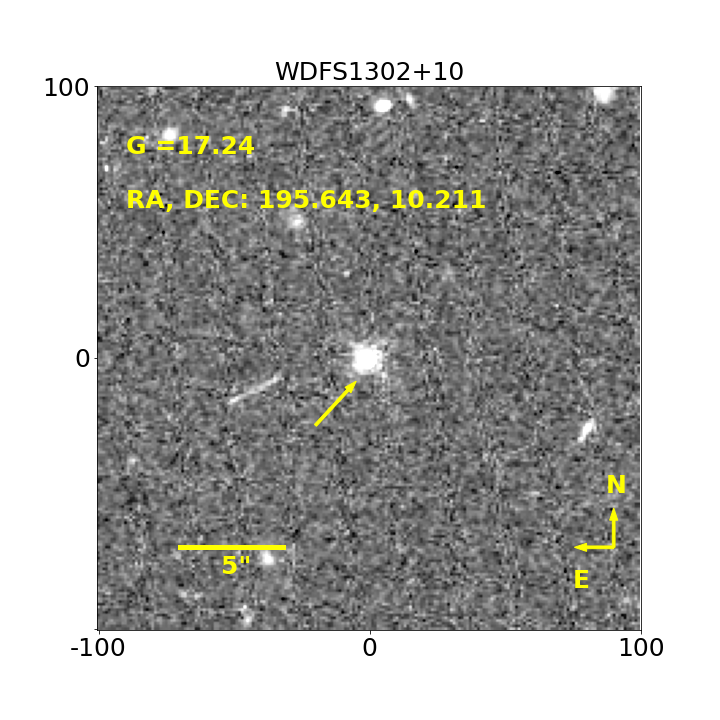} 
\end{minipage}
\begin{minipage}{0.24\textwidth}
\includegraphics[height=0.2\textheight,width=1.15\textwidth]{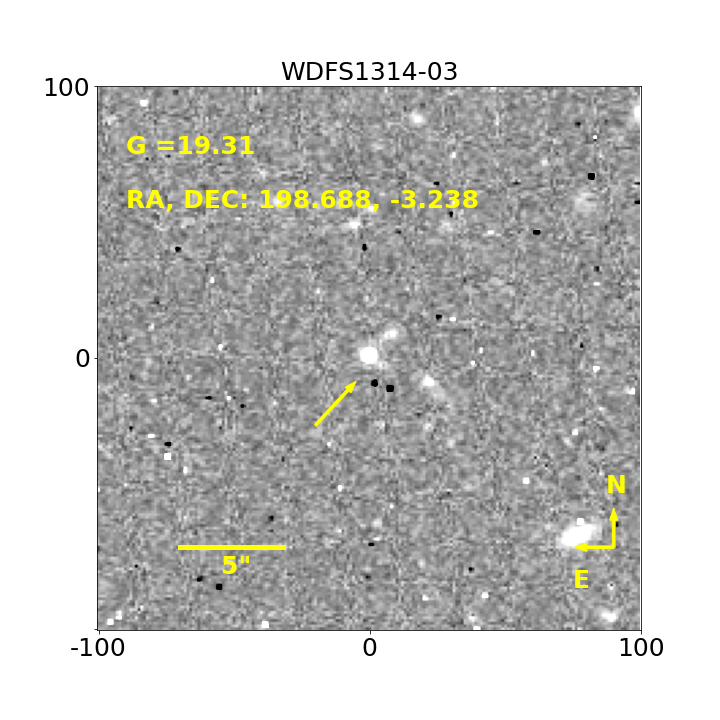} 
\end{minipage}
\begin{minipage}{0.24\textwidth}
\includegraphics[height=0.2\textheight,width=1.15\textwidth]{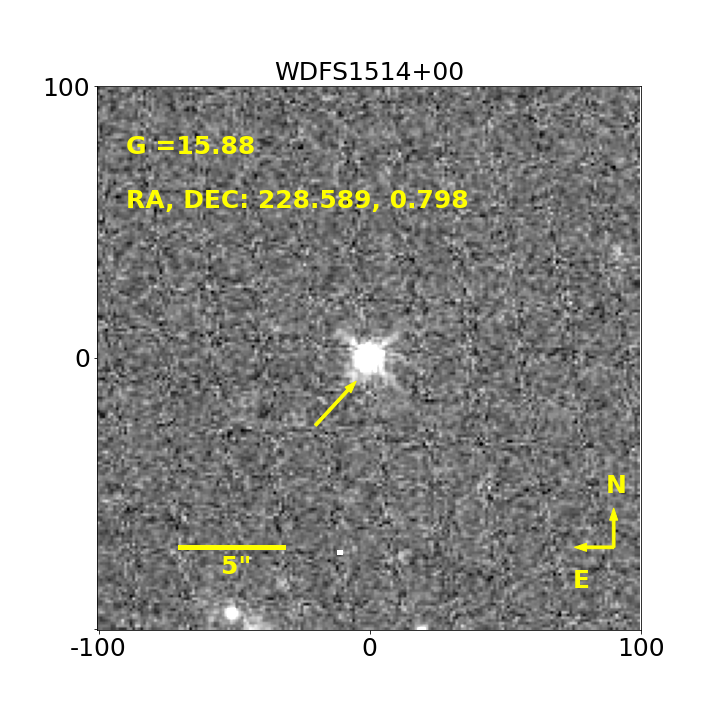} 
\end{minipage}
\begin{minipage}{0.24\textwidth}
\includegraphics[height=0.2\textheight,width=1.15\textwidth]{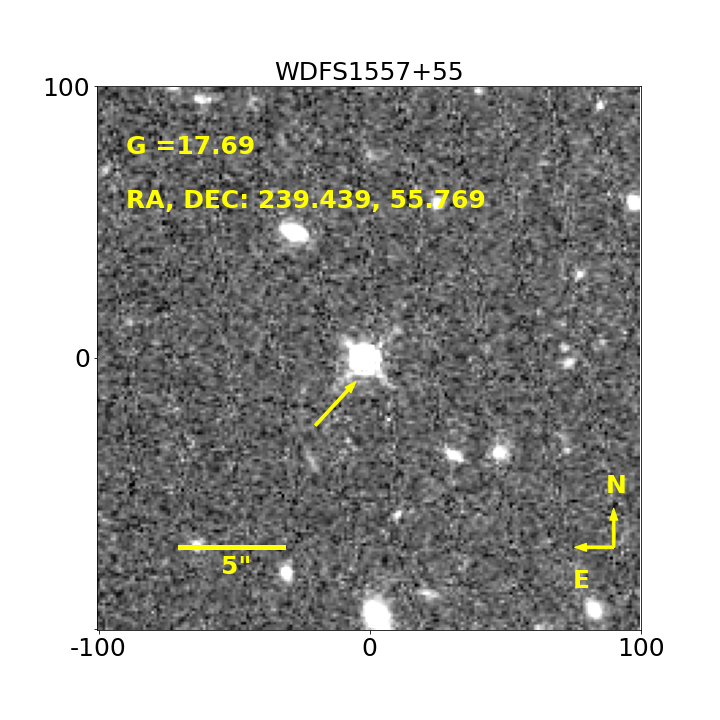} 
\end{minipage}
\begin{minipage}{0.24\textwidth}
\end{minipage}
\caption{Finding charts for 16 candidate DAWDs based on WFC3 drizzled $F160W$ image cut-outs centered on the star and covering a FoV of 26\arcsec$\times$26\arcsec. Star \emph{Gaia} DR3 magnitude and coordinates are labeled on the image. 
The yellow solid arrow points at the DAWD and the yellow line indicates 5\arcsec on the image. 
The North and East directions are also shown. \label{fig:chart_north1}}
\end{figure*}

\begin{figure*}[!h]
\begin{minipage}{0.24\textwidth}
\includegraphics[height=0.2\textheight,width=1.15\textwidth]{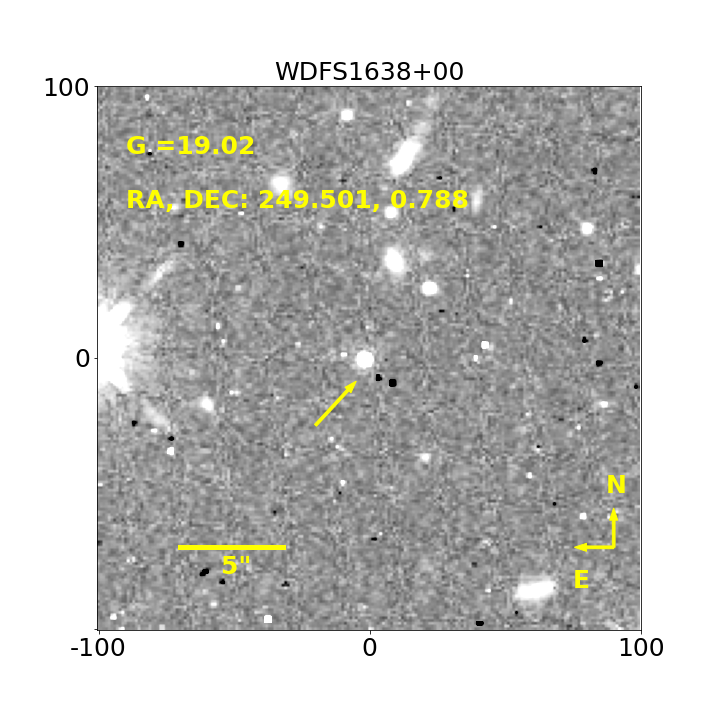} 
\end{minipage}
\begin{minipage}{0.24\textwidth}
\includegraphics[height=0.2\textheight,width=1.15\textwidth]{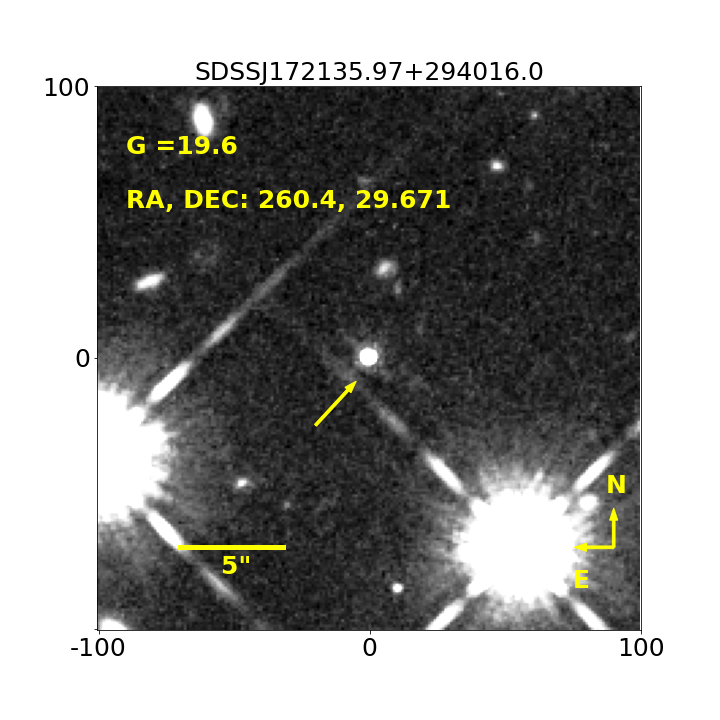} 
\end{minipage}
\begin{minipage}{0.24\textwidth}
\includegraphics[height=0.2\textheight,width=1.15\textwidth]{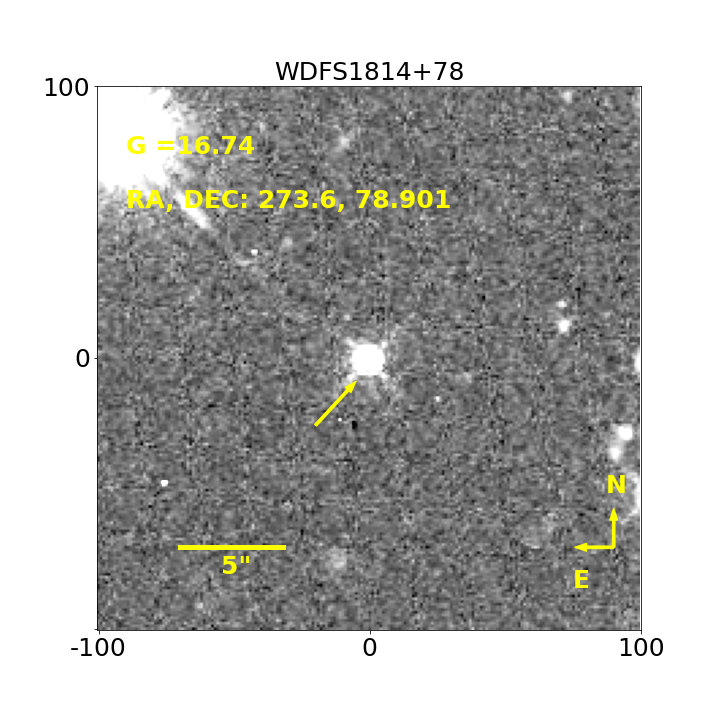} 
\end{minipage}
\begin{minipage}{0.24\textwidth}
\includegraphics[height=0.2\textheight,width=1.15\textwidth]{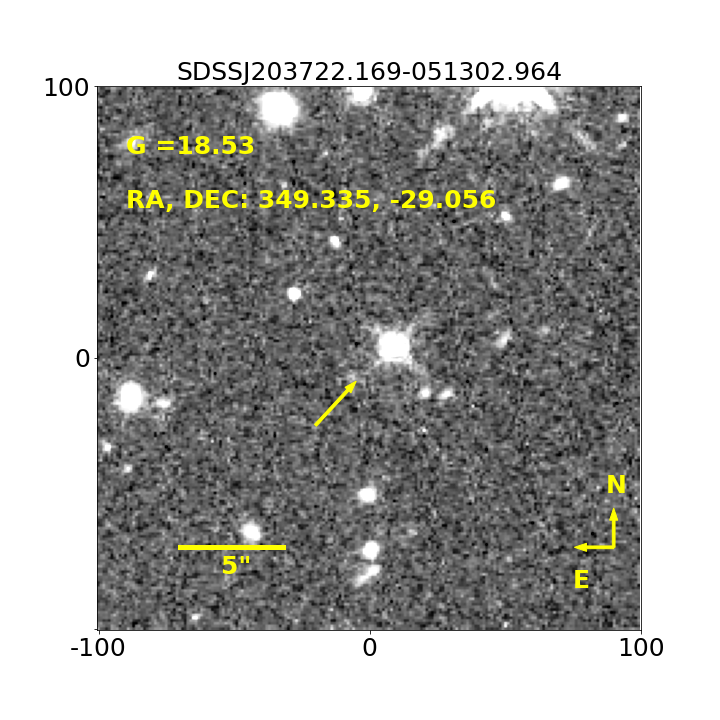} 
\end{minipage}
\begin{minipage}{0.24\textwidth}
\includegraphics[height=0.2\textheight,width=1.15\textwidth]{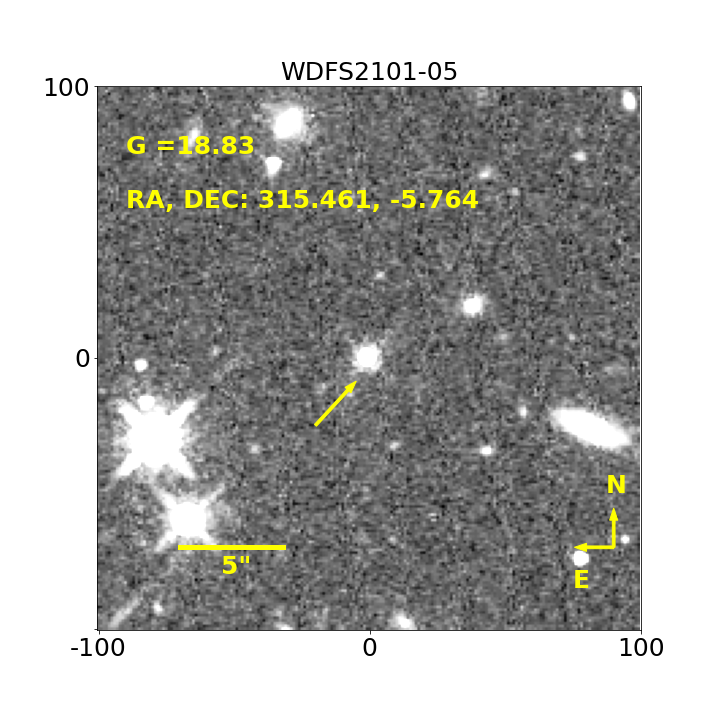} 
\end{minipage}
\begin{minipage}{0.24\textwidth}
\includegraphics[height=0.2\textheight,width=1.15\textwidth]{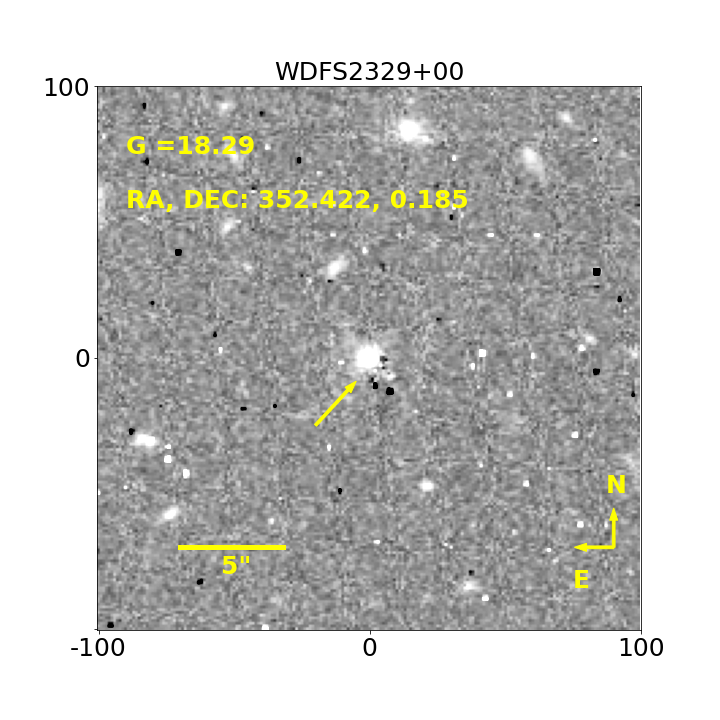} 
\end{minipage}
\begin{minipage}{0.24\textwidth}
\includegraphics[height=0.2\textheight,width=1.15\textwidth]{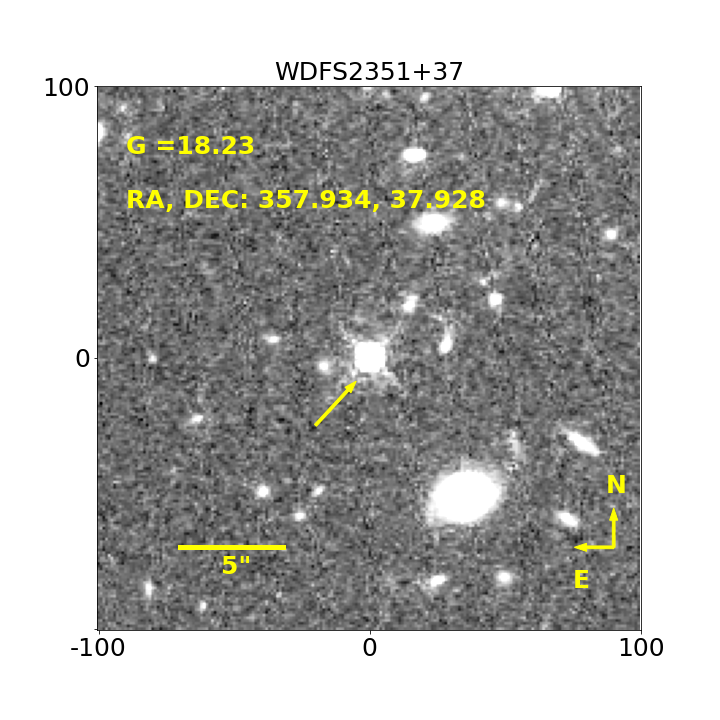} 
\end{minipage}
\newline
\caption{Finding charts for 7 candidate DAWDs based on WFC3 drizzled $F160W$ image cut-outs centered on the star and covering a FoV of 26\arcsec$\times$26\arcsec. Star \emph{Gaia} DR3 magnitude and coordinates are labeled on the image. 
The yellow solid arrow points at the DAWD and the yellow line indicates 5\arcsec on the image. 
The North and East directions are also shown. \label{fig:chart_north2}}
\end{figure*}

\begin{figure*}[!h]
\begin{minipage}{0.24\textwidth}
\includegraphics[height=0.2\textheight,width=1.15\textwidth]{WDFS0103-00_cutout_NorthUp.png} 
\end{minipage}
\begin{minipage}{0.24\textwidth}
\includegraphics[height=0.2\textheight,width=1.15\textwidth]{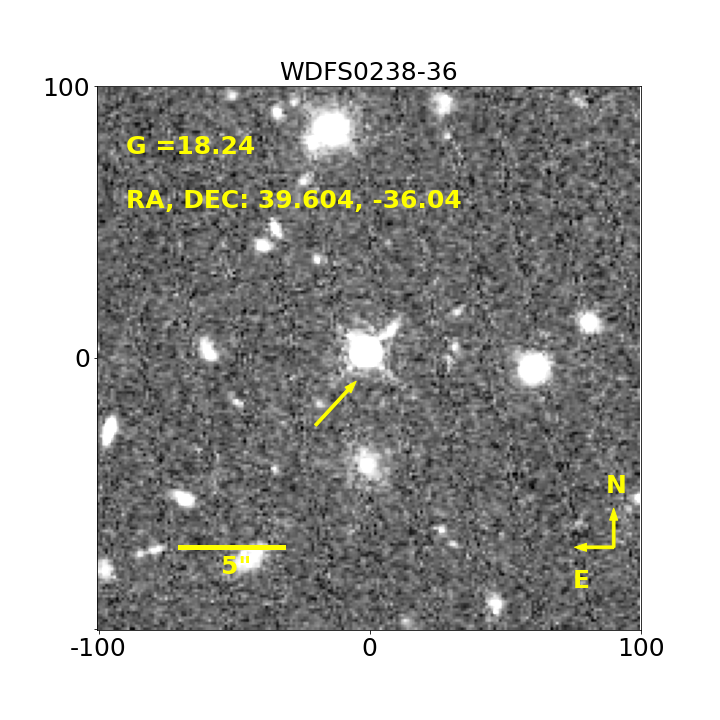} 
\end{minipage}
\begin{minipage}{0.24\textwidth}
\includegraphics[height=0.2\textheight,width=1.15\textwidth]{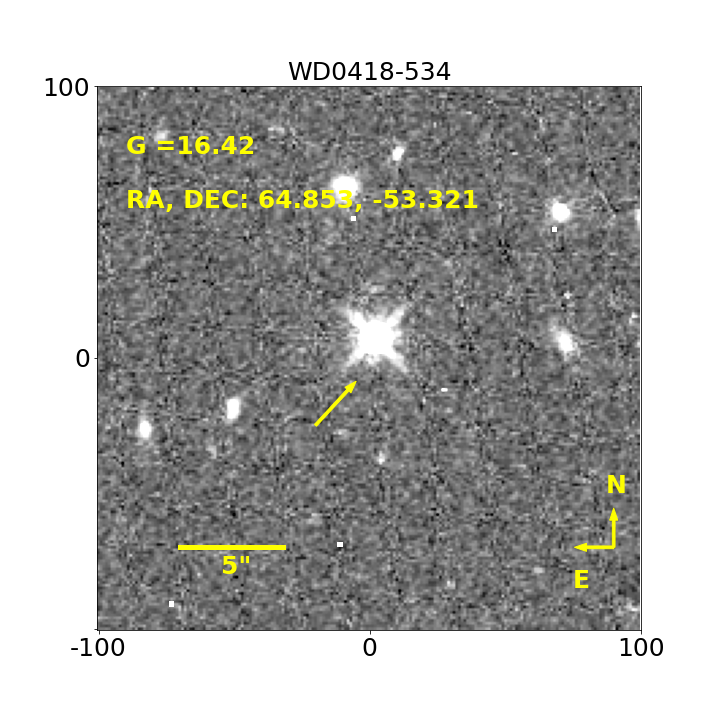} 
\end{minipage}
\begin{minipage}{0.24\textwidth}
\includegraphics[height=0.2\textheight,width=1.15\textwidth]{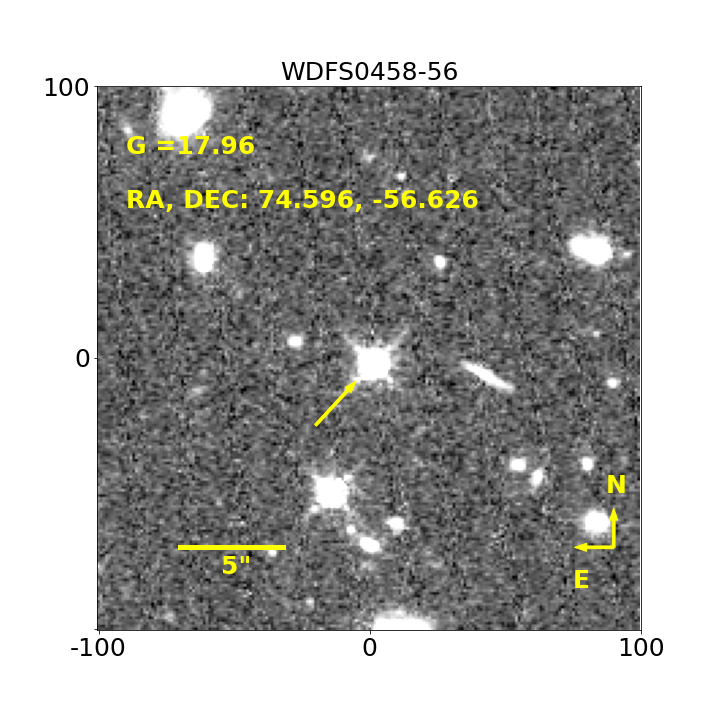} 
\end{minipage}
\begin{minipage}{0.24\textwidth}
\includegraphics[height=0.2\textheight,width=1.15\textwidth]{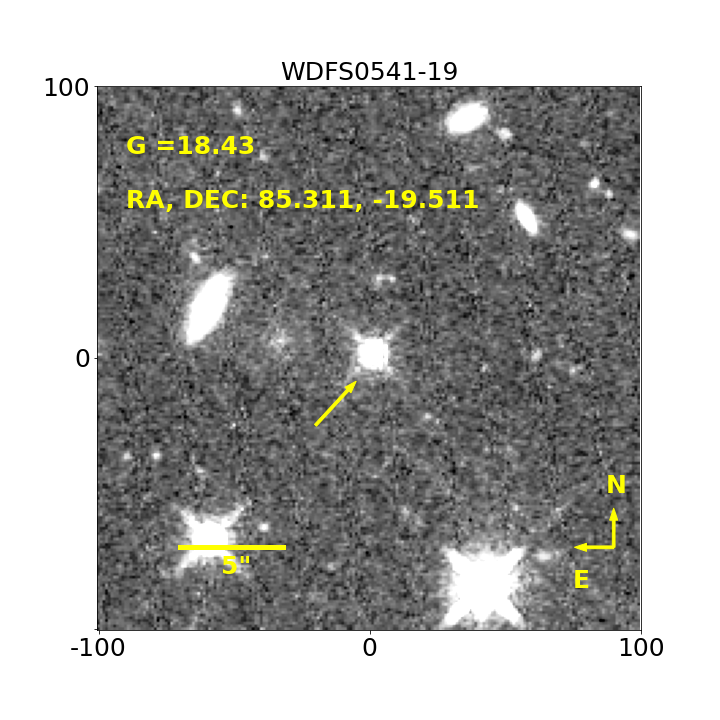} 
\end{minipage}
\begin{minipage}{0.24\textwidth}
\includegraphics[height=0.2\textheight,width=1.15\textwidth]{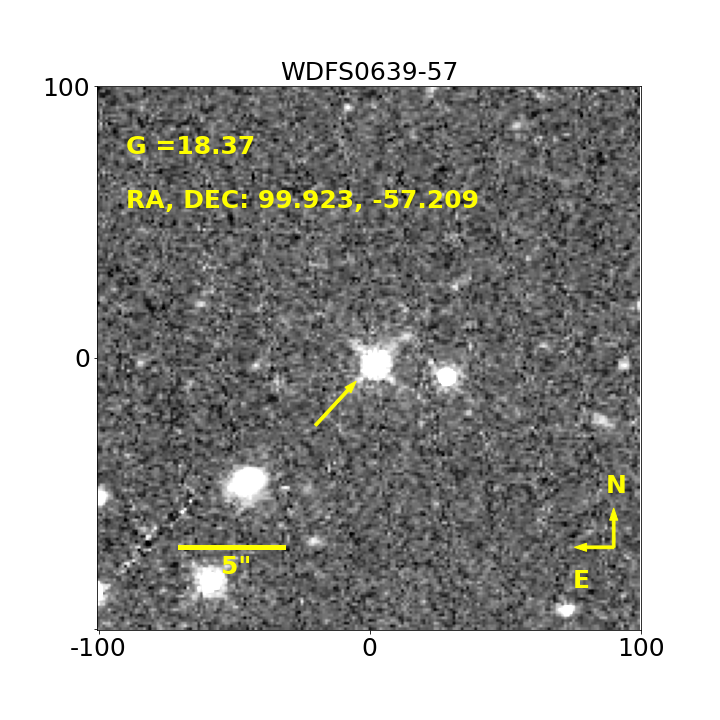} 
\end{minipage}
\begin{minipage}{0.24\textwidth}
\includegraphics[height=0.2\textheight,width=1.15\textwidth]{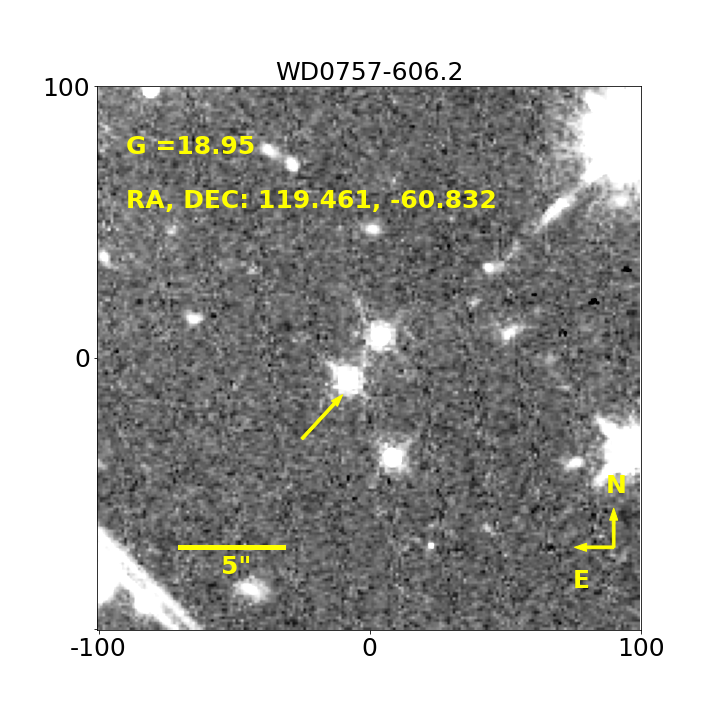} 
\end{minipage}
\begin{minipage}{0.24\textwidth}
\includegraphics[height=0.2\textheight,width=1.15\textwidth]{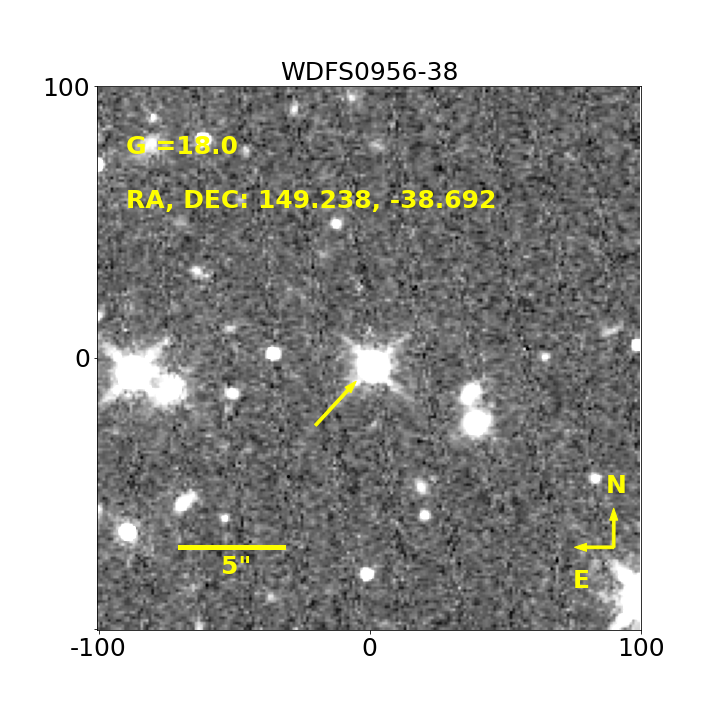} 
\end{minipage}
\begin{minipage}{0.24\textwidth}
\includegraphics[height=0.2\textheight,width=1.15\textwidth]{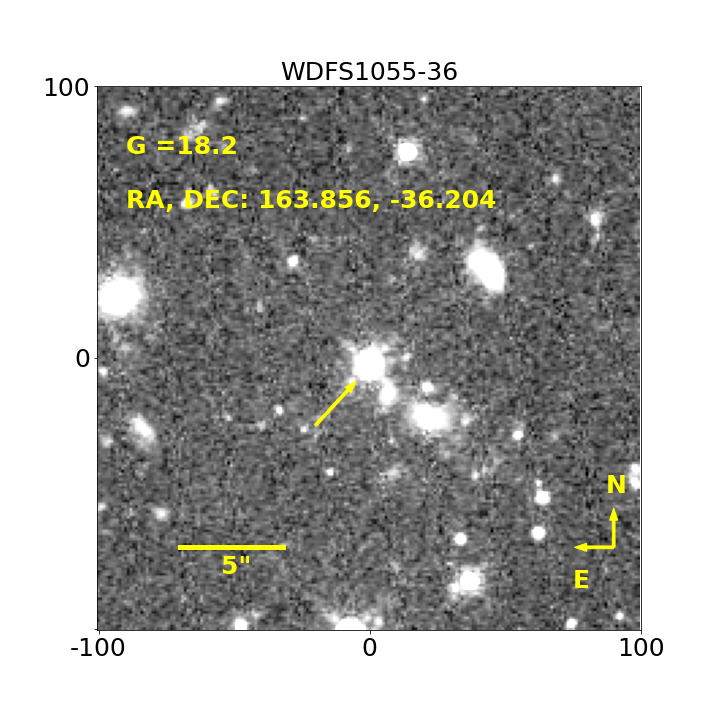} 
\end{minipage}
\begin{minipage}{0.24\textwidth}
\includegraphics[height=0.2\textheight,width=1.15\textwidth]{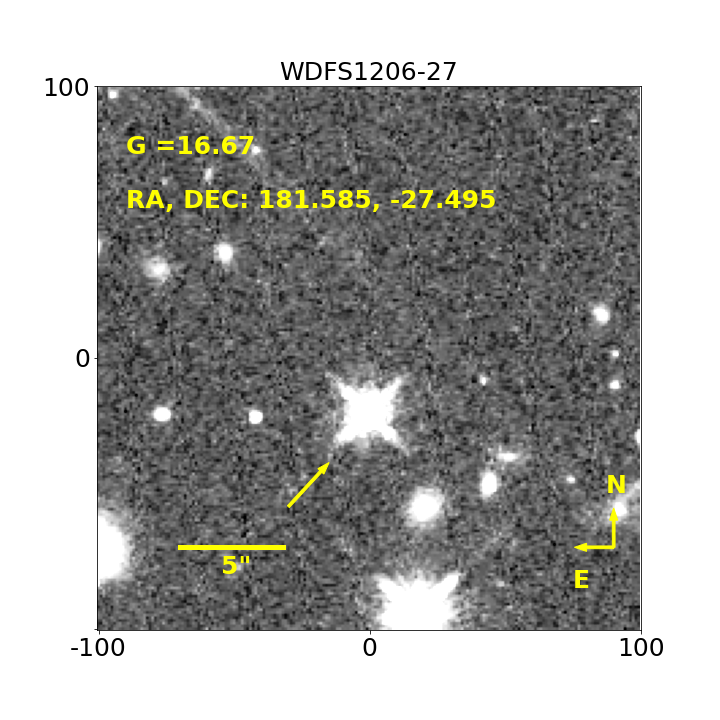} 
\end{minipage}
\begin{minipage}{0.24\textwidth}
\includegraphics[height=0.2\textheight,width=1.15\textwidth]{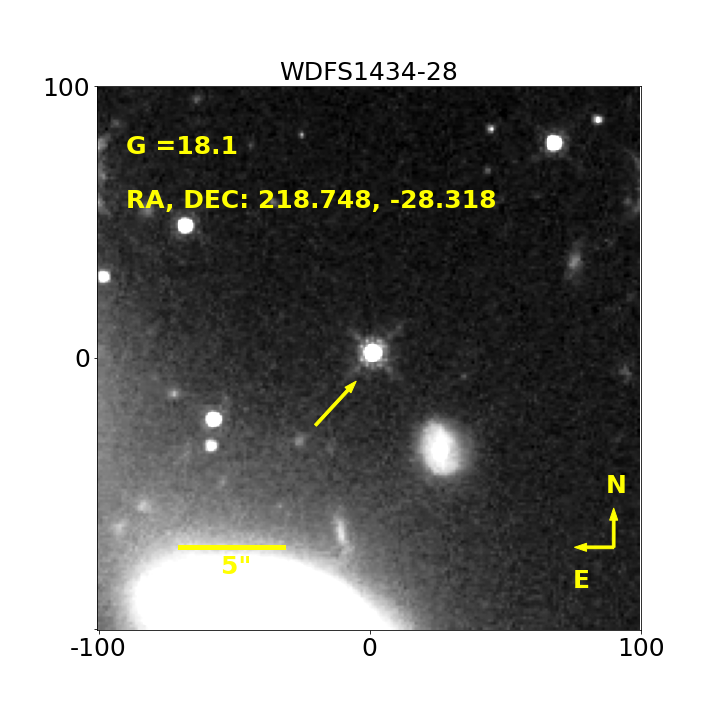} 
\end{minipage}
\begin{minipage}{0.24\textwidth}
\includegraphics[height=0.2\textheight,width=1.15\textwidth]{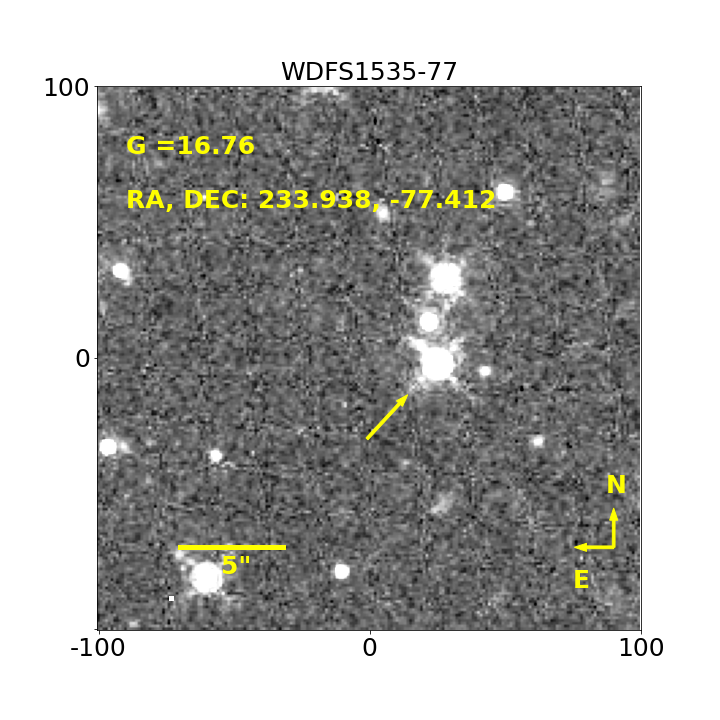} 
\end{minipage}
\begin{minipage}{0.24\textwidth}
\includegraphics[height=0.2\textheight,width=1.15\textwidth]{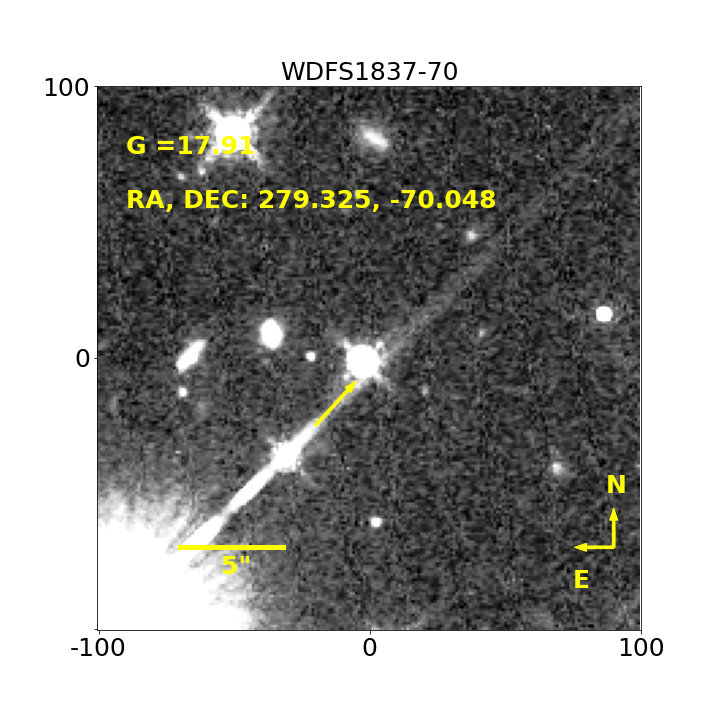} 
\end{minipage}
\begin{minipage}{0.24\textwidth}
\includegraphics[height=0.2\textheight,width=1.15\textwidth]{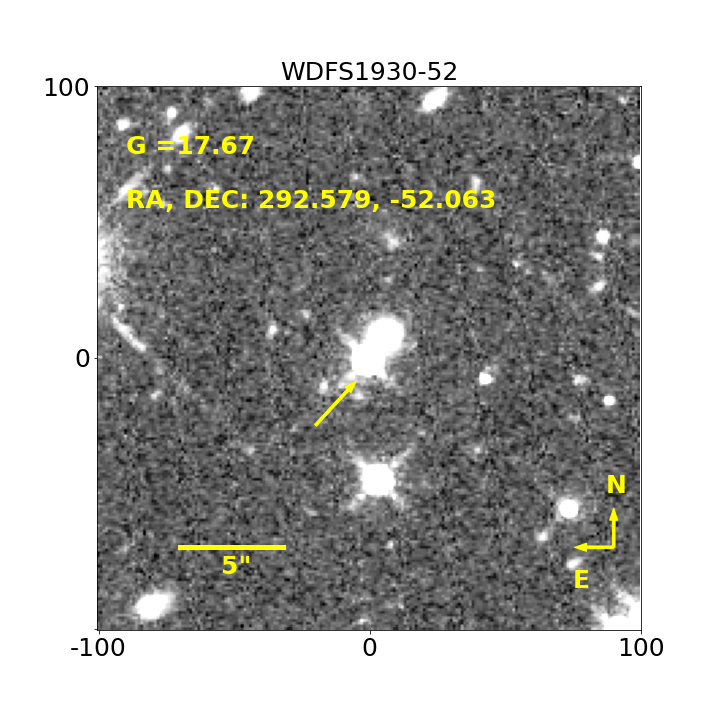} 
\end{minipage}
\begin{minipage}{0.24\textwidth}
\includegraphics[height=0.2\textheight,width=1.15\textwidth]{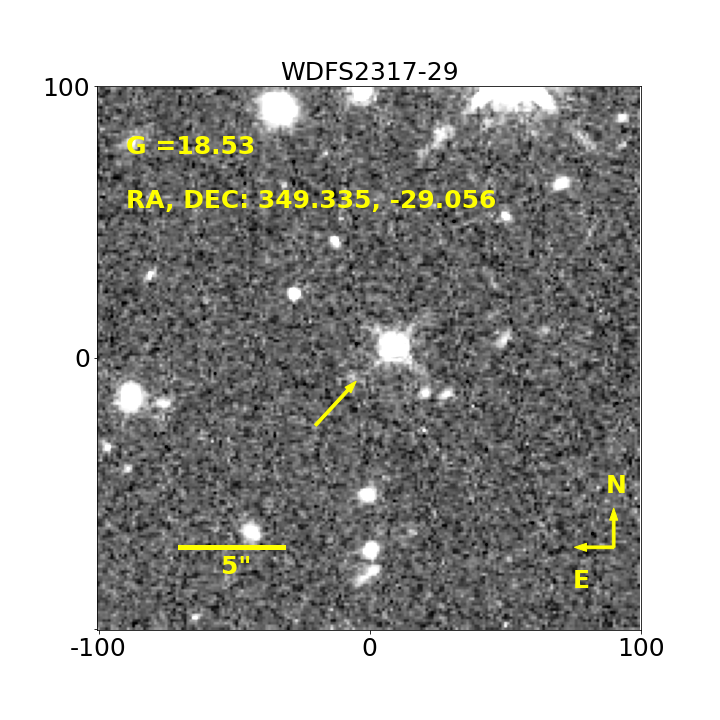}
\end{minipage}
\newline
\caption{Finding charts for 15 candidate DAWDs based on WFC3 drizzled $F160W$ image cut-outs centered on the star and covering a FoV of 26\arcsec$\times$26\arcsec. Star \emph{Gaia} DR3 magnitude and coordinates are labeled on the image. 
The yellow solid arrow points at the DAWD and the yellow line indicates 5\arcsec on the image. 
The North and East directions are also shown. \label{fig:chart_south}}
\end{figure*}

\end{document}